\documentclass[longauth]{aa}
\usepackage[]{graphicx}
\usepackage{natbib}
\usepackage[varg]{txfonts}
\usepackage{enumerate}
\usepackage{float}
\usepackage{xspace}

\usepackage{balance}
\usepackage{upgreek}
\usepackage{array}
\hypersetup{ colorlinks=true,  linkcolor=red, citecolor=blue, urlcolor=blue}

\def\upto{\!-\!}

\begin{document}

\title{The COSMOS2015 galaxy stellar mass function
  \thanks{Based on data products from observations made with ESO Telescopes at the La
Silla Paranal Observatory under ESO programme ID 179.A-2005 and on 
data products produced by TERAPIX and the Cambridge Astronomy Survey Unit on behalf
of the UltraVISTA consortium (\url{http://ultravista.org/}). 
Based on data produced by the SPLASH team from observations made 
with the Spitzer Space Telescope (\url{http://splash.caltech.edu}).}  
  }
\subtitle{13 billion years of stellar mass assembly in 10 snapshots}
  
\author{I.~Davidzon\inst{1,2}
\and O.~Ilbert\inst{1}
\and C.~Laigle\inst{3}
\and J.~Coupon\inst{4}
\and H.~J.~McCracken\inst{5}
\and I.~Delvecchio\inst{6}
\and D.~Masters\inst{7}
\and P.~L.~Capak\inst{7}
\and B.~C.~Hsieh\inst{8}
\and O.~Le F\`{e}vre\inst{1}
\and L.~Tresse\inst{9}
\and M.~Bethermin\inst{1}
\and Y.-Y.~Chang\inst{8,10}
\and A.~L.~Faisst\inst{7}
\and E.~Le Floc'h\inst{10}
\and C.~Steinhardt\inst{11} 
\and S.~Toft\inst{11}
\and H.~Aussel\inst{10}
\and C.~Dubois\inst{1}
\and G.~Hasinger\inst{12} 
\and M.~Salvato\inst{13} 
\and D.~B.~Sanders\inst{12} 
\and N.~Scoville\inst{14} 
\and J.~D.~Silverman\inst{15} 
}

\institute{
Aix Marseille Univ, CNRS, LAM, Laboratoire d'Astrophysique de Marseille, Marseille, France  
\and INAF - Osservatorio Astronomico di Bologna, via Ranzani 1, I-40127, Bologna, Italy 
\and Sub-department of Astrophysics, Department of Physics, University of Oxford, Denys Wilkinson Building, Keble Road, Oxford OX1 3RH, UK 0000-0002-1283-8420 
\and Department of Astronomy, University of Geneva, ch.~d'Ecogia 16, 1290 Versoix, Switzerland 
\and Sorbonne Universit\'{e}s, UPMC Univ Paris 6 and CNRS, UMR 7095, Institut d'Astrophysique de Paris, 98 bis bd Arago, 75014 Paris, France 
\and Department of Physics, University of Zagreb, Bijeni\v{c}ka cesta 32, HR-10002 Zagreb, Croatia 
\and Infrared Processing and Analysis Center, California Institute of Technology, Pasadena, CA 91125, USA 
\and Academia Sinica Institute of Astronomy and Astrophysics, P.O. Box 23-141, Taipei 10617, Taiwan, Republic of China 
\and Centre de Recherche Astrophysique de Lyon, Universit\'{e} de Lyon, Université Lyon 1, CNRS, Observatoire de Lyon; 9 avenue Charles André, F-69561 Saint-Genis Laval Cedex, France 
\and Laboratoire AIM Paris-Saclay, UMR 7158, CEA, CNRS, Université Paris VII, CE-SACLAY, Bat 709, F91191 Gif-sur-Yvette,  France 
\and Dark Cosmology Centre, Niels Bohr Institute, Copenhagen University, Juliane Maries Vej 30, DK-2100 Copenhagen O, Denmark 
\and Institute for Astronomy, 2680 Woodlawn Drive Honolulu, HI 96822-1839 USA 
\and Max-Planck-Institut für extraterrestrische Physik, Giessenbachstrasse, D-85748 Garching, Germany 0000-0001-7116-9303 
\and Cahill Center for Astronomy and Astrophysics, California Institute of Technology, Pasadena, CA 91125, USA 
\and Kavli Institute for the Physics and Mathematics of the universe (WPI), Todai Institutes for Advanced Study, The University of Tokyo, Kashiwa, Chiba 277-8583, Japan 
}
\date{January 2017}

\offprints{iary.davidzon@lam.fr}

\abstract{  
  We measure the stellar mass function (SMF) and stellar mass density  of galaxies in the COSMOS field up to $z\sim6$. We 
  select them in the near-IR bands of the  COSMOS2015 catalogue,  
  which includes ultra-deep photometry from UltraVISTA-DR2, SPLASH, and Subaru/Hyper Suprime-Cam. 
  At $z>2.5$ we use new precise photometric redshifts with error $\sigma_z=0.03(1+z)$ 
  and an outlier fraction of $12\%$,   
  estimated by means of the unique spectroscopic sample of COSMOS 
  ($\sim\!100\,000$ spectroscopic measurements in total, more than thousand with robust $z_\mathrm{spec}>2.5$). 
  The increased exposure time in the DR2, along with our panchromatic detection strategy,     
  allow us to improve the completeness at high $z$ with respect to previous UltraVISTA catalogues 
  (e.g., our sample is  $>\!75\%$ complete  at $10^{10}\,\mathcal{M}_\odot$ and $z=5$). 
  We also identify passive galaxies through a robust colour-colour selection, 
  extending  their SMF estimate up to $z=4$. 
  Our work provides a comprehensive view of galaxy stellar mass assembly between $z=0.1$ and 6, 
  for the first time using consistent estimates across the entire redshift range. 
  We fit these measurements with a Schechter function, correcting for Eddington bias. 
  We compare the SMF fit with the  halo mass function predicted from  $\Lambda$CDM 
  simulations, finding that at $z>3$ both functions 
  decline with a similar slope in the high-mass end. This feature could be explained assuming that    
  mechanisms quenching star formation in massive haloes 
  become less effective at high redshifts;  
  however further work needs to be done to confirm this scenario.    
  Concerning the SMF low-mass end, it shows a progressive steepening as moving towards higher redshifts, with 
  $\alpha$ decreasing from $-1.47_{-0.02}^{+0.02}$ at $z\simeq0.1$ to $-2.11_{-0.13}^{+0.30}$ at $z\simeq5$. 
  This slope depends on the characterisation of the  
  observational uncertainties,  
  which is crucial to properly remove the Eddington bias. 
  We show that   there is currently no consensus on the method to 
  quantify such errors: 
  different error models result in different best-fit Schechter
   parameters. 
}


\maketitle

\section{Introduction}

  In recent years, improvements in observational techniques and new facilities 
  have allowed us to capture images of the early universe when it was only a few  
  billion years old. 
  The Hubble Space Telescope (HST) has now provided   
   samples of high-$z$ ($z\gtrsim3$) galaxies  selected in stellar mass 
   \citep[][]{Koekemoer2011,Grogin2011,Illingworth2013}, which represent  
   a breakthrough similar to the advent of spectroscopic surveys 
   at $z\sim1$ more than a decade ago \citep{Cimatti2002a,Davis2003,Grazian2006a}. 
   Indeed, they have a statistical power comparable to those pioneering studies, 
   allowing for the same fundamental analyses such as 
   the estimate of the observed galaxy stellar mass  function (SMF). 
   This statistical tool, providing a description of stellar mass 
   assembly at a given epoch,  plays a pivotal role 
   in studying galaxy evolution. 
   
   We can distinguish between different modes of 
   galaxy growth e.g.~by comparing the  SMF of galaxies 
   divided by morphological types or environments   
   \citep[e.g.][]{Bolzonella2010,Vulcani2011,Mortlock2015,Moffett2016,Davidzon2016}.  
   Moreover, their rate of stellar mass accretion changes as a function of $z$, 
    being more vigorous at earlier epochs \citep[e.g.][]{Tasca2015,Faisst2016a}.
   Thus, the  SMF can give an overview of the whole galaxy population, at least down to the limit of  
   stellar mass completeness, over cosmic time.  Although 
   more difficult to compute than the luminosity function (LF) 
   the SMF is more closely related to the star formation history of the universe, 
   with the integral of the latter equalling the stellar mass density after accounting for  
   mass loss  \citep{Arnouts2007,Wilkins2008,Ilbert2013,Madau2014}. 
   Moreover such a direct link to star formation rate (SFR) 
   makes  the observed  SMF 
   a basic comparison point for galaxy formation models.  
   Both semi-analytical   and  hydrodynamical 
   simulations  are often  \citep[but not always, see e.g.][]{Dubois2014} 
   calibrated against the local SMF \citep[e.g.][]{Guo2011,Guo2013,Genel2014,Schaye2016}; 
   measurements at $z>0$ are then used to test theoretical predictions 
   \citep[][and many others]{Torrey2014,Furlong2015}.

   Deep HST surveys probe relatively small areas, 
   resulting in sample variance significantly greater than 
   ground-based observations conducted over larger fields  \citep[][]{Trenti&Stiavelli2008,Moster2011}. 
   Therefore it is difficult for them 
   to make measurements at low-intermediate redshifts ($z\lesssim2$) 
   where the corresponding volume is smaller.  
   As a consequence,  the literature lacks  mass functions 
   consistently measured from the local to the early universe. 
   Such a coherent set of estimates would facilitate those studies 
   probing a wide redshift range 
   \citep[e.g.][]{Moster2013,Henriques2015,Volonteri2015}, which at present 
   have to combine miscellaneous datasets. 
   
   To get a continuous view of galaxies' history, one has to combine  
   low-$z$ estimates  \citep[e.g.][]{Fontana2004,Fontana2006,Pozzetti2010,Ilbert2010} 
   with SMFs derived at $z>3$ 
   \citep[e.g.][]{McLure2009,Caputi2011,Santini2012}.
   Unfortunately linking them is not an easy task.
   In particular, samples at different redshift are built with heterogeneous 
   photometry and  selection effects. 
   For instance at high-$z$, instead  
   of using  photometric redshifts, the widespread 
   approach is based on the ``drop-out''  technique that 
   selects  Lyman-break galaxies \citep[LBGs,][]{Steidel1996}. 
   Even when photometric redshifts are used across the whole 
   redshift range, differences e.g.~in the method to fit galaxies' spectral energy distribution (SED)  
   may cause systematics  in their redshift distribution, or in following 
   steps of the analysis like the evaluation of stellar mass  
   \citep[for a comparison among various SED fitting code, see][]{Conroy2013,Mitchell2013,Mobasher2015}. 
   Eventually, such inhomogeneity among 
   the joined samples can   
   produce spurious trends in the evolution of the SMF 
   \citep[see a critical assessment of SMF systematics in][]{Marchesini2009}. 
   
   Tackling these limitations is one of the main goal of the  UltraVISTA survey 
   \citep{McCracken2012} and the Spitzer Large Area Survey with Hyper Suprime-Cam 
   \citep[SPLASH,][]{CapakSPLASH}. These surveys cover the 2 square degrees of the COSMOS
   field \citep{Scoville2007} in near and medium IR respectively 
   (hereafter, NIR and MIR). With them,  
   our collaboration built a catalogue of galaxies (COSMOS2015) 
   from  $z=0$ to $6$. 
   The COSMOS2015 catalogue has been presented in \citet{Laigle2016}, 
   where we showed the gain in terms of large-number statistics (due to the 
   large volume probed) and  improved depth (reaching  $K_\mathrm{s}=24.7$ and 
   $[3.6\mu\mathrm{m}]=25.5$, at $3\sigma$ in 3\arcsec diameter aperture).  The deeper exposure  translates 
   in a higher completeness of the sample down to lower stellar masses with respect to 
   previous versions of the catalogue. 
    
   In this paper we exploit the COSMOS2015 catalogue (together with 
   exquisite ancillary data available in COSMOS) to derive a galaxy SMF   
   up to $z\sim6$, i.e.~when the universe was about 1\,Gyr old. 
   Following galaxy mass assembly across such a large time-span allows to identify crucial  stages 
   of galaxies' life, from the reionization era \citep[see][]{Robertson2015}, through 
   the ``cosmic noon'' at $z\sim2$ \citep{Madau2014},  
   until more recent epochs when many galaxies  
   have become red and dead  \citep[e.g.][]{Faber2007}. 	
   We aim at  juxtaposing these key moments to get a global picture,  
   also separating  populations of active (i.e., star forming) and  
   passive (quiescent) galaxies.

   We organise our work as follows. 
   First, we describe the COSMOS2015 catalogue and the other datasets 
   we use, with particular attention to sample completeness  (Sect.~\ref{Dataset}). 
   At $z\leqslant2.5$  we rely on the original SED fitting estimates from  \citet{Laigle2016}, 
   while at higher $z$ we recompute photometric redshifts ($z_\mathrm{phot}$, Sect.~\ref{SED fitting}) 
   and stellar masses ($\mathcal{M}$, Sect.~\ref{SED fitting 2})  with an updated  SED fitting set-up optimised 
   for the $3\lesssim z\lesssim6$ range. 
   Since the novelty of this work is the analysis between $z=2.5$ and 6  
   we present in Sect.~\ref{Results}  the SMFs at $z>2.5$, while 
   those at lower redshifts (directly derived from L16) can be found in the Appendix.   
   The evolution of the SMF  in the full redshift range, from $z\sim6$ down to 0.2, is 
   then discussed in Sect.~\ref{Discussion}. 
    Eventually, we summarise our work in Sect.~\ref{Conclusions}. 
   
   Throughout this paper we assume a flat $\Lambda$CDM cosmology with 
    $\Omega_\mathrm{m}=0.3$, $\Omega_\mathrm{\Lambda}=0.7$, 
    and $h_{70}\equiv H_0/(70\,\mathrm{km\,s^{-1}\,Mpc^{-1}})=1$. 
   Galaxy stellar masses, when derived from SED fitting, scale as the square of the luminosity distance; 
   therefore, 
   there is a factor $h_{70}^{-2}$ kept implicit  throughout this paper 
   \citep[see][for an overview on cosmology conversions and their conventional notation]{Croton2013}. 
   Magnitudes are in the AB system \citep{Oke1974}.

\section{Dataset}
\label{Dataset}
  
  A description of our 
  dataset  is summarised in Sect.~\ref{Dataset-photometry}. 
   Sect.~\ref{Dataset-completeness} 
  offers a  comprehensive discussion about its  
  completeness as a function of  flux in  IRAC channel 1 (i.e., at $\sim\!3.6\,\mu m$).    
 Core of our analysis is the COSMOS2015 catalogue, recently published in \citet[][L16 in the following]{Laigle2016}; 
  other COSMOS datasets provide additional information. A complete list of the surveys carried out by the collaboration 
  can be found in the official COSMOS website.\footnote{ 
  \url{http://cosmos.astro.caltech.edu} }

\subsection{Photometry}
\label{Dataset-photometry}

\begin{figure}
\includegraphics[width=0.99\columnwidth]{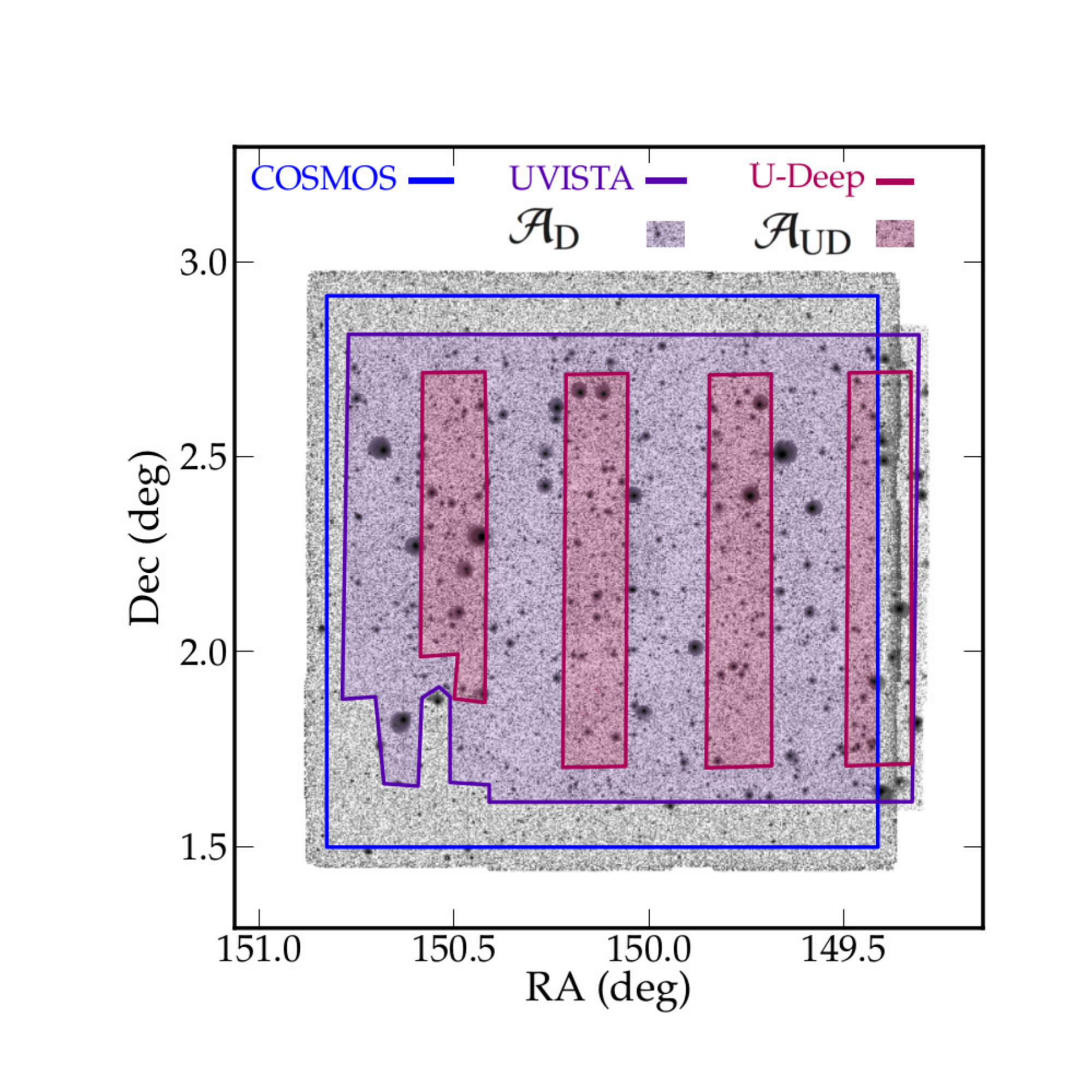}
\caption{Layout of the COSMOS field. The image in  background  is  
the $\chi^2$-stacked $zYJHK_\mathrm{s}$ image. The COSMOS 2\,deg$^2$ field is 
enclosed by a blue line, while the UltraVISTA survey is inside the purple contour. 
UltraDeep stripes, where UltraVISTA exposure time is higher, are delimited by magenta lines.  
 A purple (magenta) shaded area shows the Deep (UltraDeep) region actually used 
 in this paper.   }
\label{fig_sky}
\end{figure}
  
  One cornerstone of 
  the COSMOS2015 photometry  
  comes from the new $Y$, $J$, $H$, and $K_\mathrm{s}$  
  images from the second data release (DR2) of the UltraVISTA survey 
  \citep{McCracken2012}, along with the $z^{++}$ band from Suprime-Cam at Subaru 
  \citep[][]{Myiazaki2012}. These images were added together 
   to build a stacked  detection image, as explained below.  
   
   The catalogue also  includes  the broadband optical filters  $u^*,B,V,r,i^+,$  
   and 14 intermediate and narrow bands. 
   In NIR, UltraVISTA is complemented by the $y$ band images 
   from the Hyper Suprime-Cam (HSC), as well as $H$ and 
   $K_\mathrm{s}$ from WIRCam (at the Canada-France-Hawaii Telescope).  
   The point-spread function (PSF) in each  band  from $u^*$ to $K_\mathrm{s}$ 
    is homogenised, so that  the fraction of flux in a $3\arcsec$ diameter aperture 
    suffers from  band-to-band seeing variations by less than 5\% (see Fig.~4 in L16).  
    Space-based facilities provided data in near-UV  \citep[from the GALEX satellite,][]{Zamojski2007} and 
    MIR  (from IRAC on board the Spitzer Space Telescope), along with 
    high-resolution optical images (ACS camera on board HST, see Sect.~\ref{SED fitting-Stellar contamination}). 
    Galaxies with an X-ray counterpart from XMM \citep{Brusa2007} or Chandra \citep{Marchesi2016a}   
    are excluded from the following analysis as their photometric redshifts, or the stellar mass 
    estimates, would be likely corrupted by contamination of their active galactic nuclei (AGN). 
    They represent less than 1\% of the whole galaxy sample. 
   The entire photometric baseline of COSMOS2015 is summarised in Table 1 of L16. 
    
    Spitzer data represent another pillar of this catalogue, 
    probing the whole COSMOS area at 
   $3.6\upto8.0\,\mu$m, i.e.~the wavelength range where  the 
   redshifted optical spectrum of $z\gtrsim3$ galaxies is observed.   
    Such a crucial piece of information mainly 
     comes  from  SPLASH 
    but other surveys  are also included, 
    in particular the Spitzer-Cosmic Assembly Near-Infrared Deep  Extragalactic Legacy Survey 
   \citep[S-CANDELS,][]{Ashby2015}.  
   Further details about how Spitzer/IRAC photometry was extracted and harmonised with the other 
   datasets can be found in L16.

    Compared with the previous version of the catalogue \citep{Ilbert2013} the number of sources 
    doubled because of the longer exposure time of the UltraVISTA DR2 in the so-called 
    ``Ultra-Deep'' stripes (hereafter indicated with $\mathcal{A}_\mathrm{UD}$, see Fig.~\ref{fig_sky}). In that area of 0.62\,deg$^2$ 
    we reach a  $3\sigma$ limiting magnitude (in a 3\arcsec diameter aperture)
     $K_\mathrm{lim,UD}=24.7$, while in  the remaining 
    ``Deep'' area (dubbed $\mathcal{A}_\mathrm{D}=1.08\,\mathrm{deg}^2$)  the limit is $K_\mathrm{lim,D}=24.0$. 
    The larger number of detected sources in DR2 is also due to the new $\chi^2$ stacked image produced in L16. 
    Image stacking is a panchromatic approach for identification of galaxy sources, presented for the first time 
    in \citet{Szalay1999}. With respect to the 
    previous UltraVISTA (DR1) stacking, in L16 
    we   co-add not only NIR images but also  the  deeper $z^{++}$ band, 
    using the code \texttt{SWarp} \citep{Bertin2002}. Pixels in the resulting  
    $zYJHK$ image are the weighted mean of the flux in each stacked filter. 
    As a result,  
    the catalogue contains $\sim\!6\times10^5$ objects within 1.5 deg$^2$, 
     $190\,650$ of them in $\mathcal{A}_\mathrm{UD}$. 
    In L16 we also show the good agreement  
    of colour distributions, number counts, and clustering with other state-of-the-art surveys.

  For each entry of the COSMOS2015 catalogue we search for a counterpart 
  in the four Spitzer-IRAC channels using the code \texttt{IRACLEAN}   
  \citep{Hsieh2012}.\footnote{The wavelength range of the four channels is centred respectively at 
   $3.6$, $4.5$, $5.8$, and $8.0\,\mu$m; in the following we refer to them  
   as $[3.6]$, $[4.5]$, $[5.8]$, and $[8.0]$.} 
   The procedure is detailed in L16.   
   In brief, positional and morphological information in the $zYJHK_\mathrm{s}$ detection image 
   is used as a prior to identify IRAC sources and recover their total flux. 
   In this latest version, 
   \texttt{IRACLEAN} produces  a weighing scheme  from the surface brightness 
   of the prior, to correctly deblend  objects that are 
   near to each other less than $\sim1$ FWHM of the IRAC PSF.  
   For each  source 
   a flux error is  estimated by means of the residual map, 
   i.e.~the IRAC image obtained after subtracting the 
   flux  associated  to detections.

  \subsection{Flux limits and sample completeness at high redshift}
  \label{Dataset-completeness} 
  
     \begin{figure}
    \includegraphics[width=0.99\columnwidth]{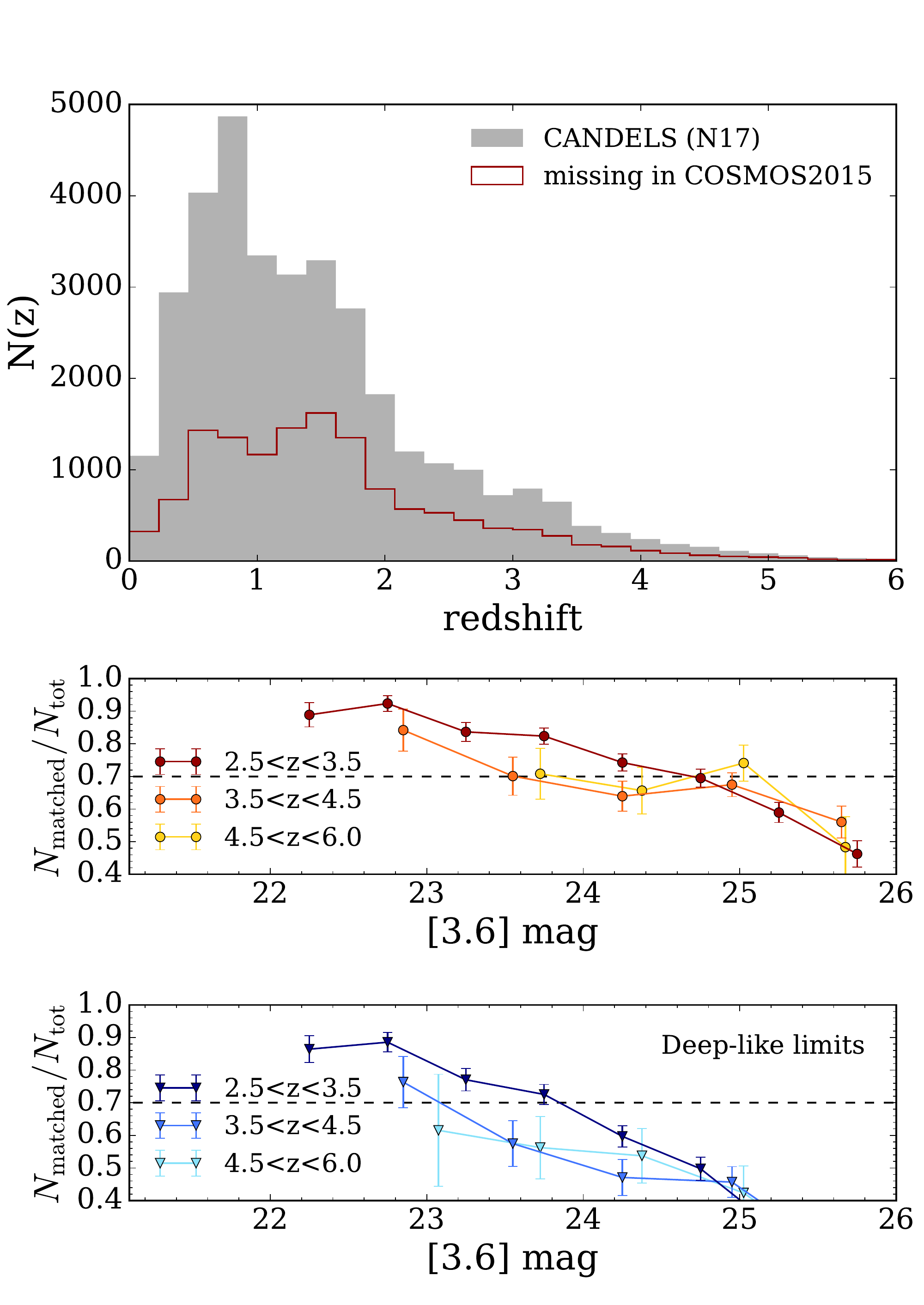}
   \caption{\textit{Upper panel:} Redshift distribution of the whole CANDELS 
   sample in the COSMOS field,  taken from  \citet[][N17, gray filled  
   histogram]{Nayyeri2017}.  We also identify the N17 objects that 
   do not have a counterpart (within $0.8\arcsec$) 
   in the COSMOS2015 catalogue, 
   showing their $N(z)$ with a red histogram. 
   \textit{Middle panel:} ratio between the CANDELS 
   objects with a match in COSMOS2015 
   ($N_\mathrm{matched}$)   and 
   all the CANDELS entries ($N_\mathrm{tot}$) 
   in bins of $[3.6]$ magnitude  (filled circles). These  
   estimates  are divided into three redshift bins 
   in the range $2.5< z_\mathrm{phot,N17}<6$ (see colour code in the legend); 
   a dashed line marks the 
   70\% completeness. \textit{Lower panel:} Similar to the 
   middle panel, but the $N_\mathrm{matched}/N_\mathrm{tot}$ ratio 
    is estimated to reproduce the sensitivity depth 
   of UltraVISTA-Deep.   }
   \label{fig_compl-cand2}
   \end{figure}
  
   We aim to work with a flux-limited sample 
   to restrict the analysis 
   to a  sample sufficiently complete with   
   photometric errors  sufficiently small. 
   In L16 the completeness as a function of stellar mass 
   has been derived in bins of  $K_\mathrm{s}$ magnitudes, but this choice is not 
    suitable for the present analysis, which extends to $z\sim6$.  
   Up to $z\sim4$, a $K_\mathrm{s}$-band selection is commonly used to derive 
   a completeness limit in stellar  mass
   \citep[e.g.][]{Ilbert2013,Muzzin2013b,Tomczak2014}, but at higher redshifts  
    this filter probes a rest-frame range of the galaxy spectrum particularly sensitive to recent star formation. 
   Indeed, the Balmer break moves to wavelengths 
   larger than $2\,\mu$m and  
   most of the stellar light coming from K- and M-class stars 
   is observed in the IRAC channels. This 
   makes  a $[3.6]$ selection suitable at $z>4$. 
   In this paper we will apply a 
   selection in $K_\mathrm{s}$ or $[3.6]$ depending on the redshift, always choosing the most  
   direct link between stellar light and mass.  
   Moreover, we will show in Appendix \ref{APP3}  
   that even between $z\sim 2$ and 4, where in principle both bands 
   can be used, a cut in $[3.6]$ is recommended.   
   Thus, for our analysis at $2.5<z<6$, we will extract from the parent catalogue 
   a sample of  galaxies with magnitude $[3.6]<[3.6]_\mathrm{lim}$.

    Determining $[3.6]_\mathrm{lim}$ is not as straightforward 
    as for the $K_\mathrm{s}$ band. 
    The nominal  3$\sigma$ depth  (equal to 25.5\,mag for 
    a $3\arcsec$ diameter aperture)  has been calculated  
    by means of the rms map of the $[3.6]$ mosaic, after removing detected objects. 
    However our sources were originally found in the co-added image, so 
    the completeness of the final sample depends not only on possible issues in the IRAC photometric extraction 
    (due e.g.~to confusion noise) but also in  $zYJHK_\mathrm{s}$. 
    For instance, we expect to miss 
    red galaxies with $[3.6]\ll 25.5$ but too faint to be detected in NIR. 
    Their impact should not be underestimated, 
    given the mounting evidence of 
    strong dust extinction in  high-$z$  galaxies 
    \citep[e.g.][]{Casey2014a,Mancini2015}.   
    This is a limitation in any  analysis that uses optical/NIR images   as a prior
    to deblend IR sources \citep[e.g.][]{Ashby2013,Ashby2015}. 
    Such an approach is somehow necessary, given the lower resolution of the IRAC camera, 
    but exceptions do exist \citep[e.g.][where IR photometry is extracted directly 
    from $4.5\,\mu$m images without any prior]{Caputi2011}.

   To estimate $[3.6]_\mathrm{lim}$
   we make use of the  catalogue built by \citet[][hereafter referred as  N17]{Nayyeri2017} 
   in the $216\,\mathrm{arcmin}^2$ of the CANDELS-COSMOS field  \citep{Grogin2011,Koekemoer2011}. 
   Since CANDELS falls  entirely in our Ultra-Deep area, N17 can be used to directly constrain $[3.6]_\mathrm{lim,UD}$. 
   The authors rely on $F160W$  images ($\sim1.6\,\mu$m) from the HST/WFC3  camera 
   and  extract IRAC sources using the software  \texttt{TFIT} \citep{Papovich2001,Laidler2007}. 
   Their approach is similar to \citet{Galametz2013a}, who derive the UV-to-IR photometry 
   in another CANDELS field, overlapping with the Ultra-Deep Survey (UDS) of UKIDSS. 
   The $5\sigma$ limiting magnitude in the F160W band is 26.5, while the data from Spitzer are the same 
   as in L16. Then, we can  test the effects of a different extraction 
   algorithm and sensitivity depth of the prior.

   First, we match galaxies from COSMOS2015 and N17  
   (within a searching radius of  $0.8\arcsec$) and compare their  photometric redshift estimates 
   and $[3.6]$ magnitudes to check for  possible bias. 
   We confirm the absence of significant offsets ($[3.6]_\mathrm{L16}-[3.6]_\mathrm{N17}<0.03$\,mag) 
   as previously verified by  
   \citet{Steinhardt2014}.  
   The $[3.6]$ number counts 
   of COSMOS2015 are in excellent agreement with CANDELS,  
   for magnitudes $\lesssim\!24.5$; after restricting the comparison inside the  $\mathcal{A}_\mathrm{UD}$ region, 
   counts agree with $<\!20\%$ difference  until reaching $[3.6]=25$, where the number of UltraDeep sources starts to decline compared  to CANDELS. 
   Despite such a small fraction of missing sources, 
   our $z\gtrsim3$ statistical analysis would nonetheless suffer from severe incompleteness 
   if most of them turned out to be at high redshift. 
    For this reason,  we inspect the $z_\mathrm{phot}$ 
    distribution of   11\,761 galaxies (out of 38\,671) in N17
    not matching any COSMOS2015 entry.    
    They are extremely faint  objects  
     with  $F160W\gtrsim26$,  most of them 
    without  a counterpart  
    even in our IRAC residual maps (see below).    
 
     The CANDELS photometric redshifts ($z_\mathrm{phot,N17}$) 
     have been computed independently by several authors, 
     by means of different codes \citep[see][]{Dahlen2013}.  
     Here we use 
     the median  of those estimates, 
     which is generally in good agreement with our $z_\mathrm{phot}$  
     for the objects in common. 
     The upper panel of Fig.~\ref{fig_compl-cand2}  shows  that 
     galaxies excluded from the CANDELS-COSMOS matching   
      have a redshift distribution $N(z)$ similar to the  
     whole N17 sample. 
     Restricting the analysis to $2.5<z_\mathrm{phot,N17}\leqslant3.5$ galaxies, 
     we clearly see that the fraction of  N17 galaxies not detected in COSMOS2015 
     increases towards fainter magnitudes. 
     The same trend, despite a larger shot noise due to the small-number statistics, 
     is visible at higher redshifts. 
     Taking CANDELS as a reference ``parent sample'',  the fraction of   sources we recover is a
      proxy of the global completeness of COSMOS2015. 
      As shown in  Fig.~\ref{fig_compl-cand2} (middle panel)
     we can assume $[3.6]_\mathrm{lim,UD}=25$ 
     as a reliable $>\!70\%$ completeness limit for our $\mathcal{A}_\mathrm{UD}$ sample up to $z=6$.
     
     We can also evaluate  such a limit in $\mathcal{A}_\mathrm{D}$ ($[3.6]_\mathrm{lim,D}$)  
     although that region does not overlap CANDELS. 
     We repeat the procedure described above after applying a cut in $z^{+}$, 
     $Y$, $J$, $H$, and $K_\mathrm{s}$ bands of N17,  
      corresponding  to the $3\sigma$ limiting magnitudes in the  $\mathcal{A}_\mathrm{D}$ area. 
     The resulting threshold  is almost 1\,mag brighter than $[3.6]_\mathrm{lim,UD}$,  
     with a large scatter at $z_\mathrm{phot,N17}>3.5$ 
     (Fig.~\ref{fig_compl-cand2}, bottom panel). 
     However, we warn that such a ``mimicked'' selection is just an 
     approximation,   less efficient  than the actual $\mathcal{A}_\mathrm{D}$ extraction.\footnote{
     The Subaru $z^{+}$ band available in N17 
     is similar to $z^{++}$, albeit shallower; 
      $Y$, $J$, $H$, $K_\mathrm{s}$ photometry is taken from UltraVISTA DR1 and therefore its depth is 
      comparable to our Deep sample.  We restrict the N17 sample to be ``Deep-like''  
      by simply considering  as detected objects whose flux is above the 
      sensitivity limits in at least one of these five bands. 
      We stress that such an approach differs from the actual $\mathcal{A}_\mathrm{D}$  
      also because doing this approximation  we did not take into account the correction 
      related to PSF homogenisation. 
      }   
      With this caveat in mind, we suggest to assume $[3.6]_\mathrm{lim,D}=24$ up to 
      $z\sim4$. However, the analysis in the present paper is restricted to the $\mathcal{A}_\mathrm{UD}$ 
      sample and an accurate evaluation of $[3.6]_\mathrm{lim,D}$ is beyond 
      our goals.

     In addition, we run  \texttt{SExtractor} \citep{Bertin&Arnouts1996}  
     on the $[3.6]$ and $[4.5]$ residual maps to check 
     whether the recovered sources coincide with those in CANDELS. 
     Most of the latter ones are not found in the residual maps, because they 
     are fainter than the SPLASH background noise ($\gtrsim\!25.5$\,mag), 
     with  a low signal-to-noise ratio ($S/N<2$) that prevents us to effectively identify them 
     with \texttt{SExtractor}.  
     For the  $20\%$ CANDELS unmatched objects that are  brighter than $[3.6]_\mathrm{lim,UD}$,  
     their absence in the residual maps can be explained by blending effects:  
     if a MIR source does not correspond to any COSMOS2015 detection, 
     \texttt{IRACLEAN} may associate its  flux  to a nearby extended object. 
     This highlights the capability 
     of HST/WFC3  to correct for 
     IRAC source confusion better than our ground-based images, although 
     the deeper sensitivity we shall reach with the oncoming VISTA and HSC observations  
     should dramatically reduce the gap.\footnote{ 
    As a consequence of this blending issue, the IRAC flux of some of our bright galaxies 
    (and stars) is expected to be overestimated, but less than $40\%$    
    since  the secondary blended source  is generally  $>\!1$\,mag fainter. 
    Some of the CANDELS unmatched 
    objects may also be corrupted detections, since for this test we did not apply any pre-selection 
    using \texttt{SExtractor} quality flags. } 
    Eventually, we visually inspect 22 sources at $24<[3.6]<25$    
    recovered from the IRAC residual map but not found in the N17 sample. These 
    objects are not  resolved  in the $F160W$ image, neither in UltraVISTA. 
    They appear also in the IRAC $[4.5]$ residual map suggesting that 
    they should not be artefacts, rather a peculiar type of $3<z<5$ galaxies with 
    a prominent $D4000$ break   
    (or less probably, $z\sim12$ galaxies)  that we shall investigate in a future work.

\section{Photometric redshift and galaxy classification}
\label{SED fitting}
    
  We estimate the $z_\mathrm{phot}$ of  
  COSMOS2015 sources 
  by fitting synthetic spectral energy distributions (SEDs) 
  to their multi-wavelength photometry. 
  The COSMOS2015 catalogue already provides   
  photometric redshifts and other physical quantities (e.g., galaxy stellar masses) 
  derived  through an SED fitting procedure  detailed in L16. 
  Here we follow the same approach, 
  using the code \texttt{LePhare} \citep{Arnouts2002,Ilbert2006}  
  but  with a configuration optimised for high-$z$ galaxies   (Sect.~\ref{SED fitting-method}).

  The main reasons for a new SED fitting computation at $z>2.5$ are the following:
  \begin{itemize}
  \item[--] L16 explored the parameter space between  $z=0$ and $6$, 
  but to build an accurate PDF($z$) for galaxies close to that upper limit 
  one has to enlarge the redshift range. Therefore, we scan now a grid $z=[0,8]$ to select  
  galaxies between $z_\mathrm{phot}=2.5$ and 6.
  \item[--] We have improved the method for removing stellar interlopers, 
  which is now based on a combination of different 
  star vs galaxy classifications, with particular attention to low-mass stars 
  (see Sect.~\ref{SED fitting-Stellar contamination}). 
  \item[--] With respect to L16, we included in the  library  additional 
  high-$z$  templates, i.e.~SEDs of extremely active galaxies with rising star formation history (SFH)
  and highly attenuated galaxies with $E(B-V)>0.5$.   
  \end{itemize}    
  Our results will replace the original photometric redshifts of L16 only 
  at $z>2.5$   (Sect.~\ref{SED fitting-spec validation}). Galaxies with a new $z_\mathrm{phot}<2.5$ 
  are not considered, so below $z=2.5$ the sample is the same as in L16. In any case, 
  the variation at low $z$ is negligible, given the high percentage of galaxies 
  that preserve their original redshift.  
  A comparison between 
  the original L16 SED fitting and our new results   
  can be found in Appendix \ref{APP1}.

\subsection{Photometric redshift of $z>2.5$ galaxies}
\label{SED fitting-method}

  We fit the multi-band photometry of the entire catalogue and then select 
  galaxies with $z_\mathrm{phot}>2.5$. 
  We apply zero-point offsets in all the bands as prescribed in L16. 
  Also when $S/N<1$, we consider the flux measured in that filter 
  (and its uncertainty) without replacing it with an upper limit.   
  This choice allows us to take into account non-detections 
  without modifying the way in which the likelihood function is computed 
  \citep[whereas a different implementation is required to use upper limits, see][]{Sawicki2012}.

  Our  SED fitting library 
  includes early- and late-type galaxy templates from 
  \citet{Polletta2007}, together with 
  14 SEDs of star-forming galaxies  
  from   
  \texttt{GALAXEV} \citep[][see also Sect.~\ref{SED fitting 2-stellar mass}]{Bruzual2003}. 
  With this code we also produce templates of passive galaxies 
  at 22 different ages (from 0.5 to 13\,Gyr). These are the same 
  templates used in L16. In addition, as mentioned above, 
  we use two \texttt{GALAXEV} templates 
  that represent starburst galaxies with an increasing SFH 
 \citep[][]{Behroozi2013b,daCunha2015,Sparre2015}. 
  The age of both templates  is 100\,Myr. 
  Instead of using an exponentially increasing SFH   \citep[][]{Maraston2010} 
  we opt for a multi-component parametrisation  \citep{Stark2014}: 
  a constant SFH is superimposed to 
   a delayed $\tau$ model with $\mathrm{SFR}\propto \tau^{-2} t e^{-t/\tau}$ 
  \citep[see][]{Simha2014} 
  where the $e$-folding time $\tau$ is equal to 0.5\,Gyr 
  and $t=10\upto40$\,Myr \citep{Papovich2001,Papovich2011,Smit2014}.

  We add to each synthetic SED the principal 
  nebular emission lines: Ly$\alpha\,\lambda1216$, [OII]\,$\lambda3727$, 
  H$\beta\,\lambda4861$, [OIII]$\lambda\lambda\,4959,5007$, H$\alpha\,\lambda6563$. 
  We calibrate the lines  starting from the UV-[OII] relation 
  of \citet{Kennicutt1998}, but we let the [OII] equivalent width vary by 
  $\pm50\%$ with respect to what the equation prescribes. 
  The approach is fully empirical, with line strength ratios based on 
  \citet{Anders2003} and \citet{Moustakas2006}. 
  The addition of nebular emission lines 
  has been discussed in several studies (see Sect.~\ref{Results-Sources of uncertainty}). 
  In general, it is considered as 
  an improvement: e.g.~\citet[][]{Ilbert2009}, by including templates with emission lines, increase 
  the  $z_\mathrm{phot}$  accuracy  
  by a factor $\sim2.5$. 
  Such a gain is due to the fact that strong optical lines (like [OII] or H$\beta$-[OIII])
  can boost the measured flux 
  and alter galaxy colours \citep[e.g.][]{Labbe2013}.

        \begin{figure}[]
        \centering
   \includegraphics[width=0.78\columnwidth]{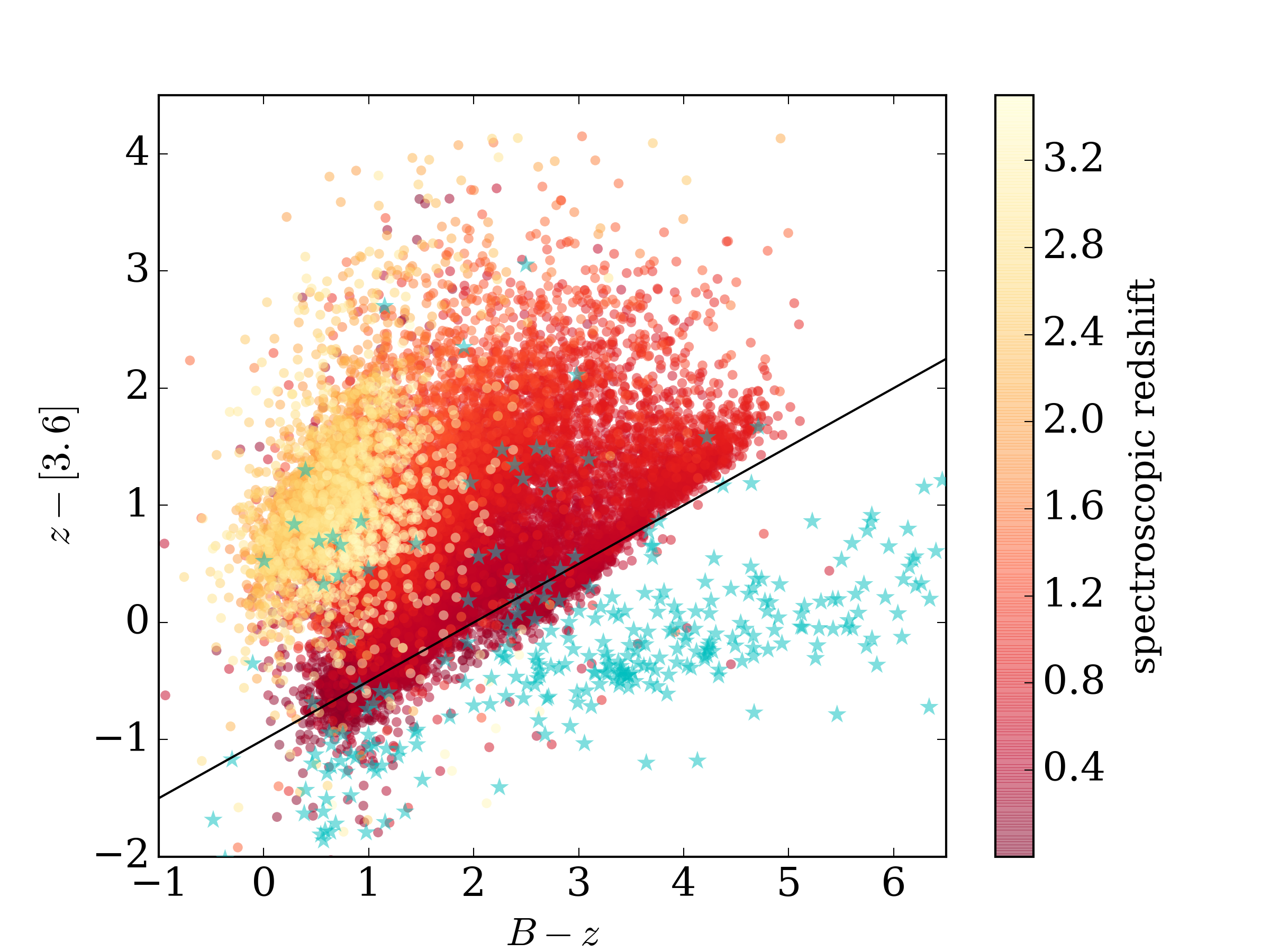} \\ 
      \includegraphics[width=0.78\columnwidth]{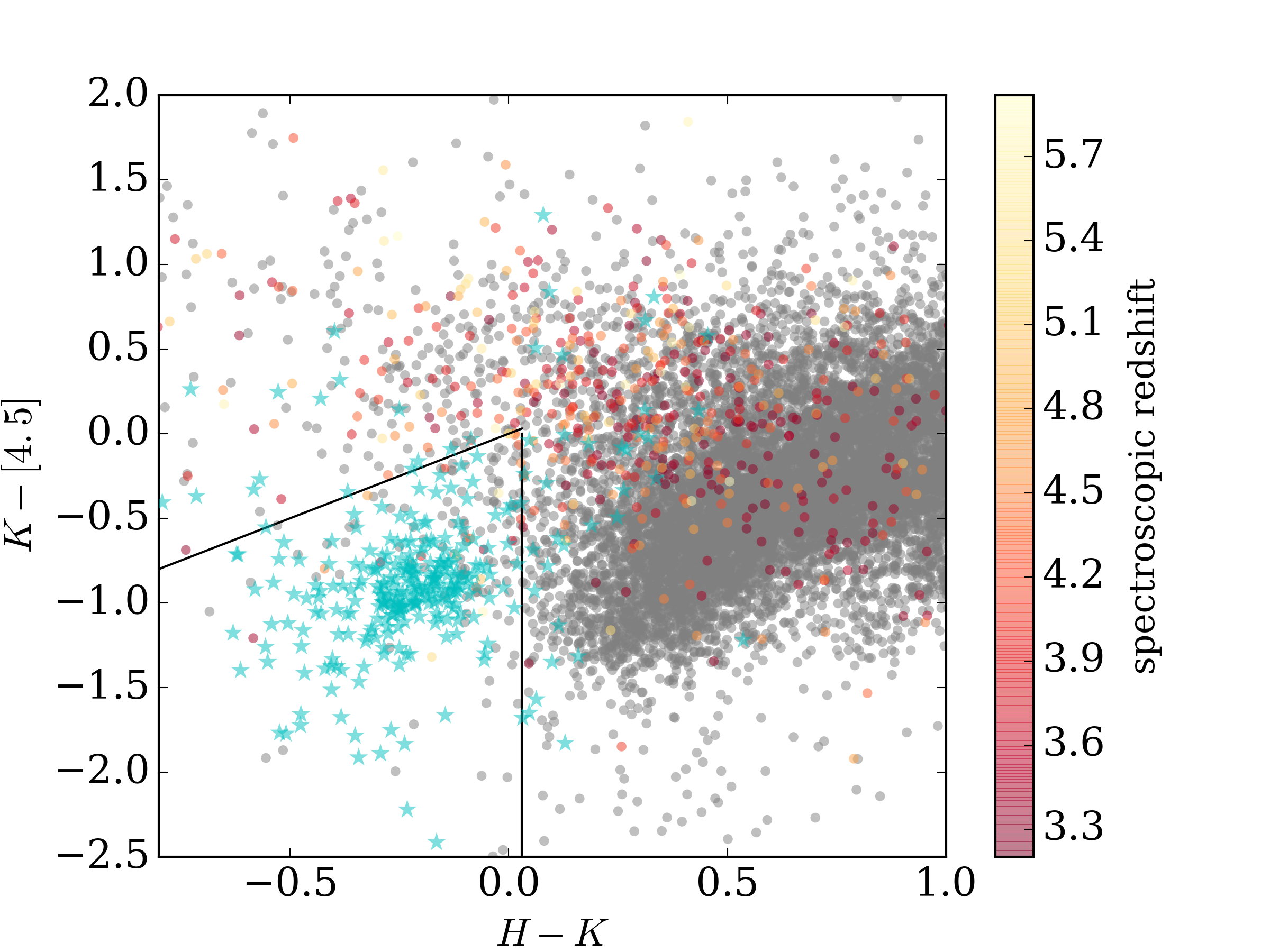} 
      \caption{Colour-colour diagrams 
      for removing stellar contaminants. Only spectroscopic measurements with quality flags 3 or 4 
      (CL $>\!95\%$) are shown in the figure. 
      \textit{Upper panel:} $(B-z^{++})$ vs $(z^{++} -[3.6])$.  Galaxies detected in the $B$ band 
      are shown with filled circles coloured from red to yellow according to their $z_\mathrm{spec}$. 
      \textit{Lower panel:} $(H-K_\mathrm{s})$ vs $(K_\mathrm{s}-[4.5])$.  
       Circles with red-to-yellow colours are $B$ drop-outs, while grey circles are the remaining spectroscopic galaxies 
       at $z\lesssim3$.  
      In both panels, solid lines delimit the 
      conservative boundaries we chose for the stellar locus. 
      These are described by the following equations:  
     $(z^{++}-[3.6])<0.5(B-z^{++})-1$ in the upper panel, and 
      $(K_\mathrm{s}-[3.6])<(H-K_\mathrm{s}) \land (H-K_\mathrm{s})<0.03$ in the lower panel.    
      Stars spectroscopically confirmed are plotted with cyan symbols.  
      Typical photometric errors are $\lesssim\!0.05$\,mag for object with $[3.6]<24$, and increase up to 
      $0.08\upto0.12$\,mag for fainter ones.   
      }
      \label{fig_hki}
   \end{figure}

  We assume for nebular emission  
  the same dust attenuation as for stars 
  \citep[see][]{Reddy2010,Kashino2013} 
  although the issue is still debated 
  \citep{ForsterSchreiber2009,Wuyts2013}.  
  Moreover, we do not implement any specific prior to control 
  the level of emission line fluxes, 
  although recent studies indicate that their equivalent width (EW) and strength ratio evolve with 
  redshift \citep[e.g.][]{Khostovan2016,Faisst2016a}. 
  Nevertheless,  
  a stronger bias in the computation is produced   
  by neglecting these lines, rather than by their rough modelling 
  \citep[see][]{Gonzalez2011,Stark2013,Wilkins2013a}.

  Attenuation by dust is implemented in the SED fitting choosing  
  among the following  extinction laws:  \citet[][SMC-like]{Prevot1984}, 
  \citet{Calzetti2000}, and two modified versions of Calzetti's law 
  that include the characteristic absorbing feature 
  at 2175\,\smash{\AA} \citep[the so-called ``graphite bump'',][]{Fitzpatrick1986} 
  with different strength. 
  The optical depth is free to vary from $E(B-V )=0$  to $0.8$, 
  to take into account  
  massive and heavily obscured galaxies   
  at $z>3$ \citep[up to $A_V\simeq3$, e.g.][]{Spitler2014}.

  Our initial sample at $2.5<z\leqslant6$ includes 92\,559 galaxies. 
  The photometric redshift assigned to each of them 
  is  the median  of the probability 
  distribution function (PDF) obtained after scanning the whole template library. 
  Hereafter, for sake of simplicity, for the reduced chi squared  of a fit (often  referred as 
  $\chi^2_\mathrm{red}$) 
  we will use the short notation $\chi^2$.\footnote{
  That is $\chi^2\equiv\chi^2_o /(N_\mathrm{filt}-N_\mathrm{DOF})$, 
  where $\chi^2_o$ is the original definition of goodness of fit, based on the comparison between the 
   fluxes observed in $N_\mathrm{filt}$ filters and those predicted by the template \citep[$N_\mathrm{DOF}$ 
   being the degrees of freedom in the fitting, see definitions e.g.~in][]{Bolzonella2000}.}   
   The  $z_\mathrm{phot}$ error ($\sigma_z$) corresponds to the redshift 
   interval around the median that delimits 68\% of the integrated PDF area.
  The same definition of 1$\sigma$ error  is adopted  
  for stellar mass estimates as well as SFR, age, and rest-frame colours (see Sect.~\ref{SED fitting 2}).    
  As an exception, we prefer to use the best-fit redshift when the 
   PDF($z$) is excessively broad or spiky 
   (i.e.~there are a few peaks with similar likelihood) 
   and the location of the median is thus highly uncertain; 
   we identify   2\,442 galaxy in  this peculiar  condition, such that  
   $\vert z_\mathrm{median}-z_\mathrm{best}\vert > 0.3(1+z_\mathrm{best})$.

  To secure our $z_\mathrm{phot}>2.5$ sample, we apply additional selection criteria. 
  In the redshift range of $u$-to-$V$ drop-outs ($z_\mathrm{phot}\gtrsim3.2$) we require galaxies 
  not to be  detected in those optical bands centred at  $<\!912(1+z_\mathrm{phot})$\,\smash{\AA}. 
  This condition is naturally satisfied by 97.3\% of the sample.  
  We also remove 249  sources with  $\chi^2>10$.  
  We prefer not to implement criteria based on visual inspection of the 
  high-$z$ candidates,  to avoid subjective 
  selections difficult to control.

\subsection{Stellar contamination}
\label{SED fitting-Stellar contamination}

  To remove stars from the $z_\mathrm{phot}>2.5$ sample  
  we adopt an approach similar to \citet{Moutard2016b},   
  combining  multiple selection criteria. 
  First, we fit the multi-wavelength baseline with stellar spectra  
  taken from different models and observations 
  \citep{Bixler1991,Pickles1998,Chabrier2000,Baraffe2015}. 
  We emphasise that the library contains a large number of 
  low-mass  stars of spectral classes from M to T, mainly from  \citet[][]{Baraffe2015}.  
 Unlike dwarf star spectra used in  previous work \citep[e.g.][]{Ouchi2009b,Bouwens2011b,Bowler2014}  
  those derived from \citet{Baraffe2015}  extend to $\lambda_\mathrm{r.f.}>2.5\,\mu\mathrm{m}$ 
   and  therefore SPLASH photometry contributes to disentangle 
  their degeneracy with distant galaxies  \citep{Wilkins2014}.

  We compare the $\chi^2$ of  stellar and galaxy fits, 
  and flag an object as star if $\chi^2_\mathrm{gal}-\chi^2_\mathrm{star}>1$. 
  When the $\chi^2$ difference is smaller than this   confidence threshold 
  we  use additional indicators, namely  (i)  the  stellar locus
  in colour-colour diagrams and  (ii) 
  the maximum surface brightness ($\mu_\mathrm{max}$) 
  above the local background level.  
  For objects with $0<\chi^2_\mathrm{gal}-\chi^2_\mathrm{star}\leqslant1$ 
  we also set $z_\mathrm{phot}=0$ 
  when the criteria (i) or (ii)  indicates that the source is a star.  
 
  The diagrams adopted for the diagnostic (i) are   
   $(z^{++}-[3.6])$  vs $(B-z^{++})$  and $(H-K_\mathrm{s})$ vs $(K-[3.6])$; 
   the former  is analogous  of the $BzK$ by \citet{Daddi2004}, the latter   
   has been used e.g.~in  \citet[][]{Caputi2015}. 
   The two diagrams are devised  using the 
   predicted colours of both stars and galaxy models,  and tested 
    by means of the $z_\mathrm{spec}$ sample  (Fig.~\ref{fig_hki}). 
   The latter is used  for galaxies 
   not detected in the $B$ band (mainly drop-outs at $z\gtrsim4$)  
   for which the  $(z^{++}-[3.6])$  vs $(B-z^{++})$ diagnostic 
    breaks down (see L16, Fig.~15). 
   In each colour-colour space we trace a conservative 
   boundary for the stellar locus, 
   since  photometric uncertainties increase the dispersion in the diagram  
   and stars can be scattered out from the original sequence.  
   Method (ii) is detailed in \citet{Leauthaud2007} and \citet{Moutard2016a}. 
   The surface brightness measurements  in the wide $I$-filter of 
   HST ($F814W$)  come from the Advanced Camera  for Surveys (ACS) images 
   analysed by \citet{Leauthaud2007}. 
   Stars are segregated in the $\mu_\mathrm{max}$-$I$ plane, which  
   has been shown to be reliable at $I\lesssim25$.\footnote{
   \citeauthor{Leauthaud2007} also discuss the limitations of the ``stellarity index'', 
   another commonly used  classification  provided by \texttt{SExtractor}. 
   This  indicator is less accurate than 
    the one based on $\mu_\mathrm{max}$,  
    especially for faint compact galaxies  \citep[see][Fig.~4]{Leauthaud2007}. 
    We then decided not to add stellarity indexes to our set of criteria.}

\subsection{Validation through spectroscopy and self-organizing map}
\label{SED fitting-spec validation}

\begin{figure}
\includegraphics[width=0.99\columnwidth]{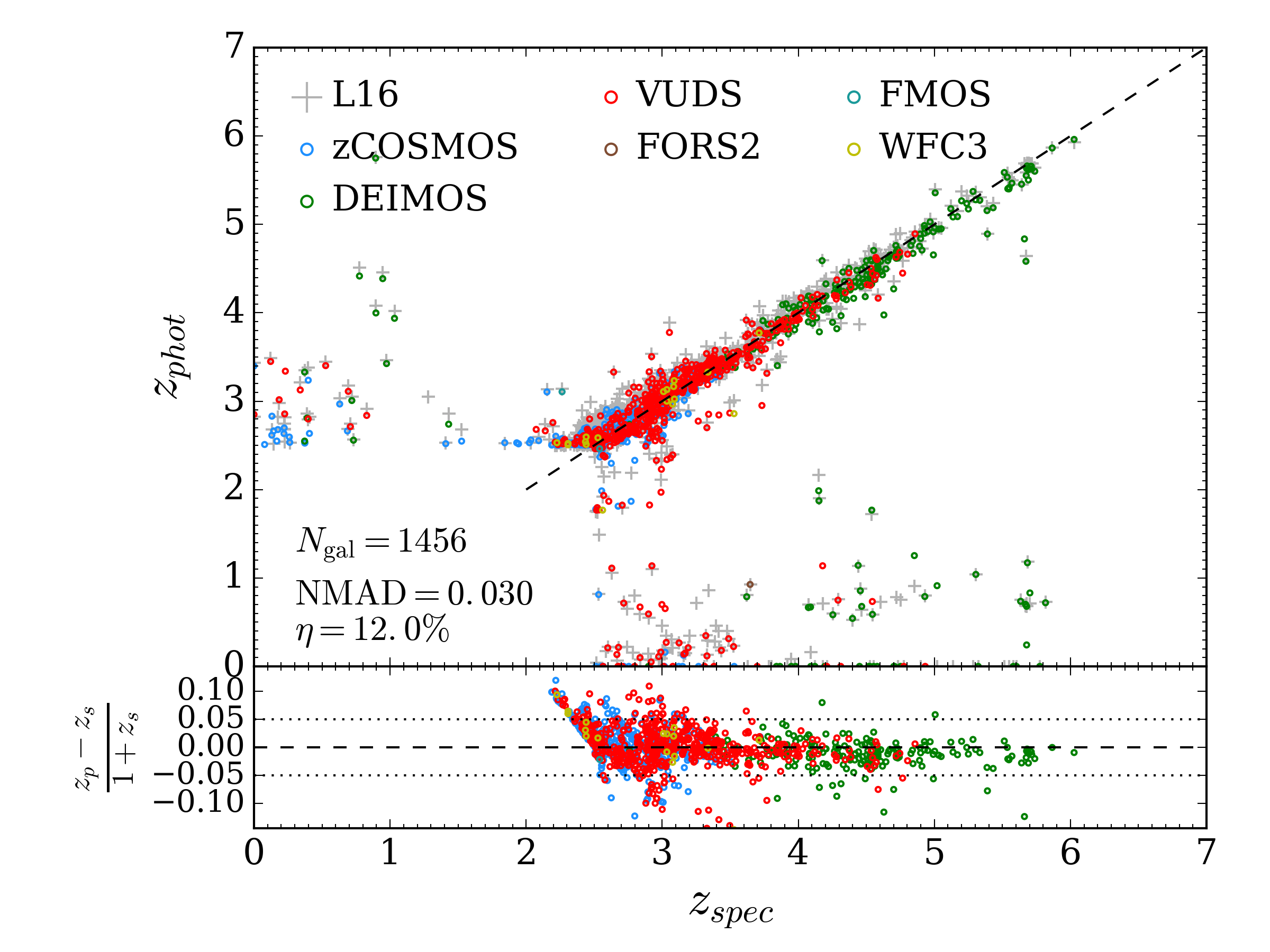}
\caption{Comparison between $z_\mathrm{spec}$ and $z_\mathrm{phot}$, for 
spectroscopic galaxies (and stars) with $[3.6]<25$ (empty circles). 
Only robust spectroscopic measurements (CL $>\!95\%$) 
are plotted, and coloured according to their survey:  zCOSMOS faint (Lilly et al., in preparation), 
VUDS \citep{LeFevre2015}, 
FMOS-COSMOS \citep{Silverman2015}, a survey  with the FORS2 spectrograph at VLT \citep{Comparat2015}, a survey with  
DEIMOS at Keck II (Capak et al., in preparation), and grism spectroscopy from HST/WFC3 \citep{Krogager2014}. 
In the background we also show the comparison between $z_\mathrm{spec}$ and the original photometric redshifts of L16   (grey crosses).  
\textit{Upper panel:}  In addition to $z_\mathrm{phot}$ vs $z_\mathrm{spec}$,  in the bottom-left corner   we report 
the number of objects considered in this test ($N_\mathrm{gal}$), the $\sigma_\mathrm{z}$ error defined as the NMAD, 
and the fraction of catastrophic outliers ($\eta$). The dashed line is the $z_\mathrm{phot}=z_\mathrm{spec}$ reference.  
\textit{Lower panel:}  Scatter of the $z_\mathrm{phot}-z_\mathrm{spec}$ values, with the same colour code 
as in the upper panel. Horizontal lines mark differences (weighed by $1+z_\mathrm{spec}$) equal to $\pm0.05$ (dotted lines)
or null (dashed line).  
}
\label{fig_zs-zp}
\end{figure}

\begin{figure*}
\includegraphics[width=0.45\textwidth]{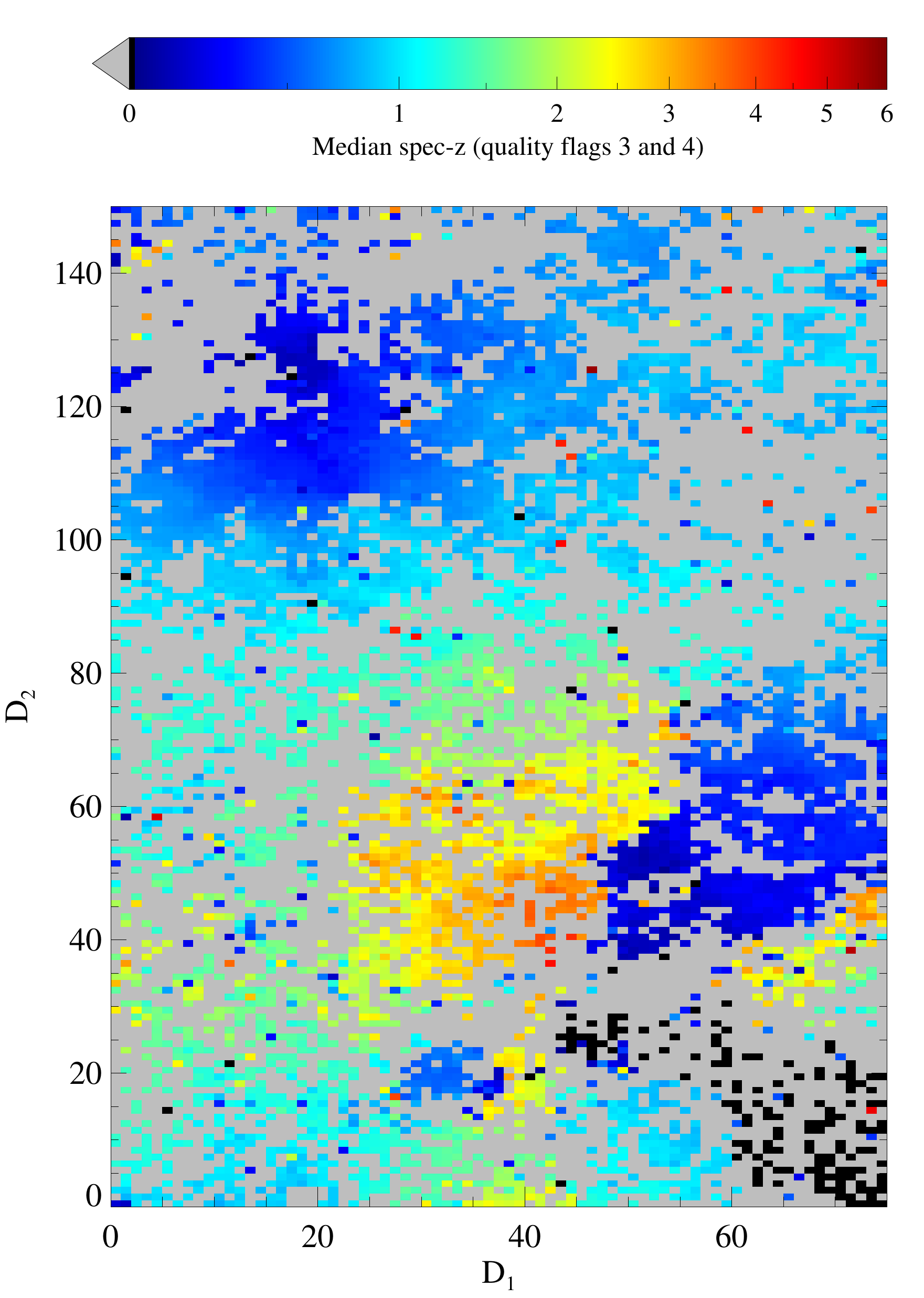} \hspace{10pt} \includegraphics[width=0.45\textwidth]{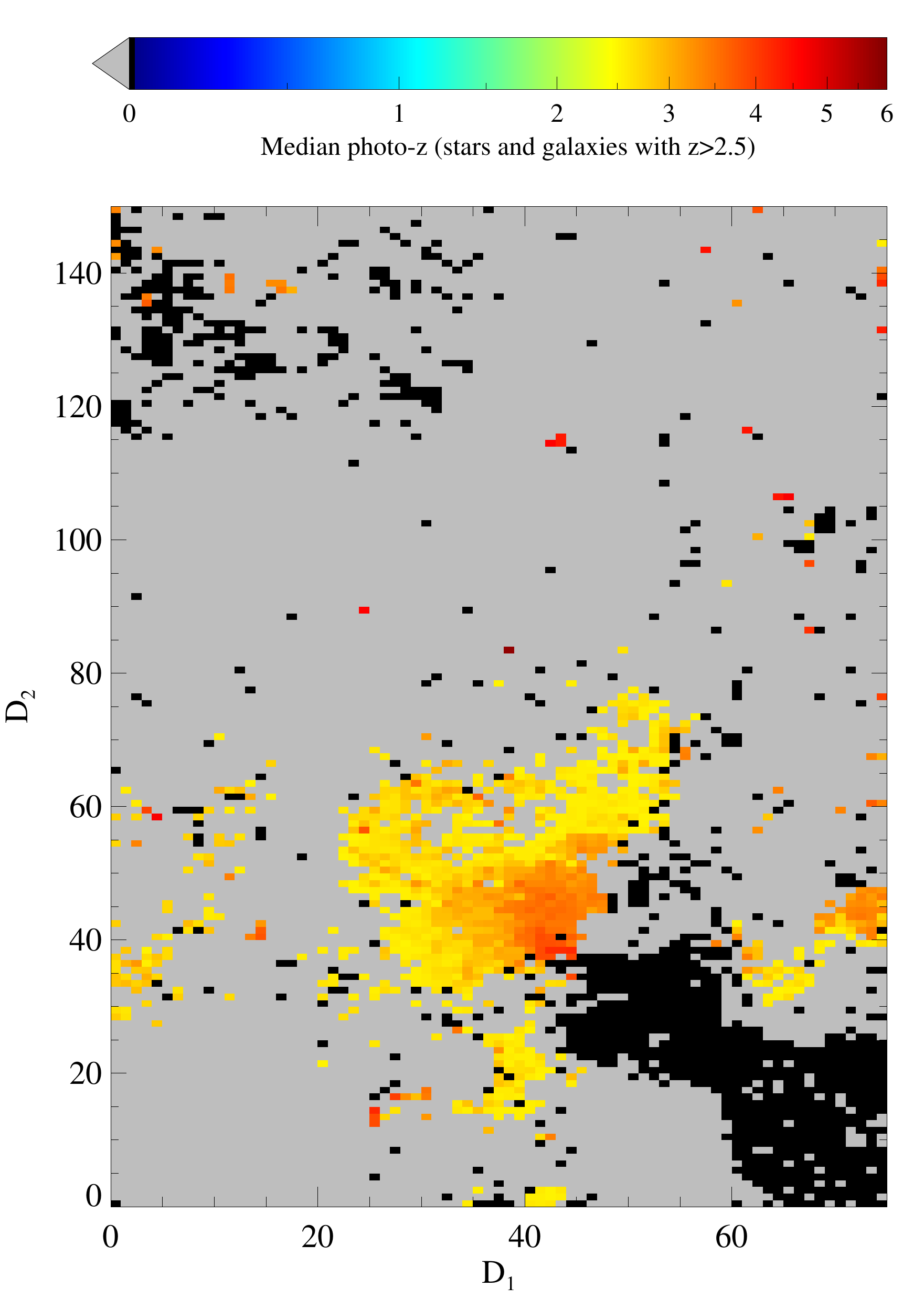}
\caption{The bi-dimensional self-organising map of the COSMOS2015 catalogue 
(the two folded dimensions have  generic labels $D_1$ and $D_2$). 
In the \textit{left panel}, only robust spectroscopic objects (CL $>\!95\%$) are shown. 
In the \textit{right panel},  the SOM is filled with photometric objects (stars and $z_\mathrm{phot}>2.5$ galaxies). 
In both cases, each cell of the map is colour-coded according to the median redshift of the objects inside the cell 
(empty cells are grey, cells filled by stars are black).  }
\label{fig_som1}
\end{figure*}

  We use a catalogue of almost 100\,000 spectroscopic redshifts to 
 quantify the uncertainties of our $z_\mathrm{phot}$ estimates. 
  These  data were obtained during several campaigns, with different instruments and 
 observing strategies (for a summary, see Table 4 of L16). They 
 have been collected and harmonised  in a single catalogue 
 by Salvato et al.~(in preparation).  
 Such a wealth of spectroscopic information  
 represents an unequalled benefit of the COSMOS field. 
 
 The $z_\mathrm{spec}$ measurements used as  a reference  
 are those with the highest reliability, i.e.~a confidence level (CL) $>\!95\%$ 
 \citep[equal to a selection of quality flags 3 and 4 according to the scheme introduced by][]{LeFevre2005}. 
 We limit the comparison to sources brighter than $[3.6]=25$,  ignoring     
 secure low-$z$ galaxies (those  having both $z_\mathrm{spec}$ and $z_\mathrm{phot}$ below $2.5$). 
 Eventually, our test sample contains 1\,456 objects. 
  The size of this sample is unique: with 350  galaxies at $z_\mathrm{spec}>3.5$,  it is 
  more than twice the number of robust spectroscopic redshifts available 
  in CANDELS GOODS-South and UDS, used in \citet{Grazian2015}. 
   
 Among the 301 spectroscopic stars considered, $>\!90\%$ of them are correctly recovered  
 by our method, with only three stellar interlopers with $z_\mathrm{phot}>2.5$. 
 On the other hand, less than 1\% of the spectroscopic galaxies are 
 misclassified as stars. The catastrophic error rate is $\eta=12\%$, 
 considering as an outlier any object with  
 $\vert z_\mathrm{phot} - z_\mathrm{spec} \vert > 0.15(1+z_\mathrm{spec})$.  
 The precision of our photometric redshifts is $\sigma_z=0.03(1+z)$, according to    
 the normalised median absolute deviation \citep[NMAD,][]{Hoaglin1983}   
 defined as $ 1.48 \times \mathrm{median} \lbrace  \vert z_\mathrm{phot} -  z_\mathrm{spec} \vert   / (1 + z_\mathrm{spec}) \rbrace$. 
 These results (Fig.~\ref{fig_zs-zp}) summarise the  
 improvement with respect to the photometric redshifts of L16: we reduce  
 the number of catastrophic errors by $\sim\!20\%$, and 
 we also observe a smaller bias at $2.5<z<3.5$ (cf.~Fig.~11 of  L16).

  The comparison between $z_\mathrm{spec}$ and $z_\mathrm{phot}$ 
  is meaningful only if the spectroscopic sample is an unbiased 
  representation of the ``parent'' photometric sample. Otherwise, we would test 
  the reliability of a subcategory of galaxies only. 
  We  introduce a self-organizing map  \citep[SOM,][]{Kohonen1982} 
  to show that our spectroscopic catalogue provides a representative 
  sample of the underlying colour and redshift distribution. 
   The algorithm version we use, specifically implemented for astronomical purposes,  
    is the one devised in  \citet{Masters2015}. 
   The SOM allows to reduce a high-dimensional dataset in 
   a bi-dimensional grid, without losing essential  topological information.   
   In our case, the starting manifold is the panchromatic space (fifteen colours) resulting from 
   the COSMOS2015 broad bands: $(\mathrm{NUV}-u)$, $(u-B)$,  ..., 
   $([4.5]-[5.8])$, $([5.8]-[8.0])$.  
   As an aside, we note that  the SOM dimensions do not necessarily 
   have to be colours: in principle the parameter space can be enlarged by including other 
  properties like galaxy size or morphological parameters. 
  Each coloured cell in the map (Fig.~\ref{fig_som1}) 
  corresponds to a point in the 15-dimensional 
  space that is non-negligibly occupied by galaxies or stars from the survey. 
  Since the topology is preserved, 
   objects with very similar SEDs  
   -- close to each other in the high-dimensional space --  will be linked  to 
   the same cell (or to adjacent cells).  
  
   We inspect the  distribution of spectroscopic objects  in the SOM, 
   using them as a training sample 
   to identify the region of high-$z$ galaxies (Fig.~\ref{fig_som1}, left panel).    
   The COSMOS spectroscopic catalogue    
    samples  well the portion of parameter space we are interested in,  except for 
    the top-left corner of the map where we expect,   according to models, the bulk of low-mass stars. 
     The lack of 
    spectroscopic measurements in that region may affect 
    the precise evaluation of the $z_\mathrm{phot}$ contaminant fraction. 
     Other cells that are  weakly constrained 
    correspond to the SED of   star-forming galaxies with $i\gtrsim23$\,mag 
    and $z<2$  \citep{Masters2015},  which however  
    are not pivotal for testing our estimates. 
    
  After the spectroscopic calibration, 
    we insert  $z_\mathrm{phot}>2.5$ galaxies in the SOM along with photometric stars  
    (Fig.~\ref{fig_som1}, right panel). Given the larger size of the photometric 
    catalogue,  cells occupation is more continuous and extended 
    (see e.g.~the stellar region in the bottom-right corner). 
      By comparing the two panels of Fig.~\ref{fig_som1}, 
    one can see that 
    the  $z_\mathrm{phot}>2.5$ galaxies are concentrated in the  SOM 
    region  that has been identified   as high-$z$ 
    by the  spectroscopic training sample. 
    Although the latter is more sparse, about 60\% of the 
    area covered by $z_\mathrm{phot}>2.5$ galaxies 
    is also sampled by  spectroscopic 
    measurements, which are quantitatively 
    in good agreement: in 82\% of those cells that contain both photometric  
    and spectroscopic redshifts, the median of the former ones is within $1\sigma_z$ from the  
    median of the $z_\mathrm{spec}$ objects laying in the same cell. 
	Moreover, by plotting individual galaxies (not shown in the Figure) 
	one can verify that catastrophic $z_\mathrm{phot}$ errors  are randomly spread 
	across the SOM, not biasing any specific class of galaxies.

\section{Stellar mass estimate and completeness}  
 \label{SED fitting 2}
  
  After building a  $z_\mathrm{phot}>2.5$  galaxy sample, 
   we   run  \texttt{LePhare}  to estimate their stellar mass, star formation rate (SFR),  
  and other physical parameters such as rest frame colours.  
  This is described in Sect.~\ref{SED fitting 2-stellar mass}, while in 
  Sect.~\ref{SED fitting 2-stellar mass limit} we compute stellar mass 
  completeness limits and we argue  
  in favour of a $[3.6]$ selection to work with a mass-complete sample  up to 
  $z\sim6$ (see also Appendix \ref{APP3}).   
   By means of their rest-frame colours we then 
    identify reliable quiescent galaxies  (Sect.~\ref{SED fitting 2-Galaxy type}).

\subsection{Galaxy stellar mass} 
 \label{SED fitting 2-stellar mass}

    \begin{figure}
   \includegraphics[width=0.99\columnwidth]{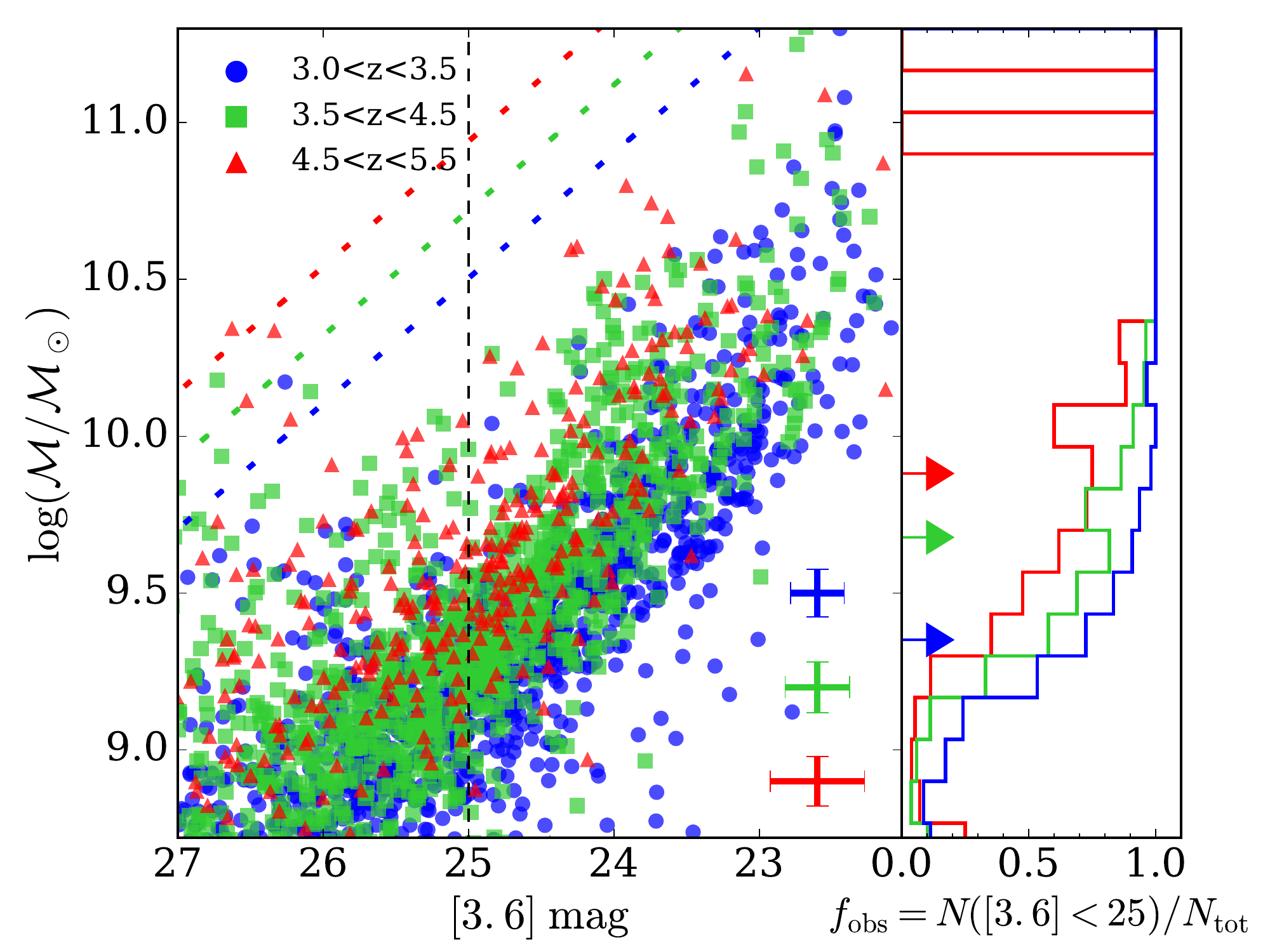}
   \caption{
    Stellar mass completeness as a function of redshift.      
    Blue circles, green squares, red triangles represent CANDELS galaxies  
    at $2.5<z_\mathrm{phot,N17}\leqslant3.5$, 
    $3.5<z_\mathrm{phot,N17}\leqslant4.5$, 
    $4.5<z_\mathrm{phot,N17}\leqslant6$ respectively.  
    The cut in apparent magnitude of our sample 
    ($[3.6]_\mathrm{lim}$) is marked with a vertical dashed line. 
    Slant dotted lines show a conservative estimate  of the  stellar mass limit, 
    corresponding to the $\mathcal{M}/L$ ratio of an old  SSP 
    with  $A_V=2$\,mag. Three crosses in the bottom-right corner 
    of the main panel show the average $x$- and $y$-axis uncertainties 
    in the corresponding bin of redshift. 
    The histograms on the right 
    (same colours and $z$-bins of the scattered points) 
    show the ratio of N17 galaxies 
    with $[3.6]<[3.6]_\mathrm{lim}$ over  
    the total N17 sample. This fraction 
    is named $f_\mathrm{obs}$ since they are the objects that would 
    be observed within the magnitude limit of COSMOS2015. 
   Below each histogram, 
   an arrow indicates     
   the stellar mass threshold $\mathcal{M}_\mathrm{lim}$ (see Sect.~\ref{SED fitting 2-stellar mass limit}). 
    }
    \label{fig_mlim}
    \end{figure}
  
  We estimate stellar mass 
  and other physical properties of the COSMOS2015 galaxies 
   always at fixed $z\equiv z_\mathrm{phot}$. We fit  
  their multi-wavelength photometry with a library  of SEDs 
  built using the stellar population 
  synthesis model of \citet[][hereafter BC03]{Bruzual2003}.  
  The $\mathcal{M}$ estimates are the 
  median of the PDF marginalised over the 
  other parameters. This kind of estimate is in good agreement 
  with the stellar mass derived from the PDF peak (i.e., the best-fit template). 
  The difference between median and best-fit values is on average 
  0.02\,dex, with  a rms of 0.11\,dex.

   The galaxy templates given in input to \texttt{LePhare} are constructed 
   by combining BC03 simple stellar populations (SSPs) 
   according to a given 
   star formation history (SFH).   
   Each SSP has an initial 
   mass function (IMF)  that follows \citet{Chabrier2003},  
   while the stellar metallicity can be $Z=0.02$, $0.008$,  
   or $0.004$.\footnote{We avoid to use $Z_\odot$ units since recent work 
   suggests that  
   solar metallicity is lower than the ``canonical'' value of 0.02 \citep[e.g., $Z_\odot=0.0134$ in][]{Asplund2009}.}
      These stellar metallicities have been chosen to encompass the range 
   observed up to $z \sim 4\upto5$ \citep[e.g.][]{Maiolino2008,Sommariva2012};  
   they are also in agreement with hydrodynamical simulations \citep{Ma2016}. 
    For each  template we combine SSPs 
   with the same metallicity (i.e.~there is no chemical enrichment in the galaxy model, 
   nor interpolation between the three given $Z$ values).

    We assumed various  SFHs, namely ``exponentially declining''  and 
    ``delayed declining''. The former ones have  $\mathrm{SFR}(t)\propto e^{-t/\tau}$, 
    while  the shape of the latter is $\propto te^{-t/\tau}$. 
   For the exponentially declining profiles, the $e$-folding time  ranges from 
   $\tau=0.1$ to 30\,Gyr,  while for delayed SFHs the $\tau$ parameter, which also marks the peak of 
   SFR, is equal to  1 or 3\,Gyr.
   We post-process the BC03 templates obtained in this way  
   by adding nebular emission lines  
   as described in Sect.~\ref{SED fitting-method}.  
   Dust extinction is implemented 
   assuming $0\leqslant E(B-V) \leqslant 0.8$.  
   We allow for only one attenuation law,  
   i.e.~\citet{Calzetti2000} with the addition of the 2175\,\smash{\AA} feature \citep[see][]{Scoville2015}. 
   
   We have tried a few alternate configurations to 
   quantify the amplitude of possible systematics 
   (see Sect.~\ref{Results-Sources of uncertainty} for 
   a detailed discussion).  
   We added  for example 
   a second attenuation curve with slope proportional to $\lambda^{-0.9}$ 
   \citep{Arnouts2013}. Such a  choice  increases the number
   of degenerate best-fit solutions without introducing any significant bias: 
   Calzetti's law is still preferred (in terms of $\chi^2$) by most of the objects at $z\gtrsim3$.  
   Other modifications,  e.g.~the expansion of the metallicity grid, have a larger impact, as also found in other studies 
   \citep[e.g.][]{Mitchell2013}.  
   Simplifying  assumptions  
   are somehow  unavoidable in the SED fitting, not only for 
   computational reasons but also because the available  information  (i.e., the multi-wavelength baseline) 
   cannot constrain the parameter space beyond a certain number of degrees of freedom. 
   This  translates into systematic offsets when comparing different SED fitting recipes.    
   The impact of these systematics on the SMF is clearly visible in \citet[][Fig.~1]{Conselice2016}, where the 
   authors overplot a wide collection of measurements from the literature: 
   already at $z<1$, where data are more precise, the various SMF estimates can 
   differ even by a factor $\sim\!3$.      
   We emphasise that one advantage of our work, 
   whose goal is to connect the SMFs at different epochs, 
   is  to be less	 affected by SED fitting uncertainties than analyses that combine measurements from different papers.  
   In our case, SED fitting systematics (unless  they have a strong redshift or galaxy-type dependence) will  
   cancel out in the differential quantities we want to derive. 
      
   As mentioned above, the 68\% of the integrated 
   PDF($\mathcal{M}$)  area gives an error to each stellar mass estimate. 
   However, the PDF is obtained from $\chi^2$-fit templates at fixed redshift. 
   To compute stellar mass errors ($\sigma_\mathrm{m}$) including  the additional uncertainty inherited 
   from $\sigma_z$,  we proceed in a   way similar to \citet{Ilbert2013}. 
   They  generate a mock galaxy catalogue by perturbing the original photometry and redshifts, 
   proportionally to their errors. 
   After recomputing   the stellar mass of each galaxy, the authors define its uncertainty as
   the difference between new $\mathcal{M}$ and the original  estimate, namely  
   $\Delta\mathcal{M}_i\equiv\log(\mathcal{M}_{\mathrm{MC},i} / \mathcal{M}_i)$ (for the $i$-th galaxy). 
   
    We implement a few modifications with respect to   \citet{Ilbert2013}. 
    Instead of adding noise to photometry and $z_\mathrm{phot}$, we exploited the PDFs.  
    Moreover, we  produce 100 mock catalogues, instead of a single one.     
    We perform a Monte Carlo simulation, re-extracting 100 times 
    the $z_\mathrm{phot}$  of each galaxy, accordingly to its PDF($z$). 
    Each time, we run again \texttt{LePhare} with  the redshift fixed 
    at the new value ($z_\mathrm{MC}$) to compute the galaxy stellar mass 
    $\mathcal{M}_\mathrm{MC}$ and the offset from the original value.  
    We group galaxies in bins of redshift and mass to obtain  an estimate of $\sigma_\mathrm{m}$ 
    from the distribution of their $\Delta\mathcal{M}$. As in \citet{Ilbert2013}, this  is well fit by 
    a Gaussian  multiplied by a Lorentzian distribution:  
    \begin{equation}
    \mathcal{D}(\mathcal{M}_0,z) = \frac{1}{\sigma\sqrt{2\pi}}\exp\left( \frac{-\mathcal{M}_0^2}{2\sigma^2}\right) 
    \times \frac{\tau}{2\pi \left[(\frac{\tau}{2})^2+\mathcal{M}_0^2 \right] },
    \label{eq_gau-lor}
   \end{equation}     
   where $\mathcal{M}_0$ is the centre of the considered stellar mass bin. 
    The parameters $\sigma$ and $\tau$ are in principle  
    functions of $\mathcal{M}_0$ and $z$, 
    left implicit in Eq.~\ref{eq_gau-lor} for sake of clarity.  
   We   find $\sigma\simeq 0.35$\,dex, without a strong dependence on 
   $\mathcal{M}$ and neither on $z$, at least for $9.5<\log(\mathcal{M/M}_\odot)<11.5$.  
   Also $\tau$  does not depend significantly on stellar mass but it increases as a function of 
   redshift. The relation assumed in \citet{Ilbert2013}, namely  $\tau(z)=0.04(1+z)$, is still 
   valid for our sample. We note that the value of $\sigma$ is instead smaller \citep[it was 0.5\,dex in][]{Ilbert2013},  
   reflecting the increased quality of the new data. 
   At face value,  
   one can assume $\sigma_\mathrm{m}(\mathcal{M})=0.35$\,dex (neglecting the Lorentzian 
   ``wings'' of $\mathcal{D}$); however a careful treatment of stellar mass uncertainties requires to 
   take into account   the whole Eq.~(\ref{eq_gau-lor}).
   This computation 
   gives us an idea about the impact of $\sigma_z$ on the stellar mass estimate: the errors resulting from 
   the PDF($\mathcal{M}$), after fixing the redshift, are usually much smaller 
   (e.g.~$<\!0.3$\,dex for 90\% of the  galaxies at $3.5<z\leqslant4.5$).  
   Further details about the impact of $\sigma_\mathrm{m}$ on the SMF 
   are provided in Sect.~\ref{Results-SMF bias}.

\subsection{Stellar mass limited sample}
\label{SED fitting 2-stellar mass limit}

  To estimate the stellar mass completeness ($\mathcal{M}_\mathrm{lim}$) as a function 
   of redshift, we apply the technique introduced by \citet{Pozzetti2010}. 
   In each $z$-bin, a set of ``boundary masses'' ($\mathcal{M}_\mathrm{resc}$) is obtained 
   by taking the most  used best-fit  templates 
   (those with the minimum $\chi^2$ 
   for 90\% of the galaxies in the $z$-bin) and rescaling them 
   to  the magnitude limit:   
   $\log \mathcal{M}_\mathrm{resc} = \log \mathcal{M} + 0.4([3.6]-[3.6]_\mathrm{lim,UD})$. 
   Then, $\mathcal{M}_\mathrm{lim}$ is defined as the 90th percentile of the  
    $\mathcal{M}_\mathrm{resc}$ distribution. 
    Taking into account also the incompleteness due to the 
    $[3.6]$ detection strategy (see Fig.~\ref{fig_compl-cand2}), 
    we expect that in  the lowest stellar mass bins of our SMF 
    (at $\mathcal{M}\geqslant\mathcal{M}_\mathrm{lim}$)  
    there could be a sample incompleteness of
    30\% at most, caused by objects previously excluded from 
    the IRAC-selection (i.e., with $[3.6]>[3.6]_\mathrm{lim,UD}$). 
    Eventually we interpolate the $\mathcal{M}_\mathrm{lim}$ values found in different $z$-bins  
    to describe the evolution of the limiting mass:  
     $\mathcal{M}_\mathrm{lim}(z)= 6.3\times10^{7}(1+z)^{2.7}\,\mathcal{M}_\odot$.
      
    To verify our calculation,  we consider the 
    $\mathcal{M}$ vs $[3.6]$ distribution of  CANDELS-COSMOS galaxies 
    (N17, see Sect.~\ref{Dataset-completeness}) after recomputing their stellar mass with \texttt{LePhare} 
    to be consistent with our catalogue.  
    By using these deeper data, 
    we verify  that the $\mathcal{M}_\mathrm{lim}$ values we found   
    correspond to a completeness of $70\upto80\%$ (Fig.~\ref{fig_mlim}).   
    By means of  the N17 sample, 
    we can account for stellar mass incompleteness  
    below $\mathcal{M}_\mathrm{lim}$ \citep[cf.][]{Fontana2004}. 
    The factor we need for such a correction in the low-mass regime  is 
     $f_\mathrm{obs}(z,\mathcal{M})$, 
     i.e.~the fraction of objects at a given redshift and stellar mass 
    that are brighter than $[3.6]_\mathrm{lim,UD}$. 
    This is obtained by fitting the histograms shown 
    in Fig.~\ref{fig_mlim} (right-hand panel).
    In doing so we assume that the 
    CANDELS sub-field, which is $\sim\!10\%$ of $\mathcal{A}_\mathrm{UD}$, 
    is large enough  to represent the parent COSMOS2015 volume 
    and sufficiently deeper to probe the full $\mathcal{M}/L$ 
    range.   
    The function $f_\mathrm{obs}(z,\mathcal{M})$ can be used 
    to correct the SMF of COSMOS2015 at   $\mathcal{M}<\mathcal{M}_\mathrm{lim}$, 
    where  the $1/V_\mathrm{max}$ weights start  to be biased 
    (see Sect.~\ref{Results}).

    We also include in Fig.~\ref{fig_mlim}  a comparison between our method and 
     a more conservative one,  
    based on the maximum   $\mathcal{M}/L$  physically  allowed at a given flux limit 
    and cosmic time    \citep[see e.g.][]{PerezGonzalez2008}.  
    The SED used to this purpose   is the one 
    of a galaxy that formed stars in a single initial burst at $z=20$, and passively evolves 
    until the desired redshift.  
    Substantial extinction ($A_V=2$\,mag) 
    is added to further enlarge $\mathcal{M}/L$. 
    However, in our redshift range,  the statistical relevance of 
    such an extreme galaxy type is small, 
    as one infers from the few CANDELS objects sparse in the upper-left corner of Fig.~\ref{fig_mlim} 
    \citep[see also the discussion in][Appendix C]{Marchesini2009}. 
    With  the maximal $\mathcal{M}/L$ ratio we would overestimate the  stellar mass completeness threshold 
    by at least a factor 5.

   At redshifts between  $2.5$ and $4$ we could in principle 
   evaluate $\mathcal{M}_\mathrm{lim}$ also as a function 
    of $K_\mathrm{s}$, with the same empirical approach used above. 
    The cut to be used in this case is   $K_\mathrm{s}<24.7$  (see Fig.~16 of L16). 
    However, the $\mathcal{M}_\mathrm{lim}$ resulting from the $K_\mathrm{s}$-band  selection 
    is  $0.2\upto0.4$\,dex higher  (depending on the redshift) than the threshold derived  
    using the $[3.6]$ band. 
    Such an offset reflects a real difference in the stellar mass distribution of 
    the $K_\mathrm{s}$-selected sample with respect to the $[3.6]$-selected one.
    The latter is anchored to a $\chi^2$-stacked image that includes bands deeper than $K_\mathrm{s}$, 
    so that the sample is unbiased down to lower masses 
    (more details are provided in Appendix \ref{APP3}).

\begin{figure}
\includegraphics[width=0.99\columnwidth]{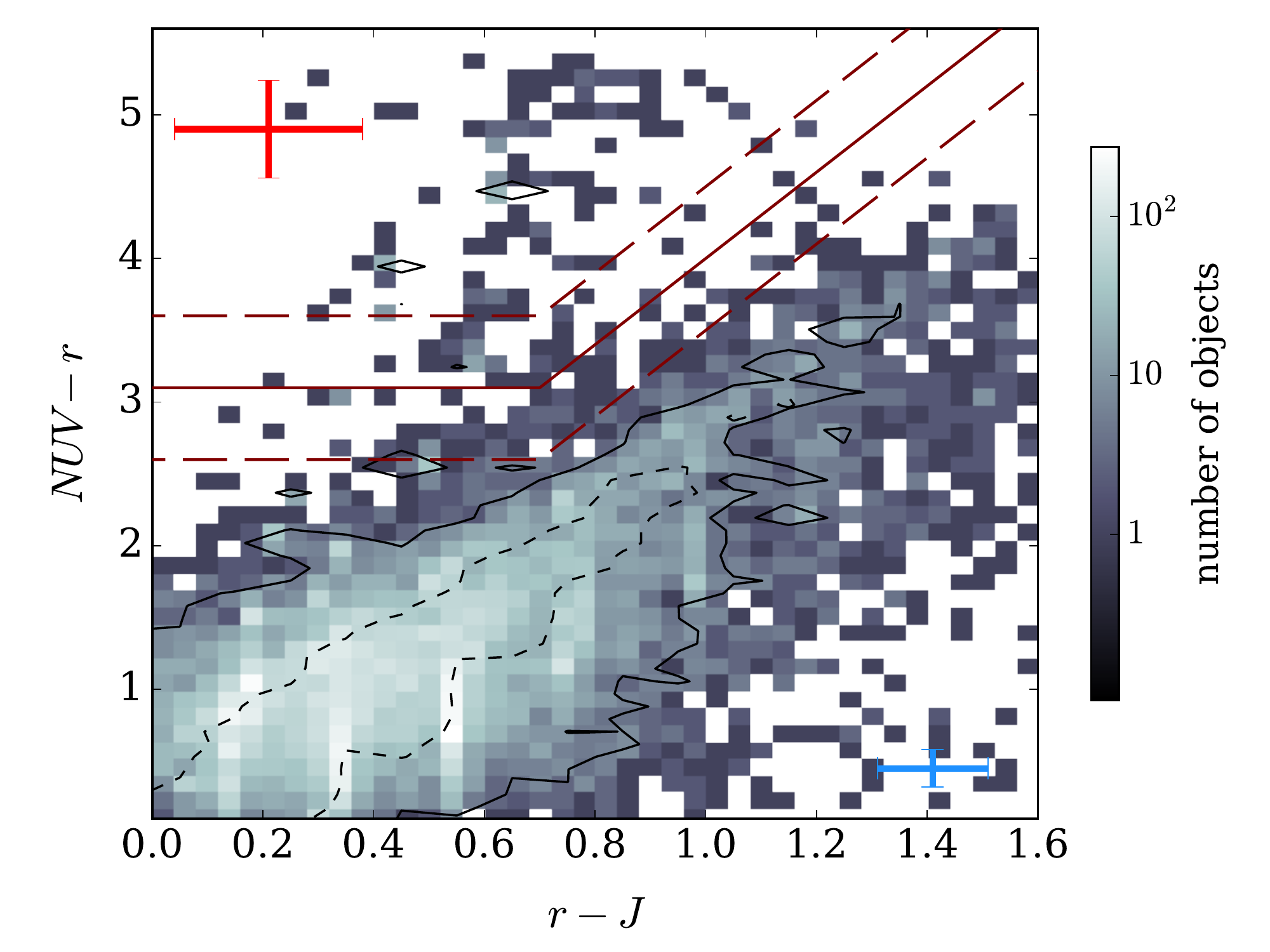}
\caption{The $NUVrJ$ diagram of galaxies between $z=2.5$ and 6 (with their density distribution  
colour coded in grey shades). The red solid line divides active and passive regions (see Eq.~\ref{eq_nuvrj}), 
while red dashed lines are shifted by $\pm0.5$\,mag from that border, to give a rough estimate of the width 
of the green valley, i.e.~the separation between active and passive clumps. 
The red (blue) cross in the top-left (bottom-right) corner shows the typical uncertainties $\sigma_{NUV-r}$ 
and $\sigma_{r-j}$  of passive (active) galaxies at $3<z<\leqslant4$. We also compare our fiducial estimates (based on the nearest observed filter) 
with colours directly derived from the template SEDs: a solid contour encloses 90\% of the former distribution, 
while a dashed line represents the 90\% envelope for the latter.   }
\label{fig_nrj}
\end{figure}

  \subsection{Quiescent galaxies classification}
\label{SED fitting 2-Galaxy type}

  We also aim to derive the SMF of passive  and active galaxies separately. 
  As shown in the following, 
  the quality of our dataset allows us to extend the classification up to $z=4$. 
  Galaxies that  stopped their star formation 
  occupy a specific region in the  
  colour-colour diagrams  $(U-V)$ vs $(V-J)$, $(NUV-r)$ vs $(r-K)$, or  $(NUV-r)$ vs $(r-J)$, 
  as shown 
  in previous work \citep[][respectively]{ Williams2009,Arnouts2013,Ilbert2013}. 
  We adopted the last one, dubbed hereafter $NUVrJ$. 
  Compared to  $(U-V)$ vs $(V-J)$, which is often referred as $UVJ$, the use of  $(NUV-r)$ makes 
  $NUVrJ$  more sensitive to recent star formation   \citep[on $10^6\upto10^8$\,yr scales][]{Salim2005,Arnouts2007}. This property results in a better distinction between fully quiescent galaxies 
  ($\mathrm{sSFR}<10^{-11}\,\mathrm{yr}^{-1}$) and those with residual star formation  (typically with $\mathrm{sSFR}\simeq10^{-10}\,\mathrm{yr}^{-1}$). With $UVJ$, these two kinds of galaxies occupy the same place in the diagram, as we verified using BC03 models. On the contrary,
   galaxies with negligible ``frostings'' of star formation are correctly classified as passive in the $NUVrJ$.

  The $NUVrJ$ indicator is similar to   
  $(NUV-r)$ vs $(r-K)$, with the advantage that at high redshifts 
  the  absolute  magnitude $M_J$ is more robust than  $M_{K}$, since for the latter   
  the $k$-correction is generally more uncertain.   \texttt{LePhare}  calculates 
  the absolute magnitude at a given wavelength ($\lambda_\mathrm{r.f.}$)  
  by  starting from the  apparent magnitude 
  in the band closest to  $\lambda_\mathrm{r.f.}(1+z)$. For example at $z=3$ the nearest  observed 
  filter to compute $M_J$ is $[4.5]$, whereas $M_K$ falls beyond the MIR window of the four IRAC channels. 
 
 The $NUVrJ$ diagram is shown in Fig.~\ref{fig_nrj}. The density map of our $z_\mathrm{phot}>2.5$ sample 
 highlights the ``blue cloud'' of star forming galaxies as well as an early ``red sequence''. 
 We also show in the Figure how the $NUVrJ$ distribution changes 
 when colours are derived directly from the template SEDs, without using the nearest observed filter as a proxy 
 of the absolute magnitude.  
 This alternative method is commonly used in the literature, so it is worth showing the different galaxy classification 
 that it yields.  The distribution from pure template colours is much narrower (cf.~solid and dashed lines in Fig.~\ref{fig_nrj}) 
 but potentially biased:  the  SED library spans a limited  range of slopes (colours), whereas  
 the nearest filter method -- by taking into account the observed flux -- naturally includes a larger variety of SFHs.

 The boundary of the passive locus is  
 \begin{equation}
 (NUV-r) > 3(r-J) +1 \quad \mathrm{and}  \quad (NUV-r) > 3.1,
 \label{eq_nuvrj} 
 \end{equation}
 defined empirically 
 according to the  bimodality of the 2-dimensional galaxy distribution \citep[][]{Ilbert2013}. 
 This border is also  physically justified: 
 the slant line resulting from Eq.~(\ref{eq_nuvrj}) runs perpendicular to  
 the direction of increasing specific SFR ($\mathrm{sSFR}\equiv\mathrm{SFR}/\mathcal{M}$). 
 On the other hand, dust absorption moves galaxies parallel to the border, effectively
 breaking the degeneracy between  genuine quiescent and dusty star-forming galaxies. 
 Same properties are observed in $(NUV-r)$ vs $(r-K)$ and $(U-V)$ vs $(V-J)$, as demonstrated e.g.~in  
 \citet[][]{Arnouts2013} and \citet{Forrest2016}.

 Rest-frame colour selections have been successfully used up to $z\sim3$ 
 \citep[e.g.][]{Ilbert2013,Moutard2016b,Ownsworth2016}. 
 The reliability of this technique at $z>3$ has been recently called into question by 
 \citet{Schreiber2017}: they warn that the IRAC photometry in the COSMOS field could be 
 too shallow to derive robust rest-frame optical colours for galaxies at those redshifts. 
 In order to take into account such uncertainties, our algorithm rejects a filter if its error is larger than 0.3\,mag.  
 Most importantly, we evaluate $(NUV-r)$ and 
 $(r-J)$ uncertainties, to quantify the accuracy of our $NUVrJ$ selection. 
 For each galaxy, a given rest-frame colour error  ($\sigma_\mathrm{colour}$) 
 is derived from the marginalised  PDF, by considering 68\% of the area around the 
 median  (similarly to $\sigma_z$).   
 In such a process, the main contributions to $\sigma_\mathrm{color}$ 
  are the photometric uncertainties and the model $k$-correction.  
  At $3<z \leqslant4$, the mean errors are $\sigma_{NUV-r}=0.13$ and 
  $\sigma_{r-J}=0.10$ for the active sample, while  $\sigma_{NUV-r}=0.34$ and 
  $\sigma_{r-J}=0.17$ for objects in the passive locus (see Fig.~\ref{fig_nrj}). 
  These values confirm that our classification is reliable up to  $z=4$: 
  even the relatively large $(NUV-r)$ uncertainty does not affect it, 
  given the scale of  the $y$ axis in Fig.~\ref{fig_nrj}. 
  Despite the larger uncertainties, we also apply the $NUVrJ$ classification at $z>4$, 
  identifying  13 potential passive galaxies (4 of them at $z_\mathrm{phot}>4.5$). In that redshift range, a strong 
  H$\alpha$ line can  contaminate the $[3.6]$ band (which is used to estimate $M_r$).   
  Comparing their observed $([3.6]-[4.5])$ colour with BC03 models  \citep[as done in][]{Faisst2016a} 
  we find that  6 passive candidates at $z>4$ could have non-negligible H$\alpha$ emission 
  ($\mathrm{EW}=100\upto500$\,\emph{\AA}) while for the others 
  their $([3.6]-[4.5])$ is compatible with  $\mathrm{EW}<50$\,\emph{\AA}. 
  Nonetheless these $z>4$ galaxies need deeper data to be confirmed as passive.

  Galaxies close to the  border defined by Eq.~(\ref{eq_nuvrj}) may be misclassified, but 
  the fraction of objects in  that intermediate corridor 
  (encompassed by dashed lines in Fig.~\ref{fig_nrj}) is small    
  and the bulk of the passive sample should not be  significantly contaminated. 
  We emphasise that by using $(NUV-r)$ instead of $(U-V)$ 
  the larger dynamical scale on the $y$ 
  axis reduces drastically  the contamination at the edge of the passive locus. 
  Inside that intermediate region  
  galaxies are expected to be in transition 
  from the blue cloud to the red sequence \citep{Moutard2016b}. 
  Therefore, their classification within the active vs passive scheme is not straightforward 
  even from   a physical point of view.   
  We will come back to discussing green valley galaxies  
  in Sect.~\ref{Results-SMF active/passive}.

   \section{Results}
   \label{Results}

  \subsection{Galaxy stellar mass function at $z>2.5$}
 \label{Results-SMF highz}

\begin{figure*}
    \centering
 \includegraphics[width=0.8\textwidth]{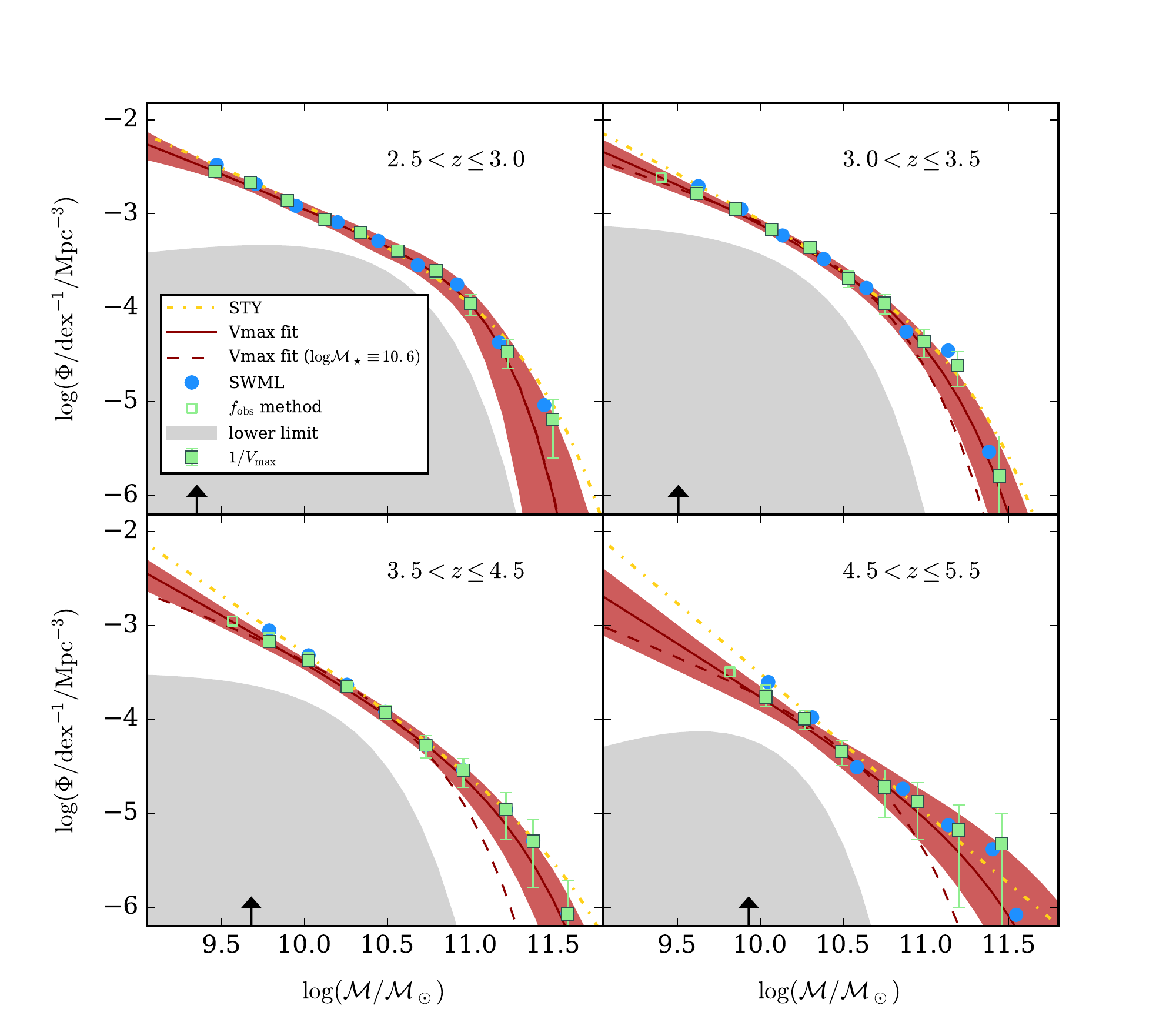}
  \caption{Stellar mass function of COSMOS2015 galaxies, in four redshift bins between  
  $z=2.5$ and 5.5. In each panel, 
  the $1/V_\mathrm{max}$ determination is shown by green squares, while 
  blue circles represent the SWML.  
  Error bars of  $1/V_\mathrm{max}$  points include Poisson noise, sample variance, 
  and the scatter due to SED fitting uncertainties 
  (see definition of $\sigma_\mathrm{\Phi}$ in the text).    
  The yellow dot-dashed line represents the STY fitting function, 
  which is a Schechter function at $z<4.5$ and a power law in the bottom right panel. 
  Empty squares are obtained from an empirical method where, instead of 
  the $V_\mathrm{max}$ correction, we 
  apply to each galaxy the statistical weight $f_\mathrm{obs}(z,\mathcal{M})$ obtained from a deeper reference sample 
  (see Sect.\ref{SED fitting 2-stellar mass limit}). At $\mathcal{M}>\mathcal{M}_\mathrm{lim}$  
  the empty squares are not visible since the $f_\mathrm{obs}$ method coincides with $1/V_\mathrm{max}$.   
  The $1/V_\mathrm{max}$ determinations 
  are  fit by a double Schechter function (Eq.~\ref{eq_dubschfun}) at $2.5<z\leqslant3$, 
  and a single Schechter  (Eq.~\ref{eq_schfun}) 
  in the other bins (in all the cases, the fit is shown by a red solid line, 
  while the red  shaded area is its $1\sigma$ uncertainty). 
  Another Schechter fit (red dashed line) to the   $1/V_\mathrm{max}$ points is made by assuming  that the parameter $\log(\mathcal{M}_\star/\mathcal{M}_\odot)$ is equal to 10.6. 
  By considering only the most secure $z_\mathrm{phot}$, 
  we compute  a lower limit  for the SMF, below which we colour the plot area in grey. An arrow 
  in the bottom  part of each panel marks the  observational limit in stellar mass 
  (see Sect.~\ref{SED fitting 2-stellar mass limit}).}
  \label{fig_mf1}
  \end{figure*}

  We estimate the galaxy SMF 
  by means of three different methods \citep[as  implemented in 
  the code \texttt{ALF},][]{Ilbert2005} 
  and compare the results in Fig.~\ref{fig_mf1}. 
  The $1/V_\mathrm{max}$   \citep{Schmidt1968} and the 
  stepwise maximum-likelihood  \citep[SWML,][]{Efstathiou1988} 
  are two techniques  
  that do not impose any  \textit{a priori} 
  model for the SMF, while the maximum likelihood method devised by 
  \citet[][hereafter STY]{Sandage1979} is parametric and 
   assumes that the SMF is described by a 
  \citet[][]{Schechter1976} function: 
  \begin{equation}
  \Phi(\mathcal{M})\mathrm{d}\mathcal{M} =
  \Phi_\star \left( \frac{\mathcal{M}}{\mathcal{M}_\star} \right)^{\alpha} 
  \exp\left( - \frac{\mathcal{M}}{\mathcal{M}_\star} \right) 
  \frac{\mathrm{d}\mathcal{M}}{\mathcal{M}_\star}  \: .
  \label{eq_schfun}
  \end{equation}

   A detailed description of the three methods, with their strengths 
   and weaknesses,     can be found e.g.~in 
    \citet{Takeuchi2000} and \citet{Weigel2016}. 
    Our principal estimator is the  $1/V_\mathrm{max}$, 
    widespread used in the literature (see Sect.~\ref{Results-SMF literature}) 
    because of its simplicity. In particular, the  $1/V_\mathrm{max}$ technique 
    makes the assumption of uniform spatial distribution 
    of the sources, which is expected to be more robust in our high-redshift bins. 
       
    In addition to these methods, 
    at $z>3$ we experiment an empirical  approach that corrects  
    for source incompleteness by means of the statistical weight $f_\mathrm{obs}$ 
    (see Sect.~\ref{SED fitting 2-stellar mass limit}). This weight plays the same role of 
    the $V_\mathrm{max}$ correction, as it  accounts for the fraction of missing objects 
    below the $[3.6]$ detection limit. The difference  is that  $f_\mathrm{obs}$ 
    is recovered empirically from a deeper parent sample (N17), instead of 
    the accessible observable volume  as in the case of  $V_\mathrm{max}$.  
    Obviously, the empirical method works under the hypothesis that the parent sample is complete. 
    Not to rely  on   $f_\mathrm{obs}(z,\mathcal{M})$ in the range 
    where is too uncertain, we use it only  
    until the correction exceeds a factor two, i.e.~$0.2\upto0.3$\,dex 
    below $\mathcal{M}_\mathrm{lim}$, whereas the previous methods stop at  that threshold.  
    For sake of clarity,  in Fig.~\ref{fig_mf1} we add error bars only to the 
    $1/V_\mathrm{max}$ estimates, noticing that they are of the same order of magnitude as the SWML 
    uncertainties. 
    A description of how these errors have been calculated is provided  
    in Sect.~\ref{Results-Sources of uncertainty}. 
    The three classical estimators coincide  
    in the whole stellar mass range.     
   Such an agreement  
   validates the completeness limits 
   we have chosen, because the estimators would diverge at 
   $\mathcal{M>M}_\mathrm{lim}$ 
   if some galaxy population were missing  \citep[see][]{Ilbert2004}.  
   The $f_\mathrm{obs}$ method is also consistent with the others, in the stellar mass range 
   where they overlap, confirming 
   the validity of our empirical approach.

   We also fit a Schechter function to the $1/V_\mathrm{max}$ points,   
   accounting for  the so-called Eddington  bias \citep{Eddington1913}.
   This systematic bias is caused by stellar mass uncertainties,  which make 
    galaxies scatter from one bin to another in the observed SMF. 
    We remove the Eddington bias 
   as done in \citet{Ilbert2013}, by convolving Eq.~(\ref{eq_schfun}) with 
   a description of the observational uncertainties (Eq.~\ref{eq_gau-lor}) 
   and using the resulting function to fit data points.  
   At $2.5<z\leqslant3.0$, we find that a double Schechter fit, i.e.
   \begin{equation}
  \Phi(\mathcal{M})\mathrm{d}\mathcal{M} = \left[ 
  \Phi^{\star}_1 \left( \frac{\mathcal{M}}{\mathcal{M}_\star} \right)^{\alpha_1} 
  +   \Phi^{\star}_2 \left( \frac{\mathcal{M}}{\mathcal{M}_\star} \right)^{\alpha_2} 
  \right]
  \exp\left( - \frac{\mathcal{M}}{\mathcal{M}_\star} \right) 
  \frac{\mathrm{d}\mathcal{M}}{\mathcal{M}_\star}  \: ,
  \label{eq_dubschfun}
  \end{equation}
    is preferred in the $\chi^2$ fitting. 
   At  $4.5<z\leqslant5.5$ 
    the STY  algorithm does not converge unless assuming unreasonable  values for the turnover mass 
    ($\mathcal{M}_\star\gg10^{11}\,\mathcal{M}_\odot$). 
   A simple power law fits well through the points, so for the STY calculation 
   in this $z$-bin we replace the Schechter function with 
      \begin{equation}
      \Phi(\mathcal{M})=A\left(\frac{\mathcal{M}}{10^{10}\,\mathcal{M}_\odot}\right)^B,
      \label{eq_powlaw} 
      \end{equation}
   where $A=-3.42\pm0.06$ and $B=-1.57\pm0.13$ (given the logarithmic scale,  
   the latter coefficient corresponds to 
   $\alpha=-2.57$ for   a  Schechter function). 
   This should be considered as an upper limit of the $z\simeq5$ SMF. 
   On the other hand, the fit to the $1/V_\mathrm{max}$ points,  
   taking into account stellar mass errors,  recovers a  Schechter profile.

    We report  in Table \ref{tab_schpar}  
    the  Schechter parameters  fitting the $1/V_\mathrm{max}$ points at various redshifts. 
    Those are the fiducial values obtained without imposing any constraint 
    (i.e., the fit assumes a flat prior on $\alpha$, $\mathrm{M}_\star$, and $\Phi_\star$). 
   To deal with the SMF uncertainties, 
   \citet{Song2016} impose in their fitting algorithm a prior for each Schechter parameter, 
    in particular a lognormal PDF($\log\mathcal{M}_\star$) 
     centred at $10.75$ with $\sigma=0.3$\,dex.  \citet{Duncan2014} perform a fit at $z\sim5\upto7$ with
     $\mathcal{M}_\star$ fixed to the value they find at $z\simeq4$. 
    In the same vein,  
    we fit again Eq.~(\ref{eq_schfun}) to the $1/V_\mathrm{max}$  points, 
     but this time with $\log(\mathcal{M}_\star/\mathcal{M}_\odot)=10.6$, 
      in accordance with the Schechter parameter we find at  $z<3$ (see Table \ref{tab_schparMfix}).  
   We adopt this solution consistent with phenomenological models   
   claiming that $\mathcal{M}_\star$ is a  redshift independent parameter \citep[e.g.][]{Peng2010}. 
   Observations at $z<2$ have confirmed this turnover mass to be 
   between  $2\times10^{10}$ and $10^{11}\,\mathcal{M}_\odot$ 
   \citep[e.g.][]{Kauffmann2003b,Bundy2006,Haines2017}. 
   This second fitting function shows how $\alpha$ and $\mathcal{M}_\star$ are coupled: 
   once the ``knee'' of the SMF is fixed, the low-mass slope is forced to 
   be shallower ($\alpha$ increases by $\sim\!~0.15$). 
   We discuss in detail a few caveats related to the fitting procedure   
   in Sect.~\ref{Results-SMF bias}. 
    
   In addition, we compute a  lower limit for the SMF by considering 
   only the most robust $z_\mathrm{phot}$ (Fig.~\ref{fig_mf1}). 
   The selection of  such a ``pure sample'' is done by removing: 
   (i) objects with bimodal 
   PDF($z$), i.e.~with a secondary  (often low-$z$) solution having 
   a non-negligible  Bayesian probability; (ii) objects for which 
   the stellar fit, albeit worse than the galaxy best fit, 
   is still a reasonable interpolation of their photometry 
   ($0<\chi^2_\mathrm{star}-\chi^2_\mathrm{gal}<1$).

\subsection{Sources of uncertainty} 
\label{Results-Sources of uncertainty}
 
  In the statistical error budget of the COSMOS2015 mass function,  
  we take into account 
 Poisson noise ($\sigma_\mathrm{Poi}$), 
 cosmic variance ($\sigma_\mathrm{cv}$),  
  and the scatter due to SED fitting uncertainties ($\sigma_\mathrm{fit}$, not to be confused with 
  the SED fitting systematics discussed below). 
    The total statistical error is  $\sigma_\Phi=(\sigma_\mathrm{Poi}^2 +
  \sigma_\mathrm{cv}^2 + \sigma_\mathrm{fit}^2)^{1/2}$. 
  
\begin{figure}
 \includegraphics[width=0.99\columnwidth]{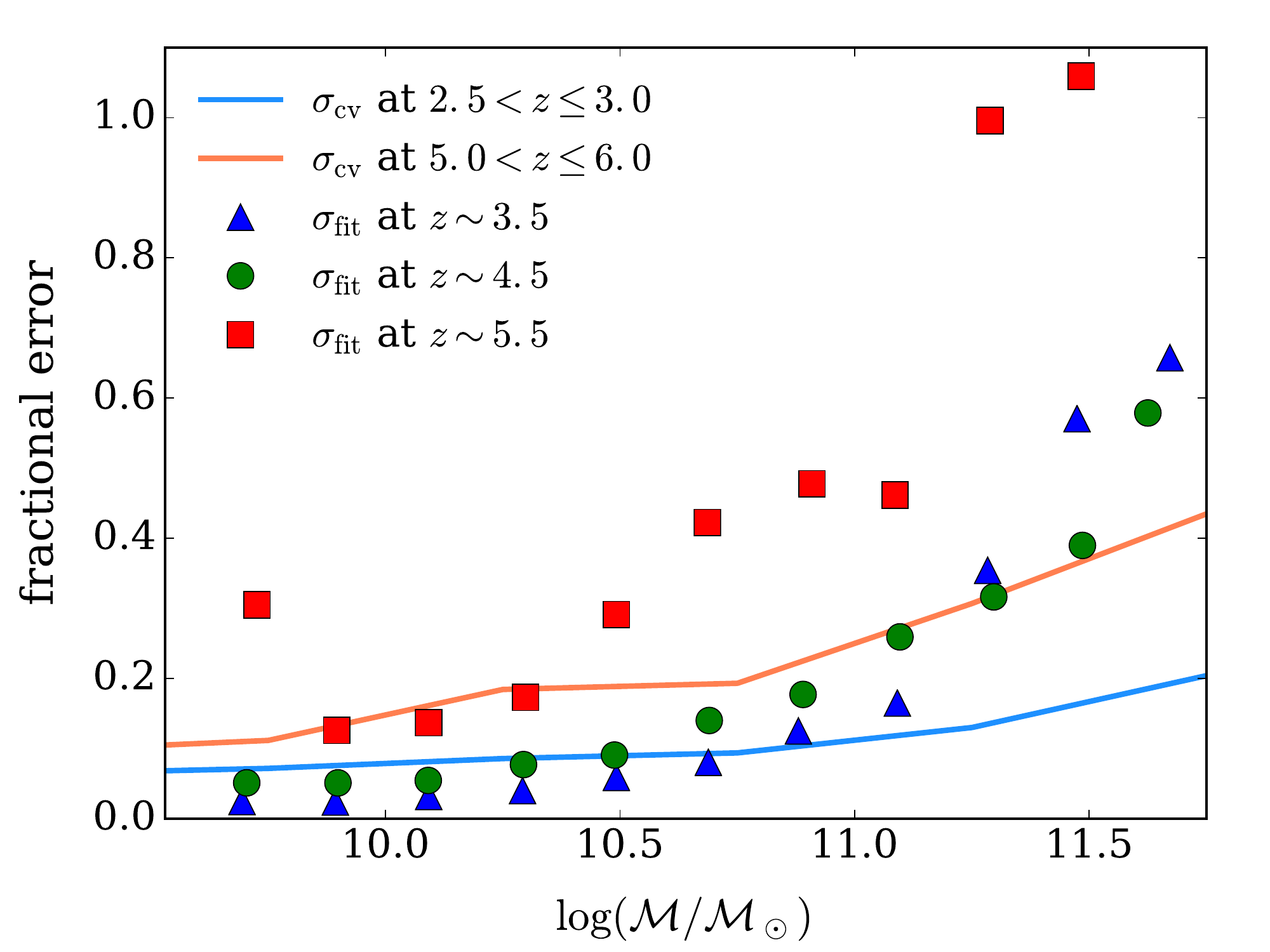}
 \caption{SMF uncertainties due to cosmic variance 
 (expressed as a fractional error: $\sigma_\mathrm{cv}/\Phi$) and SED fitting ($\sigma_\mathrm{fit}/\Phi$),  
 as a function of redshift and stellar mass. We show the impact of cosmic variance for the two extreme 
 bins $2.5<z\leqslant3$ and $5<z\leqslant6$ (blue and red line, respectively). 
 For $\sigma_\mathrm{fit}$ we take as example three 
 redshift  bins: $2.5<z\leqslant3.5$ (blue triangles), $3.5<z\leqslant4.5$ (green circles), and $4.5<z\leqslant5.5$ (red squares). }
 \label{fig_errs}
 \end{figure}

  Cosmic variance is estimated by means of a modified 
  version of the ``cosmic variance calculator'' by \citet{Moster2011}, for the geometry 
  of our survey and the cosmology we assumed. 
  The main contribution to $\sigma_\mathrm{fit}$ comes from photometric  errors  
  and degeneracies  between different  SEDs.  
  Starting from the Monte Carlo mock samples described in Sect.~\ref{SED fitting 2-stellar mass},
  we obtain 100 realisations of our SMF,  whose dispersion in a given stellar mass bin  
  is taken as the $\sigma_\mathrm{fit}$ at that mass.  
  A summary of these sources of uncertainty at $z\geqslant3$  is shown in Fig.~\ref{fig_errs}.

 In addition to random errors, the SED fitting may introduce systematic offsets in the measured SMF, depending on 
 the adopted recipe. An example is the IMF: assuming \citet{Salpeter1955} the logarithmic stellar masses will be on average 
 $0.24$\,dex larger than the ones obtained with an IMF as in \citet{Chabrier2003}. 
 At $z>3$, the largest biases are generally expected from  the  $z_\mathrm{phot}$ estimates, since an entire 
 galaxy class may be systematically put in a different $z$-bins if the code misinterprets their SED 
 (e.g.~because of the degeneracy between Lyman and Balmer breaks).

\begin{figure}
 \includegraphics[width=0.99\columnwidth]{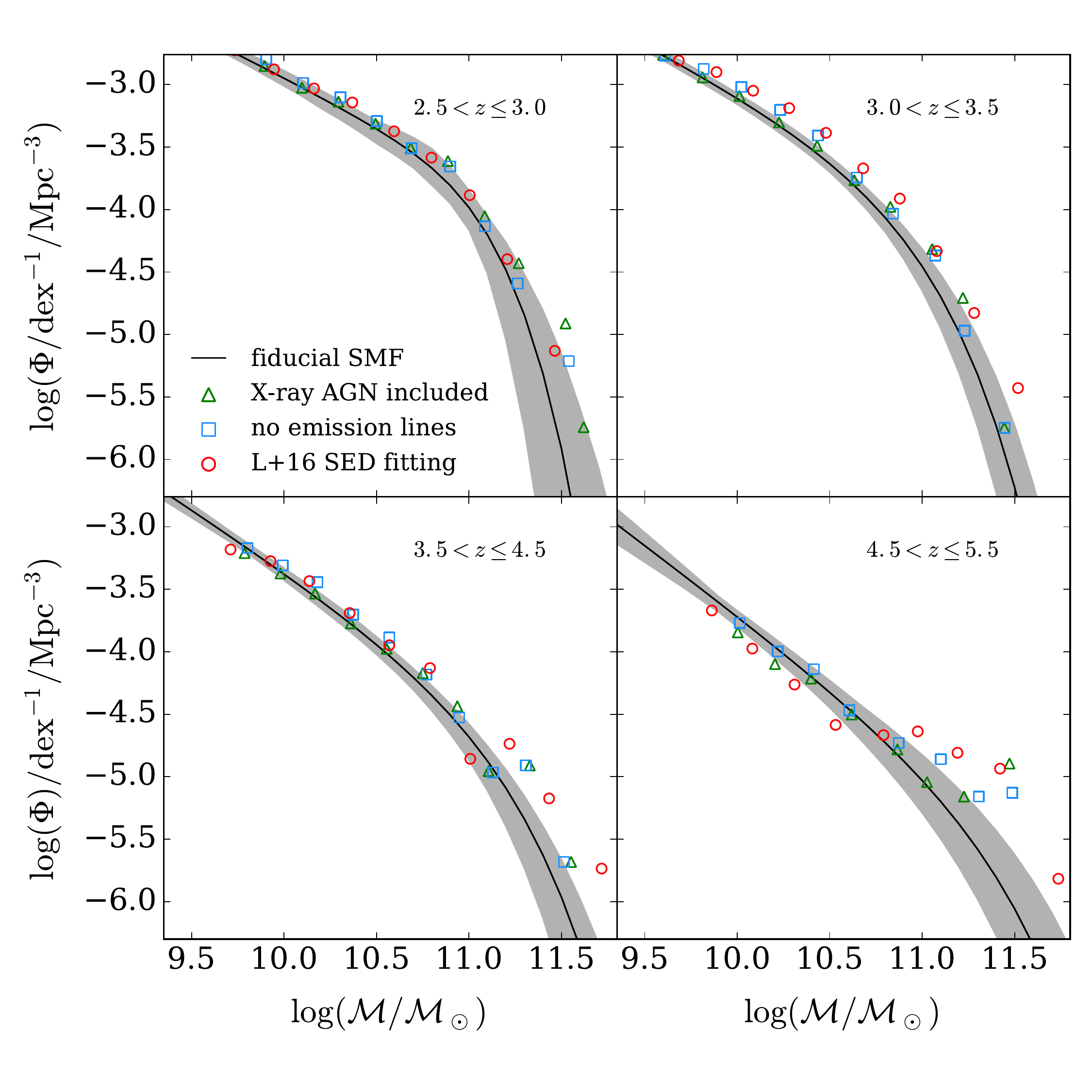}
 \caption{Alternate versions of the COSMSO2015 galaxy stellar mass function: 
 the SMF after the reintroduction of X-ray sources (green triangles), 
  a modified version without emission lines in the synthetic galaxy templates (blue squares), and the SMF based 
  on the original SED fitting of L16 (red circles). 
 The fiducial estimates, already shown in Fig.~\ref{fig_mf1}, are reproduced here 
 with solid lines and shaded areas. 
 }
 \label{fig_mf_sys}
 \end{figure}

  Figure \ref{fig_mf_sys} contains  three flavours of the COSMOS2015 SMF obtained 
  by modifying the SED fitting recipe. 
  One of these alternate estimates is  based on the  photometric redshifts and masses  provided by L16 
  ($z_\mathrm{phot,L16}$ and $\mathcal{M}_\mathrm{L16}$). 
  For the stellar mass computation, the set-up to build BC03 templates 
  includes  not only  \citet[][this time without the graphite bump]{Calzetti2000} 
  but also the extinction law $\propto \lambda^{-0.9}$  described in \citet{Arnouts2013}. 
  The metallicity grid of L16 templates is narrower than in the present work, not including 
  $Z=0.004$. The recipe to add nebular emission lines to the synthetic SEDs, although conceptually 
  identical to the one described in Sect.~\ref{SED fitting-method}, assumed slightly different values   for the line strength ratios. 
  The most evident feature in this version, contrasting it with our fiducial SMF, 
  is the excess of massive galaxies at $z>4$  
  because of the higher stellar contamination (we remove more interlopers with the  additional criteria described in Sect.~\ref{SED fitting-Stellar contamination}). 
  We also observe an enhanced  number density 
  (but less than a factor of two) at  $3<z\leqslant3.5$; the origin of this offset likely resides 
  in the different $z_\mathrm{phot}$  estimates, as discussed 
  in Appendix \ref{APP1}.    
  
  Another source of systematic errors  is the addition of nebular emission lines 
  to the BC03  templates. The issue is debated in the literature,  
  with various authors finding from marginal to substantial variations, depending 
  on the code and galaxy sample used. 
  For instance  \citet{Stark2013} find that with 
  the addition of emission lines $\mathcal{M}$ decreases by a factor from 1.2 to 2, from $z\simeq4$ to 6. 
 The offset found by \citet{deBarros2014} in  a similar redshift range is on average 
  0.4\,dex, up to 0.9  for the stronger LBG emitters.\footnote{
  However \citeauthor{deBarros2014} find 
  such a large difference   not only by introducing nebular emission lines 
  but also changing other SED fitting parameters like the SFH. 
  When considering only the impact 
  of emission lines the stellar mass offset is much smaller, especially at 
  $\log(\mathcal{M/M}_\odot)>10$ \citep[][top-right panel of Fig.~13]{deBarros2014}. 
  } 
  On the other hand, a recent study 
  on H$\alpha$ emission in galaxies at $3<z_\mathrm{spec}<6$  
  suggests that previous SED-dependent analyses 
  may overestimate H$\alpha$ equivalent width  
  \citep[][their Fig.~5]{Faisst2016a}.   
   \citet{Stefanon2015} compared three different SED fitting methods: 
   using their standard $z<3$ calibration, 
   they find no tension with the $\mathcal{M}$ estimates 
   obtained without emission lines. On the contrary, assuming  EWs evolving 
   with redshift  \citep{Smit2014} their stellar masses decrease on average by 0.2\,dex,  
   although for $\mathcal{M}\lesssim10^{10}\,\mathcal{M}_\odot$ 
   galaxies at $4<z<5$ they find an opposite trend.  
   In general SED fitting stellar masses are less sensitive to this bias when a large 
   number of bands are used \citep{Whitaker2014}  especially when  
   the estimates are derived through a Bayesian approach rather than best-fit templates 
   \citep{Salmon2015}.
  Neglecting lines in our SED fitting procedure,    
  $\mathcal{M}$ estimates increase on average by $\lesssim0.05$\,dex or less, with noticeable 
  exceptions for some individual galaxy.

  We also investigate how the SMF changes 
  if we re-introduce the  X-ray sources that have been 
  excluded from the sample in Sect.~\ref{Dataset-photometry}.    
  First we verify that their $z_\mathrm{phot}$ are in sufficiently good agreement 
  with estimates derived from a more accurate fitting done  with AGN templates 
  (Salvato et al., in preparation). Then,  we recompute the stellar mass of 
  each X-ray emitter  by means of a 3-component SED fitting code \citep[\texttt{sed3fit},][]{Berta2013}.\footnote{
  \url{http://cosmos.astro.caltech.edu/page/other-tools}} 
  This tool relies on the energy balance between dust-absorbed UV stellar continuum and the reprocessed emission 
  in the IR \citep[like \texttt{MAGPHYS},][]{daCunha2008}, and  also accounts for an additional AGN component 
  from the torus library of \citet[][see also \citealp{Fritz2006}]{Feltre2012}. 
  X-ray luminous ($L_\mathrm{x}>10^{44}\,\mathrm{erg\,s}^{-1}$) AGN are usually hosted
  in massive ($\mathcal{M}\gtrsim10^{11}\,\mathcal{M}_\odot$) galaxies 
  \citep[e.g.][]{Bundy2008,Brusa2009,Hickox2009}. 
 They increase the exponential tail of our SMF at least at $z\leqslant 3$, while  
  at higher redshift the number of sources detected by   Chandra or XMM  
  is too small  (Fig.~\ref{fig_mf_sys}). 
  At $z>3$,  massive  galaxies  can 
  host as well an active black hole and disregarding its contribution  in the SED fitting 
    may cause a  stellar mass overestimate  of $0.1\upto0.3$\,dex 
  \citep{Hainline2012,Marsan2016}. The impact of different AGN populations 
  on the SMF shall be discussed in a future work (Delvecchio et al., in preparation). 
  We do not add the various systematic errors together, as done for the statistical ones, 
  because their combined effect is different from the sum of the offsets 
  measured  by changing one parameter at a time.

\begin{table*}[t!]
\caption{Schechter parameters of the COSMOS2015 galaxy SMF (also dividing the sample in active and 
passive galaxies). A double Schechter function 
(Eq.~\ref{eq_dubschfun}) is used up to $2.5<z\leqslant3$, a single Schechter function (Eq.~\ref{eq_schfun}) 
beyond that bin.  
}
\label{tab_schpar}
\setlength{\extrarowheight}{1ex}
\centering
\begin{tabular}{lccccc}
\hline\hline
redshift & $\log\mathcal{M}_\star$  &  $\alpha_1$ &$\Phi_{\star1}$  & $\alpha_2$ &  $\Phi_{\star2}$\\
 &  $[h_{70}^{-2}\,\mathcal{M}_\odot]$ & & $[10^{-3}\,h_{70}^3\,\mathrm{Mpc}^{-3}]$ & &   $[10^{-3}\,h_{70}^3\,\mathrm{Mpc}^{-3}]$ \\ 
 [1ex] \hline
\multicolumn{6}{c}{Total sample}  \\ 
 $0.2<z\leqslant0.5$ & $10.78^{+0.13}_{-0.14}$   & $-1.38^{+0.08}_{-0.25}$  & $1.187^{+0.633}_{-0.969}$  & $-0.43^{+0.62}_{-0.60}$  & $1.92^{+0.73}_{-0.78}$ \\ 
 $0.5<z\leqslant0.8$ & $10.77^{+0.09}_{-0.08}$   & $-1.36^{+0.05}_{-0.06}$  & $1.070^{+0.287}_{-0.315}$  & $0.03^{+0.43}_{-0.43}$  & $1.68^{+0.33}_{-0.33}$ \\ 
 $0.8<z\leqslant1.1$ & $10.56^{+0.05}_{-0.05}$   & $-1.31^{+0.05}_{-0.06}$  & $1.428^{+0.306}_{-0.308}$  & $0.51^{+0.35}_{-0.34}$  & $2.19^{+0.40}_{-0.41}$ \\ 
 $1.1<z\leqslant1.5$ & $10.62^{+0.08}_{-0.07}$   & $-1.28^{+0.05}_{-0.05}$  & $1.069^{+0.222}_{-0.240}$  & $0.29^{+0.40}_{-0.42}$  & $1.21^{+0.23}_{-0.22}$ \\ 
 $1.5<z\leqslant2.0$ & $10.51^{+0.08}_{-0.07}$   & $-1.28^{+0.06}_{-0.06}$  & $0.969^{+0.202}_{-0.208}$  & $0.82^{+0.48}_{-0.52}$  & $0.64^{+0.18}_{-0.17}$ \\ 
 $2.0<z\leqslant2.5$ & $10.60^{+0.15}_{-0.12}$   & $-1.57^{+0.12}_{-0.21}$  & $0.295^{+0.173}_{-0.177}$  & $0.07^{+0.70}_{-0.74}$  & $0.45^{+0.12}_{-0.12}$ \\ 
 $2.5<z\leqslant3.0$ & $10.59^{+0.36}_{-0.36}$   & $-1.67^{+0.26}_{-0.26}$  & $0.228^{+0.300}_{-0.300}$  & $-0.08^{+1.73}_{-1.73}$  & $0.21^{+0.14}_{-0.38}$ \\ 
 $3.0<z\leqslant3.5$ & $10.83^{+0.15}_{-0.15}$   & $-1.76^{+0.13}_{-0.11}$  & $0.090^{+0.064}_{-0.039}$  & &  \\ 
 $3.5<z\leqslant4.5$ & $11.10^{+0.21}_{-0.21}$   & $-1.98^{+0.14}_{-0.13}$  & $0.016^{+0.020}_{-0.009}$  & &  \\ 
 $4.5<z\leqslant5.5$ & $11.30^{+1.22}_{-1.22}$   & $-2.11^{+0.34}_{-0.22}$  & $0.003^{+0.002}_{-0.002}$  & &  \\ 
[1ex] \hline
\multicolumn{6}{c}{Active sample}  \\ 
 $0.2<z\leqslant0.5$ & $10.26^{+0.07}_{-0.06}$   & $-1.29^{+0.03}_{-0.03}$  & $2.410^{+0.341}_{-0.337}$  & $1.01^{+0.34}_{-0.36}$  & $1.30^{+0.25}_{-0.25}$ \\ 
 $0.5<z\leqslant0.8$ & $10.40^{+0.06}_{-0.06}$   & $-1.32^{+0.02}_{-0.02}$  & $1.661^{+0.188}_{-0.176}$  & $0.84^{+0.25}_{-0.26}$  & $0.86^{+0.13}_{-0.12}$ \\ 
 $0.8<z\leqslant1.1$ & $10.35^{+0.05}_{-0.05}$   & $-1.29^{+0.02}_{-0.02}$  & $1.739^{+0.166}_{-0.164}$  & $0.81^{+0.22}_{-0.24}$  & $0.95^{+0.12}_{-0.11}$ \\ 
 $1.1<z\leqslant1.5$ & $10.42^{+0.05}_{-0.05}$   & $-1.21^{+0.02}_{-0.02}$  & $1.542^{+0.127}_{-0.122}$  & $1.11^{+0.20}_{-0.21}$  & $0.49^{+0.07}_{-0.07}$ \\ 
 $1.5<z\leqslant2.0$ & $10.40^{+0.05}_{-0.05}$   & $-1.24^{+0.02}_{-0.02}$  & $1.156^{+0.107}_{-0.105}$  & $0.90^{+0.23}_{-0.24}$  & $0.46^{+0.07}_{-0.06}$ \\ 
 $2.0<z\leqslant2.5$ & $10.45^{+0.07}_{-0.07}$   & $-1.50^{+0.05}_{-0.05}$  & $0.441^{+0.093}_{-0.088}$  & $0.59^{+0.34}_{-0.35}$  & $0.38^{+0.06}_{-0.06}$ \\ 
 $2.5<z\leqslant3.0$ & $10.39^{+0.14}_{-0.10}$   & $-1.52^{+0.07}_{-0.08}$  & $0.441^{+0.149}_{-0.137}$  & $1.05^{+0.54}_{-0.61}$  & $0.13^{+0.04}_{-0.04}$ \\ 
 $3.0<z\leqslant3.5$ & $10.83^{+0.08}_{-0.08}$   & $-1.78^{+0.05}_{-0.05}$  & $0.086^{+0.027}_{-0.021}$  & &  \\ 
 $3.5<z\leqslant4.0$ & $10.77^{+0.12}_{-0.11}$   & $-1.84^{+0.10}_{-0.09}$  & $0.052^{+0.028}_{-0.020}$  & &  \\ 
 $4.0<z\leqslant6.0$ & $11.30^{+0.15}_{-0.15}$   & $-2.12^{+0.05}_{-0.05}$  & $0.003^{+0.001}_{-0.000}$  & &  \\
[1ex] \hline 
\multicolumn{6}{c}{Passive sample}  \\ 
  $0.2<z\leqslant0.5$ & $10.83^{+0.07}_{-0.08}$   & $-1.30^{+0.26}_{-0.43}$  & $0.098^{+0.177}_{-0.177}$  & $-0.39^{+0.34}_{-0.20}$  & $1.58^{+0.16}_{-0.21}$ \\ 
 $0.5<z\leqslant0.8$ & $10.83^{+0.04}_{-0.04}$   & $-1.46^{+0.36}_{-0.48}$  & $0.012^{+0.024}_{-0.024}$  & $-0.21^{+0.14}_{-0.09}$  & $1.44^{+0.08}_{-0.08}$ \\ 
 $0.8<z\leqslant1.1$ & $10.75^{+0.02}_{-0.02}$   & $-0.07^{+0.04}_{-0.04}$  & $1.724^{+0.059}_{-0.060}$  & &  \\ 
 $1.1<z\leqslant1.5$ & $10.56^{+0.03}_{-0.03}$   & $0.53^{+0.08}_{-0.08}$  & $0.757^{+0.023}_{-0.024}$  & &  \\ 
 $1.5<z\leqslant2.0$ & $10.54^{+0.03}_{-0.03}$   & $0.93^{+0.13}_{-0.12}$  & $0.251^{+0.015}_{-0.017}$  & &  \\ 
 $2.0<z\leqslant2.5$ & $10.69^{+0.07}_{-0.07}$   & $0.17^{+0.19}_{-0.16}$  & $0.068^{+0.006}_{-0.007}$  & &  \\ 
 $2.5<z\leqslant3.0$ & $10.24^{+0.11}_{-0.11}$   & $1.15^{+0.63}_{-0.50}$  & $0.028^{+0.007}_{-0.010}$  & &  \\ 
 $3.0<z\leqslant3.5$ & $10.10^{+0.10}_{-0.09}$   & $1.15$  & $0.010^{+0.002}_{-0.002}$  & &  \\ 
 $3.5<z\leqslant4.0$ & $10.10$   & $1.15$  & $0.004^{+0.001}_{-0.001}$  & &  \\ 
[1ex] \hline
\end{tabular} 
\end{table*}

 \subsection{Comparison with previous work} 
 \label{Results-SMF literature}
 
\begin{figure*}[t]
    \centering
 \includegraphics[width=0.9\textwidth]{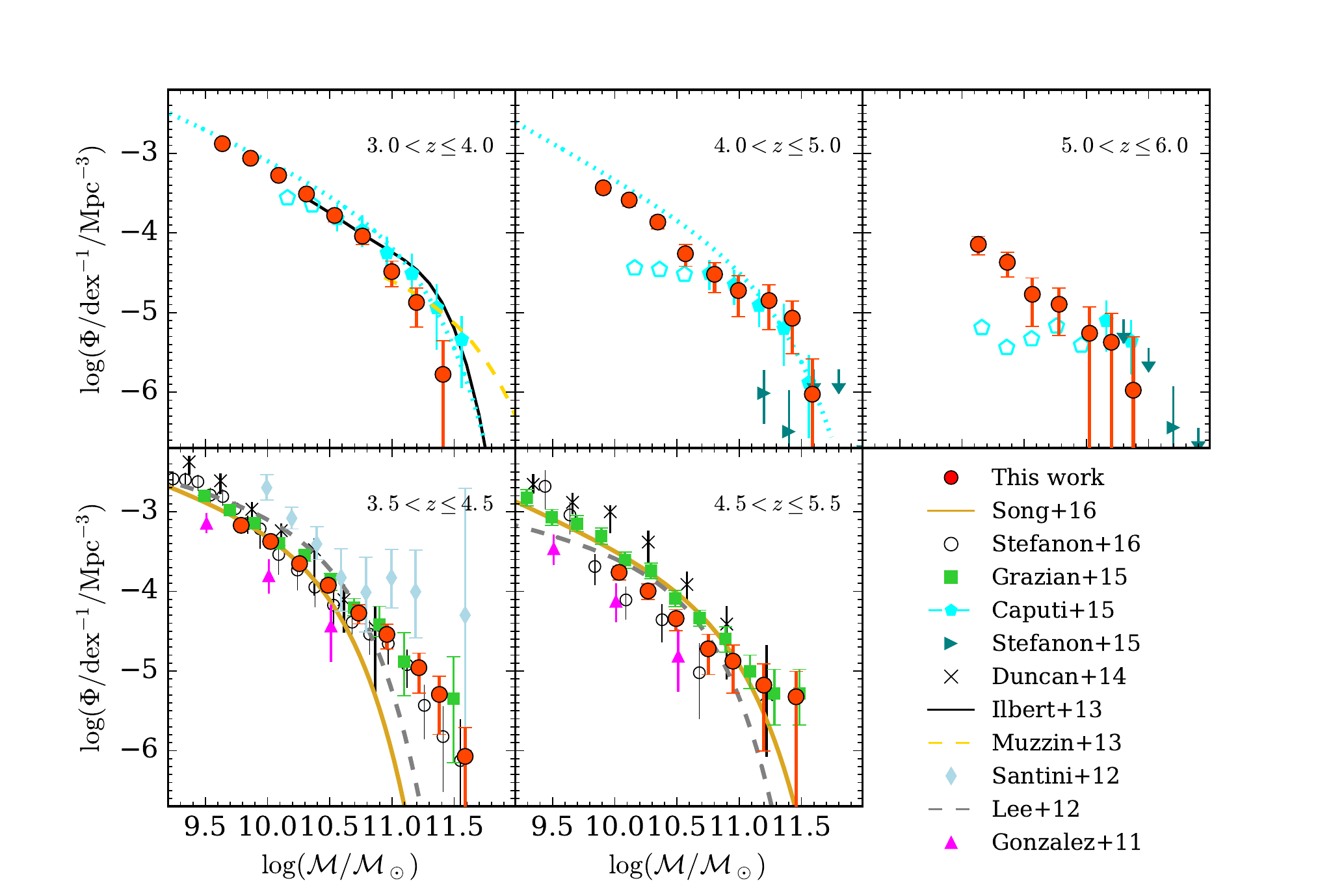}
  \caption{Comparison to galaxy SMFs from the literature. 
  Our $1/V_\mathrm{max}$  measurements are shown as red circles. 
  If needed, we converted   
   estimates from the literature to  \citet{Chabrier2003} IMF.  
    \textit{Upper panels:} We plot the SMF estimates from 
     \citet[][]{Caputi2015}, with filled (empty) blue pentagons above (below) their mass completeness limit 
    (in addition, their STY fit is shown by a blue dotted line). 
    For sake of clarity, we plot only the Schechter functions (not the  $1/V_\mathrm{max}$ points that have been fit) 
    from \citet{Ilbert2013} and  \citet{Muzzin2013b}; in both cases  the Schechter function 
    (black solid and yellow dashed line,  respectively) is truncated at their stellar mass   completeness limit. 
    Rightward triangles and downward  arrows show $1/V_\mathrm{max}$ and upper  limits  
    from a conservative estimate by  \citet{Stefanon2015}.  
    \textit{Lower panels:} $\mathcal{M}$-selected SMFs published in \citet[][light blue diamonds]{Santini2012}, 
    \citet[][black crosses]{Duncan2014}, and  \citet[][green squares]{Grazian2015}. 
    Other SMFs of LBG samples are taken from \citet[][upward triangles]{Gonzalez2011}, 
    \citet[][grey dashed lines]{Lee2012}, \citet[][orange solid line]{Song2016}, \citet[][empty circles]{Stefanon2016}.}
  \label{fig_mf_lit}
  \end{figure*}

 We  compare the SMF of COSMOS2015 galaxies with the literature 
 (Fig.~\ref{fig_mf_lit}) recomputing it 
 in the same redshift bins used by other authors (masses are 
 rescaled to Chabrier's IMF when required).
 We  plot our $1/V_\mathrm{max}$ estimates only, 
 as this is the same estimator used by most of the other authors. 
 Our error bars include $\sigma_\Phi$ errors 
 defined in Sect.~\ref{Results-Sources of uncertainty}. 
 
 From the literature, we select  papers published in the last five years. 
 Some of them, like the present work, collect a galaxy sample where 
 photometric  redshifts and stellar mass are derived via SED fitting 
 \citep[``$\mathcal{M}$-selected'' SMFs:][]{Santini2012,Ilbert2013,Muzzin2013b,Duncan2014,Caputi2015,Stefanon2015,Grazian2015}. 
 Others   
 use rest-frame optical colours to select LBGs;  
 after determining their $\mathcal{M}$ vs $L_\mathrm{UV}$  distribution, 
 they convolve it with a LF estimate in the UV  to derive the SMF 
 \citep[``$L_\mathrm{UV}$-selected'' SMFs:][]{Gonzalez2011,Lee2012,Song2016,Stefanon2016}. 
 We will show that our estimates are in excellent agreement with  $\mathcal{M}$-selected SMFs 
 derived from deep space-based surveys. 
 On the other hand there are some differences with the $L_\mathrm{UV}$-selected SMFs, as  
 LBGs samples are expected to miss 
 quiescent and dust-obscured galaxies. 
 
 Among the $\mathcal{M}$-selected SMFs, those with NIR data   
 from ground-based facilities (VISTA and UKIRT) are shown in the 
 top panels of Fig.~\ref{fig_mf_lit}. 
 Our results are in overall agreement with \citet{Caputi2015}, 
 which is an updated version of $K$-selected galaxy SMFs  in UDS and COSMOS 
  \citep[][]{Caputi2011,Ilbert2013,Muzzin2013b}. These estimates are limited to $K_\mathrm{s}<24$, 
  while \citet{Caputi2015} account for fainter NIR sources ($24<K_\mathrm{s}\leqslant24.7$) 
  bright enough in MIR to result in  $\mathcal{M}\gtrsim10^{11}\,\mathcal{M}_\odot$.
  Including them in their SMF, \citet{Caputi2015}  find a number density of massive galaxies 
   comparable to ours, except at $5<z\leqslant6$ where they have more galaxies 
  at $11 <\log(\mathcal{M}/\mathcal{M}_\odot)<11.5$. The discrepancy can be due to cosmic variance 
  (at those redshifts, \citeauthor{Caputi2015} rely mostly on UDS data) but we cannot rule out other explanations, 
  e.g.~a difference in the $z_\mathrm{phot}$ calculation (more uncertain at such high redshifts, see their Fig.~10).  
  We also note that the SMF of \citet{Caputi2011}, one of the original results revised in \citet{Caputi2015}, was  computed 
  with the 2007 model of Charlot \& Bruzual  and  converted to BC03 by applying a constant $1.24$ factor. 
  Moreover, \citet{Caputi2015} do not use SPLASH, but shallower IRAC data \citep{Sanders2007}. 
  Figure~\ref{fig_mf_lit} also shows the STY fit of \citet{Caputi2015} to demonstrate 
  that the extrapolation of the fit below their stellar mass limit is consistent with our data points.

    \begin{figure*}[t]
    \centering
 \includegraphics[width=0.9\textwidth]{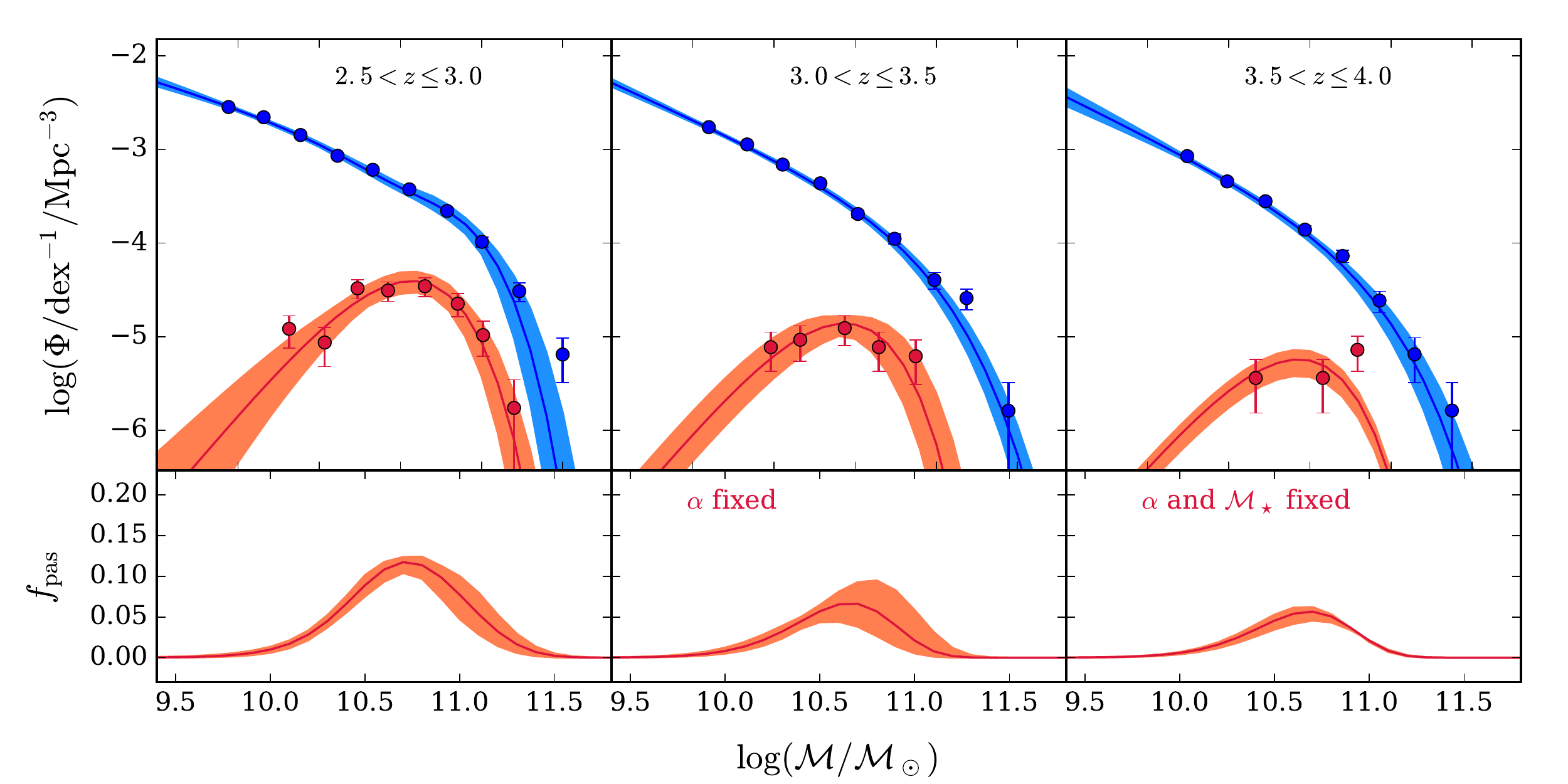}
  \caption{Active and passive SMFs (blue and red symbols, respectively) in the same  
  redshift bins of Fig.~\ref{fig_mf1}. Filled circles are 
  the $1/V_\mathrm{max}$ points, while solid lines represent the Schechter functions 
  fitting to them   
  (shaded areas being the $1\sigma$ uncertainty of the fit). In the bottom panels we show the 
  fraction of passive galaxies ($f_\mathrm{pas}$) as a function of stellar mass in the same $z$-bins. 
  }
  \label{fig_mf_typ}
  \end{figure*} 
  
    At $3<z \leqslant4$, we also compare with \citet{Ilbert2013} and \citet{Muzzin2013b}, two independent estimates 
    both based on UltraVISTA DR1 images. Surprisingly, the results found by the two collaborations, 
    despite the good agreement at lower redshifts, diverge in this bin:  massive galaxies  ($>\!10^{11}\,\mathcal{M}_\odot$) 
    are relatively  abundant in \citeauthor{Muzzin2013b}, whereas the exponential tail  in 
     \citeauthor{Ilbert2013} decreases more steeply. 
    On the other hand, the latter found more low-mass galaxies   
    and their SMF is defined down to a stellar mass limit that is 0.6\,dex smaller than the limit of \citet{Muzzin2013b}.\footnote{ 
    In \citet{Muzzin2013b} the magnitude cut is $K_\mathrm{s}<23.4$, whereas in \citet{Ilbert2013} it is 0.6\,mag brighter, but 
    this should account for a $0.24$\,dex difference only. 
    In fact, the two papers use a different  definition of $\mathcal{M}_\mathrm{lim}$
    (the one adopted by \citealp{Ilbert2013} is the same as in our analysis).  
    With respect to both papers, we reach smaller masses thanks to the deeper $\mathcal{A}_\mathrm{UD}$ images 
     and the panchromatic selection  discussed in Appendix \ref{APP3}.  
     }
    The differences between \citet{Ilbert2013} and \citet{Muzzin2013b} cannot be explained 
    by cosmic variance, since both SMFs are derived in the same field (using the same raw data). 
    On the contrary, we find that the main reason for the discrepancy between \citet{Ilbert2013} and 
    our SMF at $\mathcal{M}>10^{11}\,\mathcal{M}_\odot$ 
    is our smaller (Ultra-Deep) volume: if we derive the SMF over the whole COSMOS area  
    ($\mathcal{A}_\mathrm{D}+\mathcal{A}_\mathrm{UD}$, see Fig.~\ref{fig_sky})  
    we find a better agreement. Also the new MIR coverage from SPLASH  plays 
    a role  (see Appendix \ref{APP3}).   
    
    To get an insight on the  SED fitting uncertainties at $z>3$, we inspect 
    \citet{Ilbert2013} and \citet{Muzzin2013a} catalogues, both publicly available.  
    In the latter sample, among 165  galaxies with $3<z_\mathrm{phot,M13}<4$ and $\log(\mathcal{M/M}_\odot)>10.94$,   
    only 25\% remains in the same   $z$-bin 
     if one replaces \citeauthor{Muzzin2013a} photometric redshifts  with 
      \citeauthor{Ilbert2013} estimates.\footnote{
    Here we consider only objects cross-matched between the two catalogues, with a 0.6\arcsec searching radius.}    
   We compare  with the  COSMOS 
   spectroscopic redshifts introduced in Sect.~\ref{SED fitting-spec validation}, which were not available when both catalogues 
   have been built. 
    Among the galaxies that are at $3<z_\mathrm{phot,M13}\leqslant4$ 
    according to \citet{Muzzin2013a}, 
    94 have a match in our COSMOS spectroscopic catalogue. We find that  69 of them are catastrophic errors, i.e.~$\vert z_\mathrm{phot,M13} - z_\mathrm{spec} \vert >0.15(1+z_\mathrm{spec})$. 
    The number of robust estimates 
    in the same $z$-bin rises to 87 (out of 107 galaxies spectroscopically observed) 
    when repeating this test  using  \citet{Ilbert2013} photometric redshift.

    To show the impact of this kind of uncertainties, in Fig.~\ref{fig_mf_lit}  
    we plot also the $1/V_\mathrm{max}$ estimates 
    (and upper limits) derived by \citet[][]{Stefanon2015} from the subsample of  
    their most robust  galaxies, i.e.~those objects  (detected in UltraVISTA DR1)  for which the $z_\mathrm{phot}$ 
    satisfies a series of strict reliability criteria. The drop in number density with respect to the other measurements 
     gives an idea of the uncertainties that still affect the high-$z$ SMF. 
     
    A comparison with $\mathcal{M}$-selected SMFs derived from HST/WFC3 detections 
    is shown in  the bottom panels of Fig.~\ref{fig_mf_lit}. 
    An estimate using  the Early Release Science in the GOODS-South field (complementing 
    WFC3 data with a deep Hawk-I survey) has been provided by  \citet{Santini2012}. 
    Looking to the error bars of their $1/V_\mathrm{max}$ points, one can appreciate  
    the progress made by the latest studies in terms of  statistics. 
    HST data available to \citet{Santini2012}  covered 
    about $33$\,arcmin$^2$, with less than 50 galaxies  
    located between $z=3.5$ and 4.5.  
     \citet{Duncan2014} provide another SMF estimate in the GOODS-South field, but using more recent 
     data from CANDELS.  
     Their $1/V_\mathrm{max}$ points  are consistent with ours at $z\simeq4$, while at $z\simeq5$ their results 
     are systematically higher by $0.4\upto0.5$\,dex (although the discrepancy is smaller than $2\sigma$ because 
     of their large sample variance).

    We find a remarkably good agreement with the SMF measured  by  \citet{Grazian2015} 
    over  three CANDELS field (GOODS-South, UDS, HUDF). We underline that the authors have not 
    used the  CANDELS-COSMOS field, so their estimate is completely independent from ours. 
    At very high masses ($>\!5\times10^{11}\,\mathcal{M}_\odot$) 
    the $1/V_\mathrm{max}$ points of \citeauthor{Grazian2015} 
    seem to be affected  by a similar level of uncertainty of our SMF, 
    confirming the  excellent quality of our ground-based data.  
    Given the good agreement between CANDELS and COSMOS2015 $1/V_\mathrm{max}$ 
    mass functions,  one would expect that the  Schechter functions derived from 
    the data points are similar, too. However, the two fits differ from each other, 
    as we will discuss in Sect.~\ref{Results-SMF bias}.

   Figure~\ref{fig_mf_lit} also shows the comparison with the $L_\mathrm{UV}$-selected SMFs  
    \citep[][]{Gonzalez2011,Lee2012,Song2016,Stefanon2016}. 
    At $3.5<z<4.5$ the LBG sample of \citet{Stefanon2016} is complemented with UltraVISTA DR2 galaxies they 
    select according to their $z_\mathrm{phot}$ estimates. 
    Apart from that distinction, 
    these studies should be considered as an estimate of the abundance of UV-bright  
   active galaxy, rather than a census of the entire high-$z$ galaxy population 
   \citep[as clearly stated e.g.~in][]{Lee2012}. Moreover, LBGs are usually selected 
   by means of colour criteria that photometric errors can impair more heavily than 
   $z_\mathrm{phot}$ estimates \citep[see][]{Duncan2014}. 
   \citet{LeFevre2015} and \citet{Thomas2017} also show that   
    dust and inter-galactic medium extinction 
    move a significant fraction of galaxies outside the standard LBG regions in  colour-colour diagrams. 
    Discrepancies with $\mathcal{M}$-selected SMFs are also due to the different method to recover 
    stellar masses: galaxies that are outliers in the  $\mathcal{M}$ vs $L_\mathrm{UV}$ distribution 
    can be biased when an average $\mathcal{M}(L_\mathrm{UV})$ relation is assumed. 
    
   Nonetheless, the agreement we find at $4.5<z<5.5$ with \citet{Lee2012} and \citet{Song2016} 
   may suggest that most of the galaxies at those redshifts are going through a star-forming phase. 
   At $3.5<z<4.5$, their high-mass end is much lower than ours, an indication that the bulk of 
   dusty massive galaxies starts to form already  at  $z\sim4$. In addition, we include also the SMF of 
   \citet{Gonzalez2011}, which has a normalisation lower by a factor $\sim\!2$, 
   to show that also these analyses have to deal with severe uncertainties: in general, 
   the SMF of LBGs is  derived 
   from their UV luminosity function by assuming a $L_\mathrm{UV}$-$\mathcal{M}$ relation, 
   but such a conversion can hide a number of systematic effects \citep[see discussion in][]{Song2016,Harikane2016}.

  \subsection{Build-up of the quiescent SMF at high redshift}   
  \label{Results-SMF active/passive}

  After dividing our sample in active and passive galaxies through the $NUVrJ$ criterion 
  (Sect.~\ref{SED fitting 2-Galaxy type}),  
  we derive the SMF of each galaxy type up to $z=4$.  
  The method of \citet{Pozzetti2010} is applied to each sub-sample in order to compute the corresponding 
  limiting mass $\mathcal{M}_\mathrm{lim}$,  
  which differs for active and passive galaxies as the range of 
  $\mathcal{M}/L$ ratios is not the same \citep[see e.g.][]{Davidzon2013}. 
  The SMFs are shown in Fig.~\ref{fig_mf_typ} (top panels) 
  and their Schechter parameters are reported in Table \ref{tab_schpar}.  
   For the passive mass functions 
  at  $z>3$ we kept $\alpha$ fixed to the value found at $2.5<z\leqslant3$, and at $3.5<z<4$ 
  we also fix $\mathcal{M}_\star$ because otherwise the small number of $1/V_\mathrm{max}$ 
   points   cannot constrain the fit effectively.
   At $z>4$ the uncertainties are still too large to 
  perform a robust $NUVrJ$ classification. 
  However,  the small number of passive candidates we found 
  (which could potentially include several interlopers)  
  indicates that  the  SMF at $z>4$  is essentially composed by active galaxies only. 
  
  At $2.5<z\leqslant3$, similarly to the total sample,  the active SMF 
  shows a mild double Schechter profile (Eq.~\ref{eq_dubschfun}),  
  with the characteristic bump  between $10^{10}$ and $10^{11}\,\mathcal{M}_\odot$. 
  Most of the $NUVrJ$-passive galaxies have a stellar mass within that range, while in 
  the massive end of the SMF  red  galaxies are exclusively 
  dusty objects belonging to the star-forming population 
  (confirming the trend that \citealp{Martis2016} find at $z<3$).  
  Thus, our  $NUVrJ$ technique does not misclassify   
  galaxies reddened by dust as part of the passive sample: 
  using Spitzer/MIPS and Herschel data to identify this kind of contaminants, 
  we find only five $NUVrJ$-passive objects with far-IR emission. 
  For instance at $2.5<z\leqslant3$,  
  $\sim70\%$ of the active galaxies have $E(B-V)>0.25$ ($A_V\gtrsim1$) 
  in the bin   $10.5<\log(\mathcal{M/M}_\odot)\leqslant11$, 
  and almost the entire active sample is heavily attenuated at $\log(\mathcal{M/M}_\odot)>11$.

  To compare with  \citet[][ZFOURGE survey]{Spitler2014},  
  we can merge the bins $3<z\leqslant3.5$ and $3.5<z\leqslant4$ 
  to get the galaxy number density in their same redshift range ($3<z\leqslant4$).    
  By integrating the SMF at $\log(\mathcal{M/M}_\odot)>10$
  we find  $\rho_N=6.6^{+1.6}_{-0.4}\times10^{-6}\,\mathrm{Mpc}^{-3}$,  
   a factor $3\upto4$ lower than the quiescent galaxies of ZFOURGE,
   for which  $\rho_N = (2.3\pm0.6)\times10^{-5}\,\mathrm{Mpc}^{-3}$.   
   Also at $2.5<z<3$  ZFOURGE quiescent galaxies are more numerous 
   than ours, with their SMF being higher  especially in the low-mass regime \citep[cf.][]{Tomczak2014}.    
   The difference is mainly due to the classification method: using $UVJ$ 
   the ZFOURGE passive sample includes galaxies with 
   $\mathrm{sSFR}\simeq10^{-10}\,\mathrm{yr}^{-1}$ 
   (especially at $\mathcal{M}<\mathcal{M}_\star$) that are excluded in our selection. 
   Another reason for such a discrepancy may be the 
   AGN contamination: e.g.~in \citet{Spitler2014}  
   9 out  of  their 26 quiescent galaxies at $3<z<4$ are potential AGN.

  We also compute the passive galaxy fraction ($f_\mathrm{pas}$)  defined as 
  the ratio between passive and total  SMFs (Fig.~\ref{fig_mf_typ}, bottom panels). 
  The peak of $f_\mathrm{pas}$ is always located at $10.5<\log(\mathcal{M}/\mathcal{M}_\odot)<10.9$, with 
  values decreasing from 12 to 5\%. The evaluation at $3.5<z\leqslant4$ 
  is more uncertain if we let Schechter parameters free in the fit, nonetheless 
  $f_\mathrm{pas}$ remains below 10\%.

\begin{table}
\caption{Schechter parameters of the COSMOS2015 galaxy SMF at $z>3$, resulting from
 a fit in which $\mathcal{M}_\star$ has been fixed to $10^{10.6}\,\mathcal{M}_\odot$. 
}
\label{tab_schparMfix}
\setlength{\extrarowheight}{1ex}
\centering
\begin{tabular}{lccc}
\hline\hline
redshift & $\log\mathcal{M}_\star$  &  $\alpha_1$ &$\Phi_{\star1}$  \\
 &  $[h_{70}^{-2}\,\mathcal{M}_\odot]$ & & $[10^{-3}\,h_{70}^3\,\mathrm{Mpc}^{-3}]$   \\ 
[1ex] \hline 
 $3.0<z\leqslant3.5$ & $10.60$   & $-1.58^{+0.07}_{-0.07}$  & $0.204^{+0.024}_{-0.023}$    \\ 
 $3.5<z\leqslant4.5$ & $10.60$   & $-1.65^{+0.10}_{-0.11}$  & $0.094^{+0.013}_{-0.013}$    \\ 
 $4.5<z\leqslant5.5$ & $10.60$   & $-1.69^{+0.26}_{-0.28}$  & $0.038^{+0.011}_{-0.010}$    \\ 
[1ex] \hline
\end{tabular} 
\end{table}

  \subsection{Impact of the Eddington bias}
  \label{Results-SMF bias}

  Describing the low-mass end 
  of the  SMF (or better, its slope) by means of the  Schechter parameter $\alpha$  
  is crucial to get a comprehensive view of stellar mass assembly.   
  The probe of low-mass galaxies can be impeded by the survey depth, 
  as it can result in a  too high stellar mass limit 
  \citep[see][and \citealp{Parsa2016} for an analogue discussion on the LF]{Weigel2016}. 
  In our case, the extrapolation of the low-mass end seems reliable, 
  as the estimates below the limiting mass (derived through the $f_\mathrm{obs}$ weights) 
  are  in agreement with the  fit to $1/V_\mathrm{max}$ points, which stops at $\mathcal{M}_\mathrm{lim}$.

  Besides this caveat, another fundamental problem is related to observational errors that
  alter the galaxy  distribution, i.e.~the Eddington bias. 
  It has already been shown that the Eddington bias modifies significantly the SMF shape 
  \citep[][]{Ilbert2013,Caputi2015,Grazian2015}. 
  Observational (photometric) errors affect 
  the SMF calculation in two ways. First, they introduce an error in $z_\mathrm{phot}$ 
  that spreads out galaxies from the intrinsic $N(z)$ galaxy distribution. 
  Moreover, even if the redshift of an object is known precisely,  
  photometric errors can still affect the $\mathcal{M}$ estimates as they allow a certain number of 
  SEDs (with different  $\mathcal{M}/L$ normalisation) to fit reasonably well the data.   

  Here, we stress out that there is no unique approach to account for such effect, and    
  the shape of the final, bias-free SMF strongly depends on the capability of describing 
  accurately the observational uncertainties ($\sigma_\mathrm{m}$). 
  Despite the different implementations, 
  the basic concept is to convolve a pure Schechter function (see Eq.~\ref{eq_schfun}) 
  with $\sigma_\mathrm{m}$. In this way we construct a function 
   \begin{equation}
    \Phi_\mathrm{obs}(z,\mathcal{M}) = \int  \Phi(m)  \sigma_\mathrm{m}(z,m) \,\mathrm{d}m 
    \label{eq_convomf}
   \end{equation}    
   that can be used in the STY method, or to fit  non-parametric estimates 
   like the $1/V_\mathrm{max}$ points. 
   As noticed, the main distinction among  different correction techniques is how $\sigma_\mathrm{m}$ is 
   defined.  This difference can be summarised by comparing     
  \citet[][hereafter G15]{Grazian2015} and the present analysis. 
   We will show that the two studies, starting from 
   data points that are in excellent agreement, lead to  discrepant 
    Schechter functions after the Eddington bias correction, 
    because of different $\sigma_\mathrm{m}$ estimates.

  The method we used to estimate the stellar mass uncertainty 
  is fully described in Sect.~\ref{SED fitting 2-stellar mass}. 
  The function $\sigma_\mathrm{m}$ required in Eq.~(\ref{eq_convomf}) is 
  the one of  Eq.~(\ref{eq_gau-lor}), with a 
  Lorentzian component  that   
   increases with redshift, while the standard deviation of the 
   Gaussian distribution stays constant. Both components are independent of the 
   logarithmic stellar mass, therefore 
    $\sigma_\mathrm{m}$ is symmetric,  affecting the SMF mainly in the high-mass end 
   because of the exponential decline of number counts.

\begin{figure}
\includegraphics[width=0.99\columnwidth]{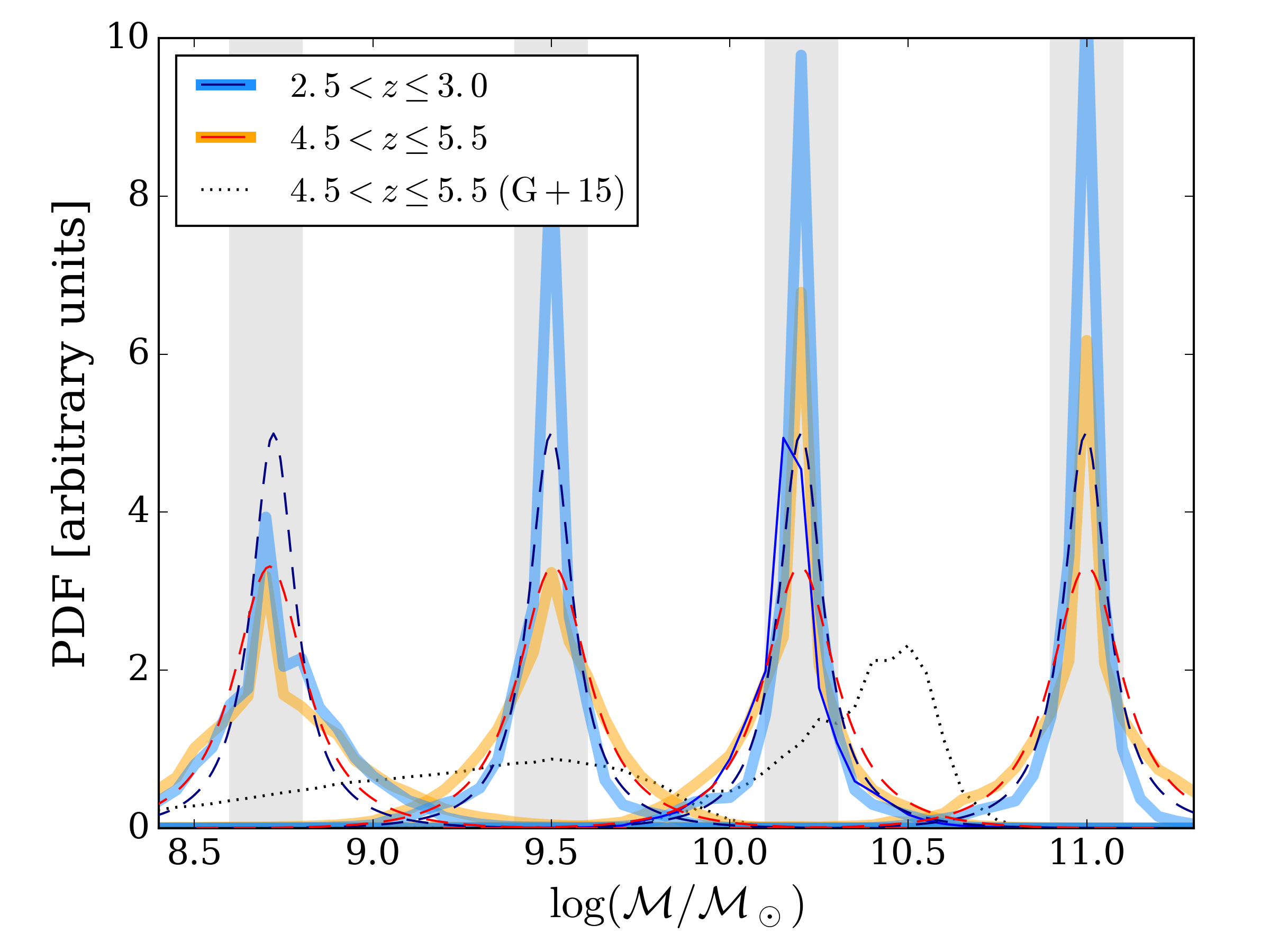}
\caption{ Average PDF($\mathcal{M}|z$) for  galaxies in four bins of stellar mass 
(highlighted with grey vertical bands), at redshift $2.5<z\leqslant3$ 
(blue solid lines) and $4.5<z\leqslant5.5$ (orange). The plot also shows  
the $\sigma_\mathrm{m}(z)$ uncertainty (see Eq.~(\ref{eq_gau-lor}) with 
dark blue and red dashed lines, in the low and high $z$-bin respectively. 
Two examples of $z\simeq5$ PDFs from \citet[][]{Grazian2015} are shown with dotted lines; they are 
obtained by stacking the PDF($\mathcal{M}\vert z$) of CANDELS galaxies at $9.4<\log(\mathcal{M/M}_\odot)<9.6$ and $10.4<\log(\mathcal{M/M}_\odot)<10.6$ respectively (Chabrier's IMF). 
All the PDFs in the figure have been normalised to unity.  }
\label{fig_errmass}
\end{figure}

    On the other hand, the correction implemented by G15 in CANDELS  is  
    derived in the following way. For the photometry of each galaxy in a given $z$ bin 
    (say, $z_1<z<z_2$)  
    they scan the whole BC03 library, after fixing the redshift, to obtain the PDF($\mathcal{M},z\!=\!z_i$).  
    The fixed redshift $z_i$ is taken from an equally-spaced grid  that covers the full range between $z_1$ and $z_2$. 
    The procedure is repeated for each step of the grid, 
     in other words the conditional PDFs are initially computed  assuming a flat prior on $z$. 
     To include the uncertainty on photometric redshift and obtain  $\mathrm{PDF}(\mathcal{M}\vert z)$, 
     G15 multiply each 
     $\mathrm{PDF}(\mathcal{M},z\!=\!z_i)$ times $\mathrm{PDF}(z_i)$, where the latter is the redshift probability 
     coming from the PDF($z$) that \citet{Dahlen2013} provide  for each CANDELS galaxy.  
    The global uncertainty $\sigma_\mathrm{m}(z_1<\!z\!<z_2,\mathcal{M}_j)$ is then obtained by adding the PDFs of 
    all the objects in the given bin of redshift and stellar mass ($\mathcal{M}_j$).\footnote{ 
    Before adding them, each PDF in the stellar mass bin is re-aligned in the centre of the bin. 
    Note that this approach implies a discretised version of Eq.~\ref{eq_convomf}: 
    $\Phi_\mathrm{obs}(\mathcal{M})= \Sigma_j \Phi(\mathcal{M}_j)\sigma_\mathrm{m}(\mathcal{M}_j)\Delta\mathcal{M}_j$, 
    that is inaccurate if the mass binning is too coarse, much larger than the average PDF (this is not the 
    case of G15, who assume $\Delta\mathcal{M}_j=0.2$\,dex).   
    } 
    Such an estimate of $\sigma_\mathrm{m}$, unlike ours, depends also on the stellar mass:  
    the larger the galaxy mass is, the narrower the error  (because of the higher $S/N$ in the photometry). 
     Moreover, the typical PDF($\mathcal{M}\vert z$), especially below $10^{10}\,\mathcal{M}_\odot$,  
     is very skewed, with a prominent tail towards lower masses (see Fig.~\ref{fig_errmass}, and Fig.~B.1 of G15).        
     
   G15 clearly illustrate  the difference between their treatment 
   and \citet[][whose approach is similar to ours]{Ilbert2013}: 
   starting from an intrinsic (i.e., unbiased) SMF and convolving it   
    by the $\sigma_\mathrm{m}(z)$ used by \citeauthor{Ilbert2013}, 
   the observed SMF will be a Schechter function 
   with an enhanced number density of massive galaxies, 
     but substantially unchanged at lower masses.  
   Conversely, by applying to the same intrinsic mass function 
   the $\sigma_\mathrm{m}(z,\mathcal{M})$ 
   computed in G15, $\Phi_\mathrm{obs}$ will have a steeper low-mass end. 
   This latter method should not modify significantly  the high-mass end, where 
   the PDF($\mathcal{M}\vert z$)  is expected to be  narrower (G15). 

     \begin{figure}
 \includegraphics[width=0.99\columnwidth]{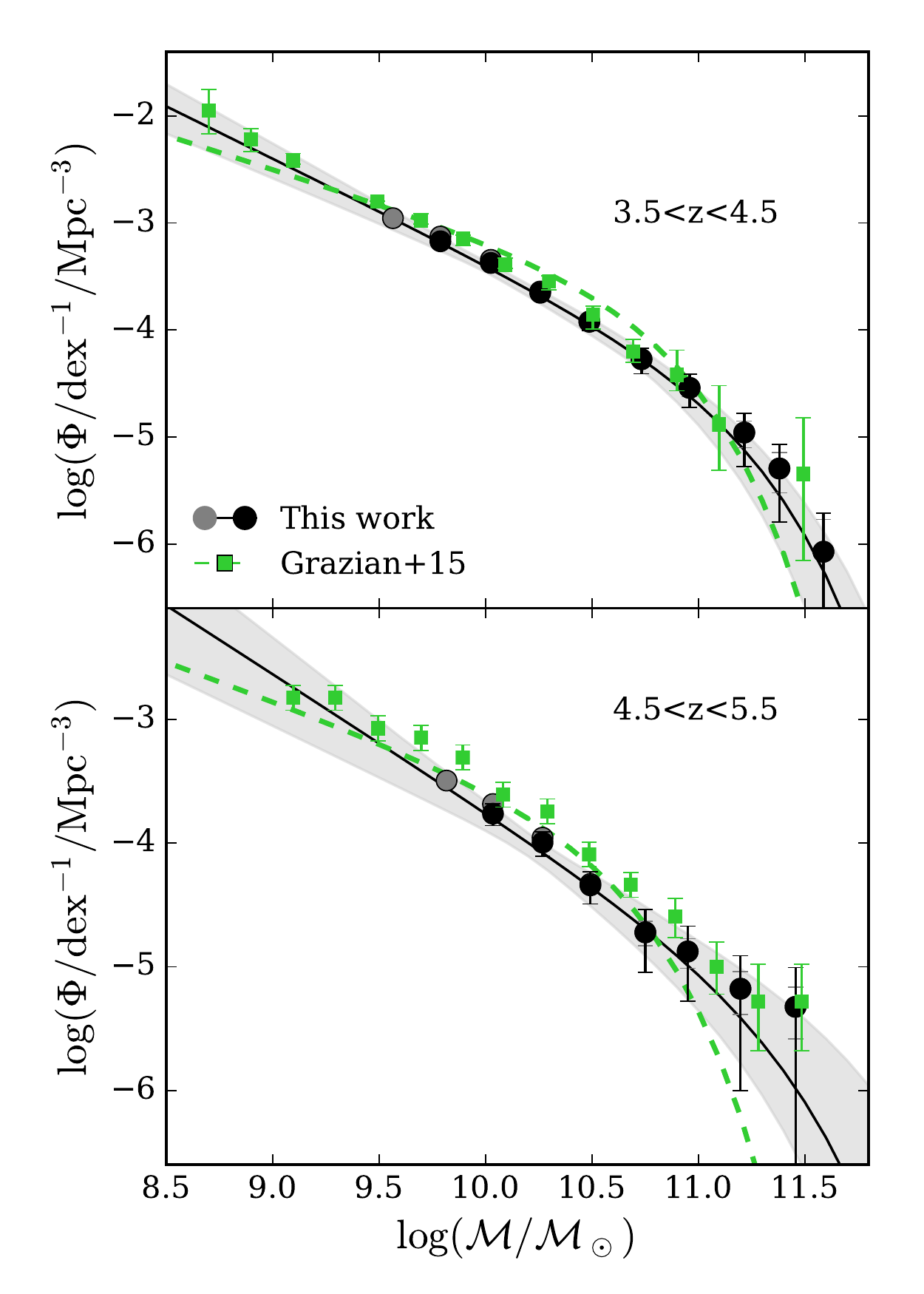}
 \caption{Comparison between the SMF of COSMOS2015 galaxies and \citet{Grazian2015}, in two redshift 
 bins between $z=3.5$ and 5.5 (as indicated in the panels). 
  Our $1/V_\mathrm{max}$ estimates and the fit to those points are shown with black filled circles and solid lines  
  \citep[green squares and dashed lines show the same quantities for][]{Grazian2015}. Grey filled circles are the 
  SMF estimates we obtain using the $f_\mathrm{obs}$ weights (see Sect.~\ref{Results}).  
  In both $z$-bins, the $1\sigma$ uncertainty of the COSMOS2015 Schechter function is encompassed in a  shaded area. 
  }
 \label{fig_mf_slope}
 \end{figure}
 
      Thus, $\sigma_\mathrm{m}$ estimates  are different in CANDELS 
      and COSMOS. 
      To find the reason of the discrepancy, 
      we recompute $\sigma_\mathrm{m}$ with a procedure   
      similar to G15, deriving the galaxy PDF($\mathcal{M}\vert z$) in bins of stellar mass (Fig.~\ref{fig_errmass}). 
      Interestingly,  our PDF($\mathcal{M}\vert z$) is well described by 
      the same distribution (Eq.~\ref{eq_gau-lor}) 
      we found with the method of \citet{Ilbert2013}. 
      There is a trend with stellar mass like in G15, but so mild that    
      for galaxies with $\mathcal{M}\gtrsim \mathcal{M}_\mathrm{lim}$  
       the PDF can be considered constant  (at a given redshift, see Fig.~\ref{fig_errmass}). 
      So, even with this alternate computation the stellar mass uncertainties of CANDELS 
      and COSMOS2015 remain at variance.   
      The fact that at a given redshift 
      our $\sigma_\mathrm{m}$ (at least above $10^{9}\,\mathcal{M}_\odot$) 
      does not depend on mass  can appear counterintuitive: 
      the SED fitting should be better constrained for massive galaxies since they 
      generally have higher $S/N$ photometry. This is true for the $\mathcal{M}$ estimates, 
      i.e.~the SED fitting performed \textit{after fixing the redshift}.      
      However the $z_\mathrm{phot}$ uncertainty, which is the dominant component 
      in $\sigma_\mathrm{m}$, is nearly constant in this stellar mass range 
      \citep[see the discussion in][]{Caputi2015}.

      Therefore, despite the  agreement between the $1/V_\mathrm{max}$ estimates 
      of G15 and ours,   
      the correspondent Schechter functions diverge below $10^{9}\,\mathcal{M}_\odot$, 
      after accounting for the Eddington bias (Fig.~\ref{fig_mf_slope}).  
     The flatter slope recovered by G15  is due to the mass-dependent $\sigma_\mathrm{m}$   
     that becomes much wider at lower $\mathcal{M}$. 
     On the other hand, 
     their fit should be anchored to the more precise stellar mass estimates 
     at $\mathcal{M}>\mathcal{M}_\star$, but in practice their Schechter exponential tail 
     results below the data points 
     because of the correlation between $\alpha$ and $\mathcal{M}_\star$.  
     Vice versa, with our $\sigma_\mathrm{m}$ 
     the error deconvolution does not modify significantly the low-mass end 
     while it decreases the number density in the high-mass end.\footnote{  
      \citet{Song2016} state that, after accounting for Eddington bias, the low-mass end of their 
      SMF does not change significantly, as in our case. However, details about how the 
      bias correction has been implemented are not provided in their paper.}     
      The bias-free Schechter function of G15 
      turns out to have a low-mass end nearly constant
      from $z\sim4$ to 6, comparable to what found at  $z\sim2\upto3$ 
      (i.e., $\alpha\simeq-1.6$). In particular they find $\alpha=-1.63\pm0.05$ at $z=4$ and 
     $\alpha=-1.63\pm0.09$ at $z=5$. 
      Without removing the bias,  the G15 Schechter function has  $\alpha=-1.77\pm0.05$ and $-1.90\pm0.20$, 
        at  $z=4$ and 5 respectively.    
      In the same redshift bins we find  $\alpha=-1.97^{+0.10}_{-0.09}$  and   $-2.11^{+0.30}_{-0.13}$. 
   
       These tests  indicate  that the  
       different characterisation of $\sigma_\mathrm{m}$ 
       is the major responsible for the contrasting Schechter fits in the two analyses. 
       The $z_\mathrm{phot}$ uncertainties are the 
       most plausible reason of discrepancy  but at present we cannot establish  whether 
       the PDFs($z$) of COSMOS2015 are underestimated, 
       or those in CANDELS overestimated. 
       We aim at investigating this issue in a future work.

\section{Discussion}
\label{Discussion}

\subsection{Stellar mass assembly between  $z\sim0$ and $6$}
\label{Discussion-SMF evolution}

\begin{figure*}[!t] 
\centering
    \includegraphics[width=0.75\textwidth]{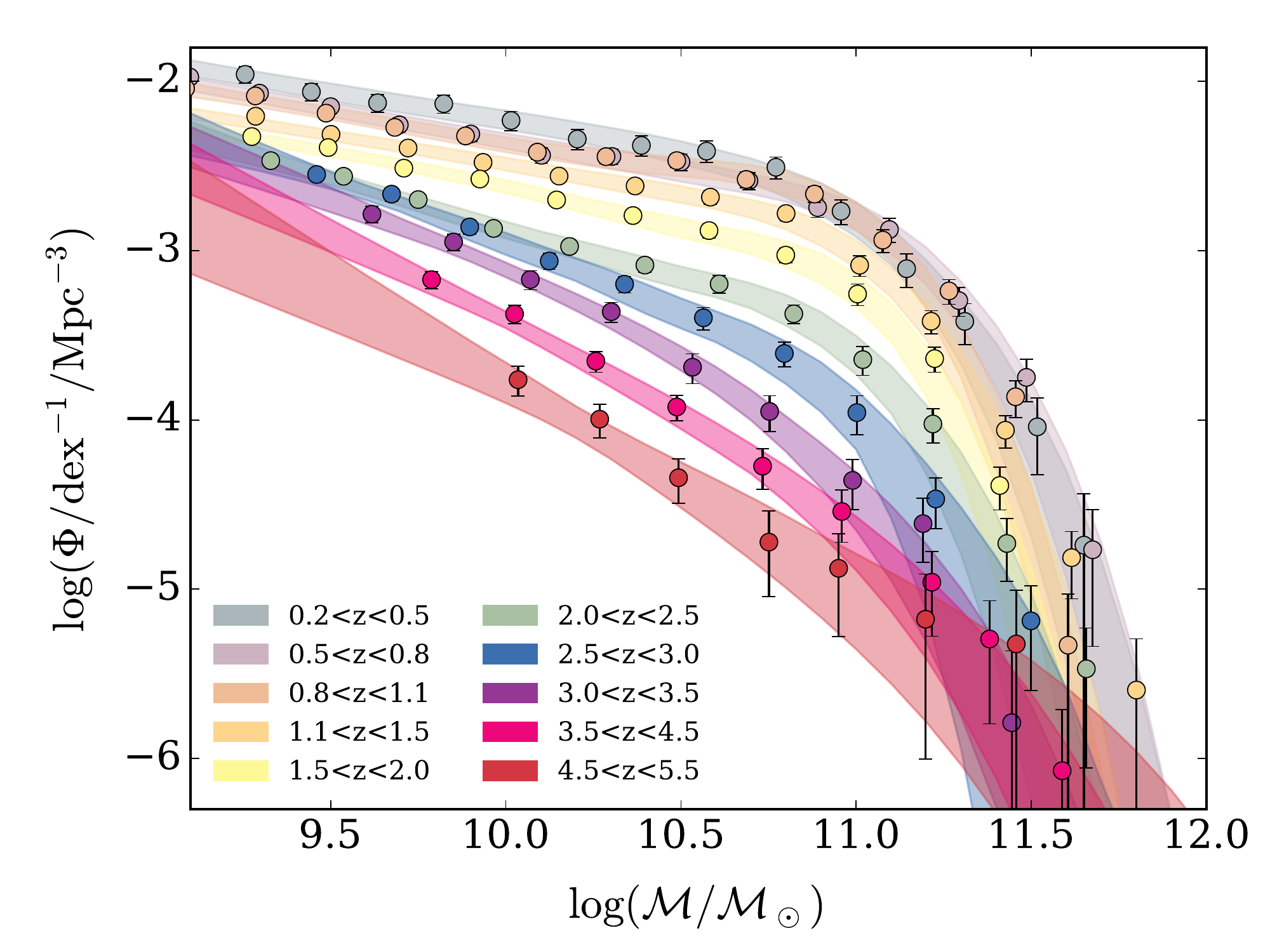}
   \caption{Evolution of the SMF between $z=0.2$ and 5.5, for  the COSMOS2015 galaxy sample.  
   Filled circles show the $1/V_\mathrm{max}$ estimates, 
    and shaded areas show the $1\sigma$ uncertainty of the best Schechter function fitting to them 
   (as in Fig.~\ref{fig_mf1} and  \ref{fig_mf_3x3}). Colours indicating the redshift bins are summarised in the  
   bottom-left corner of the plot.   }
   \label{fig_smfevo}
\end{figure*}

\begin{figure}[t]
               \includegraphics[width=0.99\columnwidth]{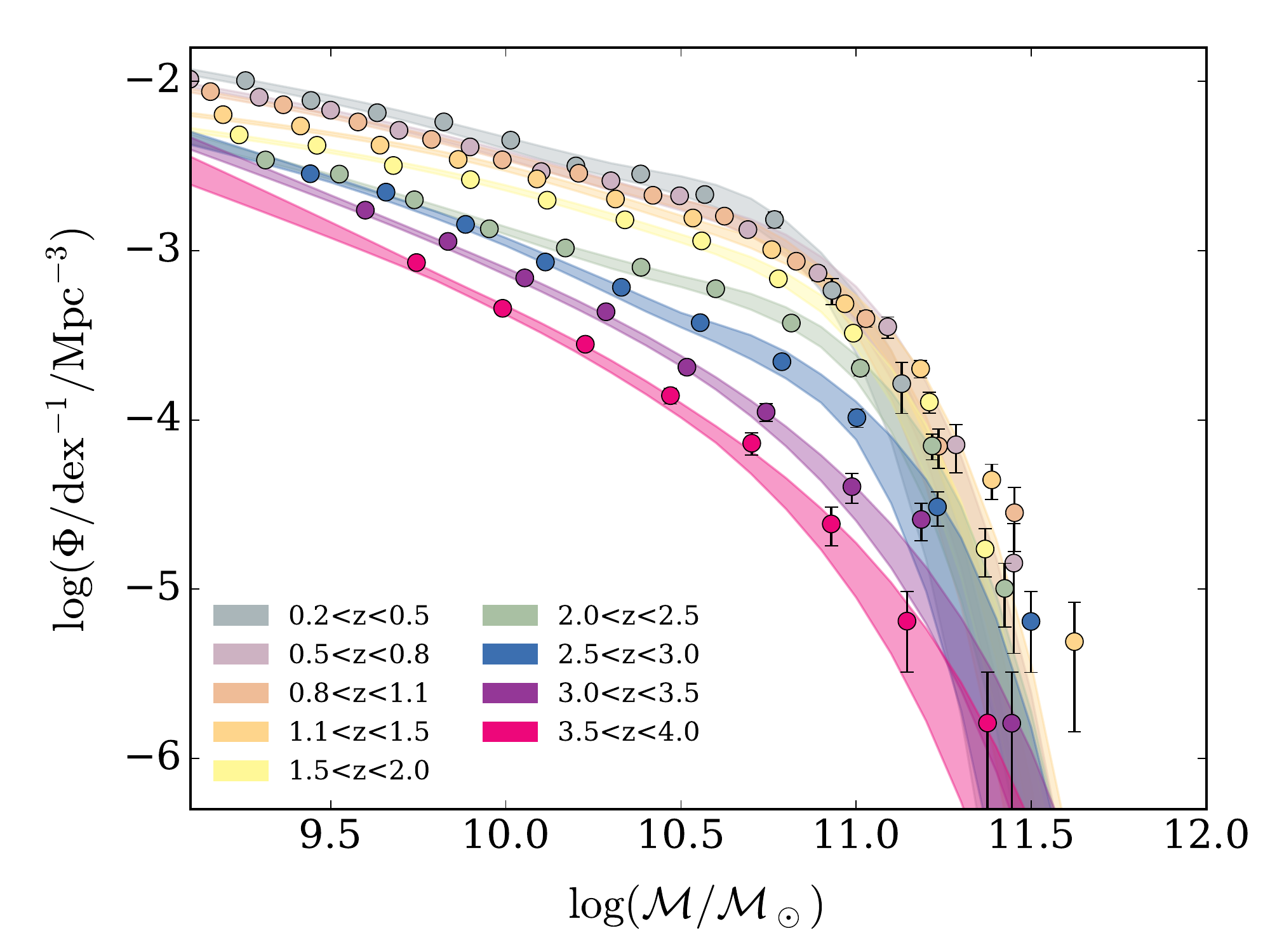} 
               
              \includegraphics[width=0.99\columnwidth]{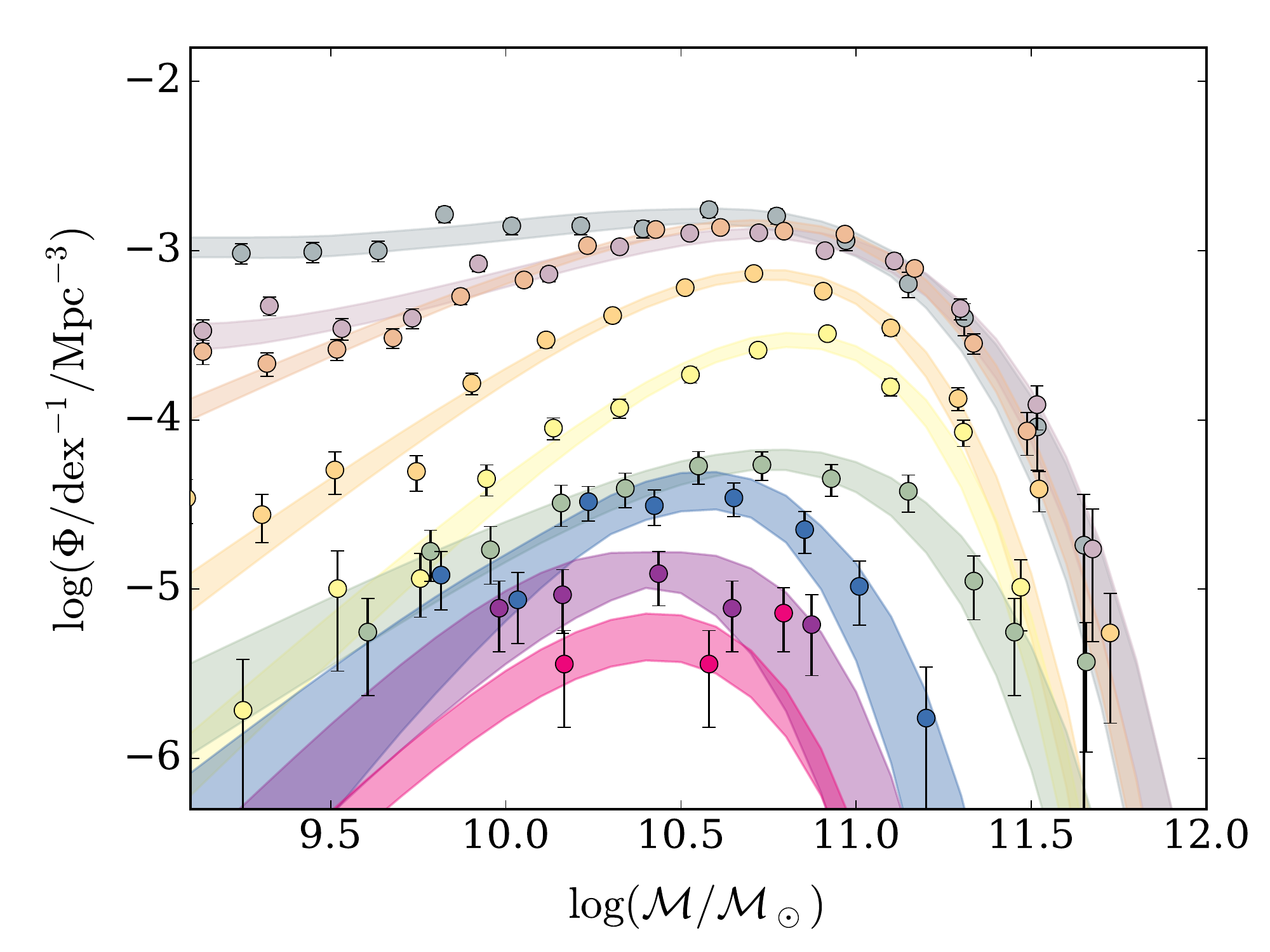}
                              \caption{Evolution of the SMF between $z=0.2$ and 4, for  active (\textit{upper panel}) 
                              and passive (\textit{lower panel}) galaxies.  Same symbols as in Fig.~\ref{fig_smfevo}. 
                              }  
\label{fig_smfevo_typ}
\end{figure}

 To study the SMF evolution over a larger time interval, 
 we combine our results at $z>2.5$ (Sect.~\ref{Results}) 
 with the SMFs at lower redshifts. 
 The latter ones are obtained using  the SED fitting estimates of L16, 
 whose main features have already been described  (see Sect.~\ref{SED fitting-method} and  
 \ref{Results-Sources of uncertainty}).     
 To have a comprehensive view, in Fig.~\ref{fig_smfevo} 
 we overplot the estimates from $z=0.2$ to 5.5. 
 In Fig.~\ref{fig_smfevo_typ} we also 
 show the SMFs of star-forming and quiescent objects, up to $z=4$.  
 Additional plots of the $0<z\leqslant4$ SMFs  are shown in Appendix~\ref{APP2}. 
 
 The resulting picture shows a progressive build-up of galaxies at 
 $10.0<\log(\mathcal{M/M}_\odot)<11.5$, sharpening the knee of the SMF as time goes by. 
 This feature becomes stable at $z\lesssim2$, since the SMF grows in normalisation 
 but the shape of the exponential tail remains nearly the same.  
 In comparison, there is  little increase 
 in the  number density 
 of galaxies with $\log(\mathcal{M/M}_\odot)\leqslant10.0$ across the whole redshift range.  
 In the bin centred at $\log(\mathcal{M/M}_\odot)=9.5$, where 
 our data provide a direct constraint in all the $z$-bins, 
 the increase in number density is $6\upto7$ times smaller than 
 at 10.5. 
 The combined effect of such a differential evolution is a flattening of the low-mass end as moving 
 to lower redshifts. In the early universe ($<\!2$\,Gyr old) 
 the best fit is a single Schechter function, with the addition of a secondary component only at $z<3$.   
 We observe that this evolutionary trend  does not  imply necessarily 
 a change of $\mathcal{M}_\star$. 
 If  galaxies below the turnover mass outnumber the most massive ones,  the ``dip'' of the Schechter 
 function is smoothed out even if $\mathcal{M}_\star$ remains constant  \citep[see][]{Tomczak2014,Song2016}.

 These observations are consistent with  models in which the suppression of star formation 
  (``quenching'') is particularly efficient when galaxies reach a given 
  stellar mass threshold ($\sim\!\mathcal{M}_\star$).    
  In this way, star-forming galaxies that accreted up to such a mass 
  cannot easily grow further, so  they accumulate at $\mathcal{M}_\star$. 
  This effect is even more evident in the active SMF (Fig.~\ref{fig_smfevo_typ}, upper panel), 
  which does not extend beyond  $\log(\mathcal{M/M}_\odot)=11.5$. 
  This kind of mass-dependent quenching 
  could be caused by internal processes, e.g.~AGN feedback or heating via stable 
  virial shocks \citep[see][and references therein]{Gabor2010}. 
  Without any assumption on the underlying physics, the empirical model of 
  \citet{Peng2010} shows how a galaxy SMF that is a power-law function at $z=10$ will  
  assume a Schechter  profile at lower $z$, mainly because of the  action of mass quenching.   
  Potentially confirming this picture, the COSMOS2015 SMF,  moving towards higher $z$,  
  starts to resemble a  power law  (Eq.~\ref{eq_powlaw}). 
  We caution that this clue may actually be the effect of galaxy interlopers on the high-mass end, 
   however it is not implausible that the SMF departs from a Schechter function at 
   $z\gtrsim6$, to reproduce more closely the shape of the underlying dark matter  (DM) distribution (see below). 
   Similarly,   \citet{Bowler2015,Bowler2017} find evidence that the UV LF 
   of $z\simeq7$ galaxies is better fit with a double power law.

  The  SMF of $NUVrJ$-passive galaxies  (Fig.~\ref{fig_smfevo_typ}, lower panel) 
   agrees with this scenario, with a distinct 
  $\log(\mathcal{M}_\star/\mathcal{M}_\odot)=10.5\upto10.8$  peak  
  even at $z>3$. 
   The most significant growth of the passive sample,  in terms of number density,  
   happens from $z=2.5$ to 1. A substantial increase (by a factor $\times 4$) 
    is observed in particular from $2<z<2.5$ to $1.5<z<2$ (i.e., in less than 1\,Gyr). 
   This is consistent with   previous studies that indicate that local early-type galaxies 
   with $11<\log(\mathcal{M/M}_\odot)<12$ 
   entered in their quiescent phase between $z\simeq0.8$ and 2.5 \citep{Thomas2010}. 
   
   The build-up of passive galaxies  corresponds to a transition of the total SMF from a single to 
   a double Schechter function. 
   This is only an approximate scheme, because 
    the emerging secondary component cannot be fully ascribed to  quenching: 
    also the active SMF  is better  fit by a  double Schechter function  at least at $z<2.5$ 
   \citep[see also][]{Ilbert2013,Tomczak2014}. 
   When  the active sample is divided in two or more classes -- e.g.~distinguishing between 
   intermediate and high sSFR, or different morphologies, as in  \citep[][]{Ilbert2010} -- each SMF is 
   well described by a single Schechter function. 
   From a morphological analysis of $z<0.06$ galaxies, \citet[][GAMA survey]{Moffett2016} find 
   that the double Schechter profile of the active SMF 
   is the sum of the SMF of Sd and irregular galaxies (dominant at the low-mass end) 
   and the one of Sa to Sc types (which creates the dip at intermediate masses). 
   With irregular galaxies being more common in earlier epochs, the result should be 
   a single Schechter at high $z$, as observed. 
    Moreover, \citeauthor{Moffett2016} \citep[but also][]{Kelvin2014b} 
   find a precise decomposition of their local SMF 
    in two Schechter functions by simply dividing disc- and bulge-dominated galaxies.  
   Without speculating further, we just remark that  a similar  morphological transformation, 
   characterised by an ``inside-out'' quenching and bulge growth, 
   is expected to  begin at $z\simeq2.5$ \citep[according to recent simulations as][]{Tacchella2016}
    i.e.~the epoch when we observe  
    a secondary low-$z$ component to emerge in the SMF.

\begin{figure}[t]
  \includegraphics[width=0.99\columnwidth]{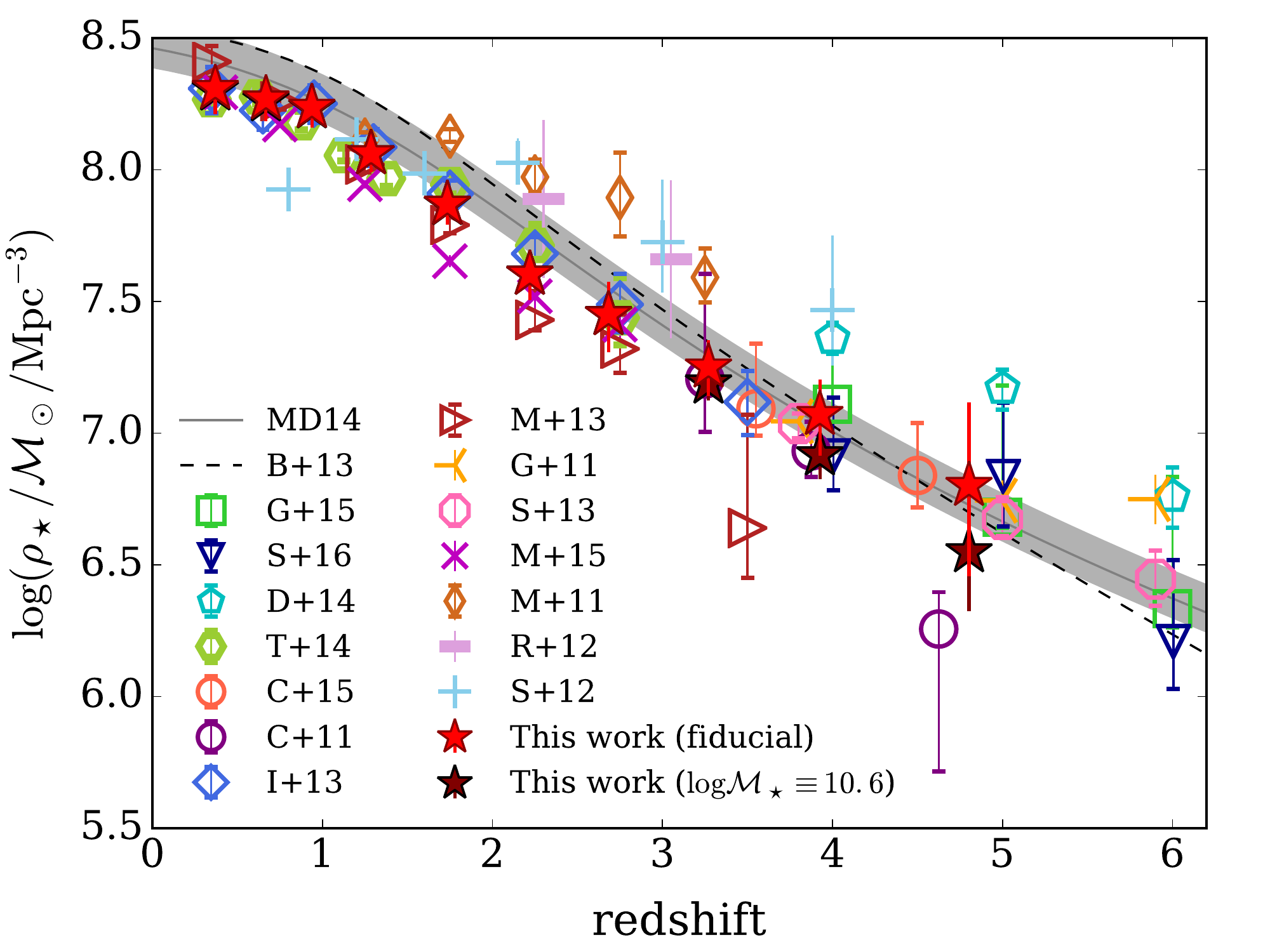}
  \caption{Redshift evolution of $\rho_\ast$, as measured in different papers 
  by integration of the SMF: 
  \citet[][C+11 in the legend]{Caputi2011}, \citet[][C+15]{Caputi2015}, \citet[][D+14]{Duncan2014}, \citet[][G+11]{Gonzalez2011}, 
  \citet[][G15]{Grazian2015}, 
  \citet[][I+13]{Ilbert2013}, \citet[][M+11]{Mortlock2011}, \citet[][M+15]{Mortlock2015}, \citet[][M+13]{Muzzin2013b}, 
  \citet[][R+12]{Reddy2012}, \citet[][S+12]{Santini2012},   \citet[][S+16]{Song2016}, and  \citet[][T+14]{Tomczak2014}. 
  If $\rho_\ast$ uncertainties are not quoted in the paper, we plot approximate error bars by considering the $1\sigma$ error of the   $\alpha$ parameter.  
  Red stars are the stellar mass density  from our fiducial Schechter,  
  brown stars are from the fit with fixed $\mathcal{M}_\star$.  
  By integrating their  SFRD functions,  we can plot $\rho_\ast(z)$ from \citet[][black dashed line]{Behroozi2013b} 
  and \citet[][grey solid line]{Madau2014}. In both integrations we assume  
  $f_\mathrm{return}=41\%$ (coherently with Chabrier's IMF). 
  For \citet{Madau2014} we also show with a shaded area the $\rho_\ast$ 
  range  enclosed by 
  $f_\mathrm{return}=50\%$ and   25\% (the latter value is similar to the one prescribed by Salpeter's IMF). 
   }
  \label{fig_rhostar}
  \end{figure}

 We also determine  the stellar mass density ($\rho_\ast$) as a function of $z$. 
  This is usually done by integrating the Schechter function  between $10^8$ and $10^{13}\,\mathcal{M}_\odot$. 
  Since our $\mathcal{M}_\mathrm{lim}$ is larger than $10^9\,\mathcal{M}_\odot$ at $z>2$, 
  the computation at high redshift is extremely sensitive to the extrapolation of the low-mass end below  our data point 
  (see Sect.~\ref{Results-SMF bias}). 
  We show in Fig.~\ref{fig_rhostar} several $\rho_\ast$ estimates 
  from COSMOS2015 and other  surveys, compared to the stellar mass density derived  
  via integration of the  star formation rate density (SFRD) function 
  \citep[as given in][]{Behroozi2013b,Madau2014}. 
  The difference between the two methods is smaller than in the analogous plot shown in 
  \citet[][]{Madau2014}, where estimates derived from SED fitting are $\sim\!0.2$\,dex 
  lower than $\rho_\ast$ from SFRD (their Fig.~11).   
  As explained in that paper, the level of consistency   
  also depends on the assumed IMF.  Time integration of the SFRD takes into account the  
  gas recycling fraction ($f_\mathrm{return}$), which is 0.41 for Chabrier's and 0.27 for Salpeter's IMF. 
  Since we use the former, the resulting stellar mass density is $\sim\!0.1$\,dex smaller than the one 
  obtained by \citet{Madau2014}  starting from the same SFRD function.

  In Fig.~\ref{fig_rhostar} we also see that 
  our fiducial SMF at $z\geqslant4$ originates higher $\rho_\ast$ values than 
  the fit with fixed $\mathcal{M}_\star$, whose main difference is indeed the flatter low-mass end. Both estimates 
  are nonetheless consistent at $1\sigma$  within each other, and in fairly good agreement with $\rho_\ast$ from \citet{Behroozi2013b} and \citet{Madau2014}. 
  The tension with the SFRD predictions starts to be evident 
   when considering e.g.~\citet{Santini2012} or \citet{Duncan2014}, whose SMFs are even steeper.    
   
     A precise determination of $\alpha$ is also pivotal in 
   the essential formalism of those empirical models \citep[e.g.][]{Peng2010,Boissier2010} 
   that try to connect the SMF evolution  to  the ``main sequence'' (MS) of star forming galaxies 
  \citep{Noeske2007,Daddi2007,Elbaz2007}.  
  Reconciling the galaxy growth predicted by the MS 
  with the redshift evolution of $\alpha$  is an effective way to constrain stellar mass assembly 
  and quenching mechanisms \citep[e.g.][]{Leja2015,Steinhardt2017}.

   
 \subsection{Dark matter connection}
\label{Discussion-HMF evolution}
 
  To better understand the evolution of the SMF  
  we investigate the relation between galaxy stellar mass and DM halo  mass  assembly. 
  As pointed out by \citet{Lilly2013}, 
  galaxy sSFR and the specific mass increase rate 
  of DM haloes \citep[$\mathrm{sMIR}\equiv \mathcal{M}_\mathrm{h}^{-1}\, \mathrm{d}\mathcal{M}_\mathrm{h}/\mathrm{d}t$, see e.g.][]{Neistein&Dekel2008} 
  evolve in a similar way, as expected if  star formation is regulated by the amount of cold gas in the  
  galaxy  reservoir, 
  which in turn depends on the inflow of DM  into the halo \citep{Lilly2013,Saintonge2013}.

  We compare the SMF of COSMOS2015 galaxies to  the halo mass function (HMF) 
  provided in \citet{Tinker2008}.\footnote{
  Starting from the analytical form of \citet{Tinker2008} the HMF has been computed 
  in our $z$-bins and cosmological framework ($\sigma_8=0.82$)  
  by means of the code \texttt{HMFCalc}  \citep{Murray2013}. 
  The code allows us to choose among several models \citep[e.g.][]{Sheth2001,Tinker2010,Angulo2012} without 
  any significant  impact on our conclusions.}
  Recently, a discrepancy between these two quantities has been highlighted by \citet{Steinhardt2016}: 
  the most massive galaxies observed at $z>4$ seem to be too numerous  
  compared to the haloes that should host them. 
  Such an excess, if confirmed, would call into question either theoretical aspects of 
  the  $\Lambda$CDM model  
  or some fundamental principle of galaxy formation 
  \citep[see the discussion about these ``impossibly early galaxies'' in][]{Steinhardt2016}.

\begin{figure*}
   \includegraphics[width=0.98\textwidth]{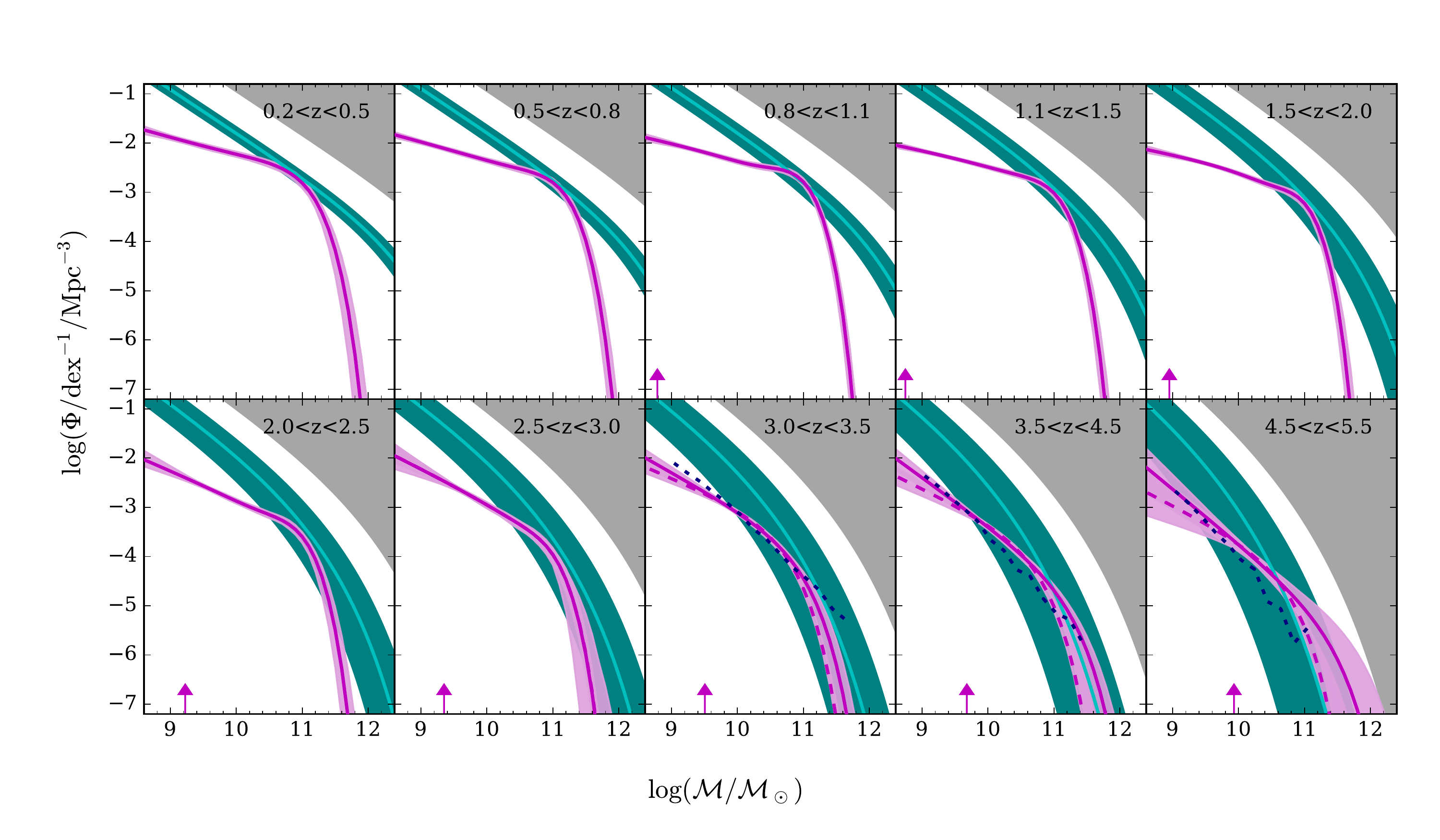}  
   \caption{Evolution of HMF (cyan line)   vs SMF (magenta)  from $z=0.2$ to $5.5$. 
    In each panel halo mass is multiplied by a factor 0.018, 
    i.e.~the SHMR at $z=0$ and $\mathcal{M}_\mathrm{h}=\mathcal{M}_\mathrm{h}^\star$ 
    \citep[according to][]{Behroozi2013b}. 
   Shaded cyan regions show the uncertainties in the HMF shift by taking the $1\sigma$ error of 
    the SHMR parametrisation. 
   At $z>2.5$, the solid magenta line is our fiducial fit of the SMF,  
   while the dashed magenta line is the Schechter function with 
   $\log(\mathcal{M}_\star/\mathcal{M}_\odot)$ fixed to $10.6$ (see Fig.~\ref{fig_mf1}). 
    The shaded magenta regions combine  the $1\sigma$ CL of the two 
    fits, to give a conservative estimate of the uncertainties. 
     A grey shaded area highlights the ``forbidden region'' where, according to the 
     $\Omega_\mathrm{b}/\Omega_\mathrm{m}$ ratio, no galaxies are expected. 
   Black dotted lines at $z>3$ are the SMFs predicted from  the semi-analytical model of \citet{Garel2016}, 
   converted to the IMF of \citet{Chabrier2003}. 
  }
   \label{fig_hmfevo}
   \end{figure*}

  The co-evolution of SMF and HMF between $z=0.2$ and 5.5 is shown in Fig.~\ref{fig_hmfevo}. 
  For a synoptic view, we superimpose the HMF on the SMF, rescaling $\mathcal{M}_\mathrm{h}$ 
   by a constant   factor equal to 0.018.  This scaling factor is 
  the stellar-to-halo mass ratio ($\mathrm{SMHR}\equiv\mathcal{M}/\mathcal{M}_\mathrm{h}$) 
  provided in  \citet[][see their Eq.~3]{Behroozi2013b} for a typical  $\mathcal{M}_\mathrm{h}^\star$ halo 
  at $z=0$.\footnote{
  $\mathcal{M}_\mathrm{h}^\star$
  is the  characteristic halo mass that marks the peak at which the integrated star formation is most efficient. 
  At $z=0$ it is about $10^{12}\,\mathcal{M}_\odot$ \citep{Behroozi2013b}. 
  Roughly speaking, $\mathcal{M}_\mathrm{h}^\star$  separates the SHMR behaviours at low and high halo masses.  
  In \citet{Behroozi2013b} the $\mathcal{M}$-$\mathcal{M}_\mathrm{h}$ relation
  is calibrated against $0\leqslant z\leqslant8$ data (several SMF, specific SFR, and cosmic SFR estimates)
  through a Markov Chain Monte Carlo.   
  We verified that adopting the SHMR of \citet{Moster2013}, 
  constrained via galaxy-halo abundance matching, differences are within the $1\sigma$ error bars. }   
  We emphasise that the same rigid translation 
  is applied in each $z$-bin, just to ease the comparison between 
  the HMF and the SMF shapes. A thorough  link between haloes and galaxies, e.g.~via 
  abundance matching, is defer to a future work. 
  
  At $z<2$, Fig.~\ref{fig_hmfevo} (upper row of panels) 
  illustrates a well-known result: the shape of stellar and halo mass functions do not coincide, 
  neither at $\mathcal{M}<\mathcal{M}_\star$  nor at $\mathcal{M}>\mathcal{M}_\star$. 
  Reconciling the observed SMF with  the DM distribution   
  has required the introduction of quenching mechanisms  in galaxy formation models   
  \citep[for a review  see][]{Baugh2006}. 
  Star formation of low-mass galaxies is assumed to be halted via stellar feedback, 
  e.g.~stellar winds or supernova explosions that heat/eject gas \citep[][]{Larson1974,Dekel&Silk1986,Leitherer1999}.  
   In galaxies at $\mathcal{M}>\mathcal{M}_\star$, 
   hot halo gas is removed or prevented from cooling 
   e.g.~by AGN outflows \citep[see][for a review]{Fabian2012}   
     or virial shocks heating \citep{Dekel&Birnboim2006}.

     As for the transition from single to double Schechter function, the epoch 
     of a key change is $z=2\upto3$.  
     In fact, the tension between SMF and HMF lessens at $z>2$ (Fig.~\ref{fig_hmfevo}, lower panels).  
     In the high-mass regime, the SMF exponential tail moves closer to the rescaled HMF, until they overlap 
     at $z>3$. Considering the crude rescaling (i.e., the 0.018 factor)  and the SMF uncertainties at high $z$, 
     the match between massive galaxies and massive DM haloes is excellent. 
     At $4.5<z\leqslant5.5$, the massive end of our fiducial SMF is slightly higher than the HMF,  
     but still compatible within the errors.  
     Such an excess of observed galaxies does not challenge the theoretical framework, since 
     small modifications e.g.~to the HMF scaling factor (which has been fixed to the SHMR at $z=0$) 
     would be enough  to reconcile the two functions. 
     
     To show that there is no substantial inconsistency between the two functions, we derive from the HMF an 
     upper limit for the SMF.  
     Starting from the present baryon density $\Omega_\mathrm{b,0}=0.0486$ \citep{Planck2015-XIII}, 
     we assume $\Omega_\mathrm{b}/\Omega_\mathrm{m}$ as a SHMR 
     with a baryon-to-stellar mass conversion of 100\%. 
     Rescaling the HMF accordingly, we obtain the maximal SMF physically allowed 
     (grey shaded area in Fig.~\ref{fig_hmfevo}). The observed SMF is always below this upper limit. 
     In other words, 
     we do not find any impossibly early galaxy, at least at $z<6$.  \citet[][]{Steinhardt2016} discuss  
     this critical issue  relying on UV LFs up to $z\sim10$ 
     \citep{Bouwens2015a,Bouwens2016}.  
     In this respect, \citet{Mancuso2016} claim that the tension between the observed 
     number density of massive galaxies and the predicted abundance of their host haloes is mainly due 
     to the dust corrections applied to UV data.  When including far-IR data to determine the  SFR 
     function, they find that the formation of stars in $z\geqslant4$ massive haloes is not required to start 
     as early as argued in \citet[][]{Steinhardt2016}. 
     This kind of bias does not affect our comparison, which however probes $z<6$. 
     Conclusive evidence on this issue will come from next-generation surveys, which 
     shall provide  direct measurements of the SMF at $z>7$ 
     \citep[i.e.~the epoch when these discrepancy should be the largest][]{Steinhardt2016}.

    The high-mass end of our SMF declines with a slope similar to the HMF. 
     Such an agreement  suggests that massive galaxies at $z>3$, which resides 
     in $\gtrsim\!\mathcal{M}_\mathrm{h}^\star$ haloes, all have similar SHMR 
      ($\sim\!2\%$). 
     However, one should discriminate between the contribution of central and satellite galaxies 
     before concluding about the star-formation efficiency in distinct haloes \citep[see][]{Coupon2015}. 
      A nearly constant SHMR for $\mathcal{M}_\mathrm{h}>\mathcal{M}_\mathrm{h}^\star$ is 
      at odds with \citet{Behroozi2013b} and \citet{Moster2013}  results,  
      but a few caveats concerning those two studies should be noticed. \citet{Moster2013} 
      relation has been calibrated by means of abundance matching technique 
       up to $z=4$. The SMFs they used \citep{Santini2012,PerezGonzalez2008} 
      are  more plagued by sample variance, and at the massive end their technique is sensitive 
       to the assumptions made for correcting observational errors.  
      \citeauthor{Behroozi2013b} used a larger set of SMFs, but at $z>4$ these are all 
      derived from LBG samples \citep{Stark2009,Lee2012,Bouwens2011b,Bradley2012}, possibly biased at high masses 
      as we discussed in Sect.~\ref{Results-SMF literature}. Interestingly, if the $\mathcal{M}$-$\mathcal{M}_\mathrm{h}$ 
      relation of \citet{Behroozi2013b} is recomputed without  $z>4$ observational priors, 
      the SHMR  at $z\simeq4$ and 5 is flatter \citep[and only marginally consistent within the $1\sigma$ errors of 
      the original fit, see][]{Behroozi2015}.  
      Thus, the COSMOS2015 catalogue represents a novel opportunity to investigate the connection between 
      DM and stellar content  
      in a redshift and mass regime where previous SHMR estimates were lacking robust observational constraint 
      (Coupon et al., in preparation).

      \subsection{A reduced impact of star formation feedback}
      \label{Discussion-AGN feedback}
   
     At $z>3$, therefore, 
     there is no need of additional quenching to reconcile the abundance of 
     $\mathcal{M}>\mathcal{M}_\star$  galaxies with that of 
     $\mathcal{M}_\mathrm{h}>\mathcal{M}_\mathrm{h}^\star$ haloes. 
     This finding  supports results from simulations, in which massive systems at high redshifts  
     are weakly affected by AGN activity.  
     An example is the cosmological hydrodynamical simulation 
     Horizon-AGN \citep[][]{Dubois2014}:  at $z>3$ 
     the  SMF of the Horizon-AGN simulated galaxies  does not change significantly if  AGN feedback is 
     switched off   \citep{Kaviraj2017}.      
     In addition, Horizon-AGN allows to follow in detail the evolution of  central black holes.  
     By tracing  their mass assembly as a function of time, 
     \citet{Volonteri2016b} find that black holes with Eddington ratio $<\!0.01$ (those 
     responsible for radio-mode feedback  in the simulation) are the dominant 
     population at $z\lesssim2$, while at $z>3$  most of the black holes are fast accretors 
      (Eddington ratio $>\!0.1$), luminous enough to trigger a radiative feedback 
      (the so-called ``quasar mode''). 
    Smaller occurrence of  radio-mode feedback  at high redshifts 
       is also suggested by observational studies on radio-loud AGN,
      whose volume density  decreases as a function of $z$ \citep{Padovani2015}.

      Radiative AGN feedback is expected to be inefficient, too: 
      in  hydrodynamical simulations of $z\sim6$ galaxies, 
      outflows generated by bright quasars tend to escape 
      from the direction of least resistance, without interacting with the dense filamentary 
      structure around the galaxy \citep{Costa2014}.  
      Interferometric  observations of high-$z$ targets are in line with  these theoretical results. 
      Their CO and [CII] mapping indicates that, although AGN  are able to remove a
      large amount of molecular gas \citep{Maiolino2012,Cicone2014},  they do not  
      prevent extended cold clouds from  fuelling  star formation \citep{Cicone2015}.  
      
     Nevertheless, $z\sim5\upto6$ radiative outflows, despite their relatively weak  impact  on  
     interstellar medium,  can perturb the cold filamentary accretion over larger time scales,  
     e.g.~quasar energy injection can cause 
     a sort of starvation in later stages of galaxy's life \citep{Dubois2013a,Curtis2016}. 
     Our results constrain the time scale of quasar-mode feedback: 
      even if central black holes are at work  in the early universe, their effect (i.e., a deviation  
     of the SMF from the HMF high-mass end) is observed only at $z\leqslant3$. This means that 
     such a  quenching mechanism is effective on time scales larger than $2$\,Gyr, 
     likely after  multiple  outflow episodes.

    We can also compare stellar mass and halo mass functions at $\mathcal{M}<\mathcal{M}_\star$ 
    (or equivalently $\mathcal{M}_\mathrm{h}<\mathcal{M}_\mathrm{h}^\star$).  
     A similar exercise has been made in \citet{Song2016} up to $z\simeq7$, finding that the low-mass end 
    of their SMF has a slope similar to the HMF at $z\gtrsim7$. 
    In our SMF, the   low-mass end  becomes steeper, and slightly closer to the HMF,  
    already at $z>3.5$.    
    In terms of  Schechter  parametrisation, we find that 
     $\alpha$ ranges between $-1.4$ and $-1.2$ at $z\leqslant2$,  becomes $\simeq\!-1.7$ at 
    $2<z\leqslant3.5$ and eventually is $\lesssim-2$ at $z\simeq4\upto5$ (while  \citealp{Song2016} find 
    $\alpha<-1.9$ only beyond this redshift range). 
    Although the difference in $\alpha$, the low-mass end of both SMFs 
     diverges from the HMF, as expected  \citep[e.g.~from the simulations in][]{Costa2014} 
     if stellar feedback remains efficient at least up to $z\sim6$. 

     In addition, we compare to  the semi-analytical model of \citet{Garel2015,Garel2016}, specifically designed 
     to study Ly$\alpha$ emitters and LBGs.  
     At face value, the  slope of the SMF  predicted by 
     \citet{Garel2016}   is similar to ours (Fig.~\ref{fig_hmfevo}), suggesting that their model  
     may be an effective description of stellar feedback in the early universe
     A more detailed comparison with simulations is deferred to another paper of this series (Laigle et al., in preparation).

\section{Summary and conclusions}
\label{Conclusions}

 Relying on the latest photometric catalogue in the COSMOS field \citep[COSMOS2015,][]{Laigle2016}, 
 we have estimated the SMF of galaxies between $z=0$ and 6.   
 A deep NIR coverage from UltraVISTA, and associated  SPLASH images in MIR, allowed us to 
 probe the high-mass end of SMF as well as $\mathcal{M}\lesssim\mathcal{M}_\star$
  across the whole redshift range. 
 In particular our SMF reaches  $5\upto10\times10^{9}\,\mathcal{M}_\odot$ at $z\simeq5$, 
 an unprecedented mass regime at that redshift for a ground-based survey covering such a large area. 
 One of the reasons for this achievement is the panchromatic detection technique we applied, 
 based on a $\chi^2$ stacking of images from $z^{++}$ to $K_\mathrm{s}$.  
 Our stellar mass completeness limit $\mathcal{M}_\mathrm{lim}(z)$ 
  is almost 0.5\,dex smaller than what would result from 
  a single (e.g.~$K$) band selection, as done in previous analyses. 
 Deeper HST data available in the overlap with the CANDELS field 
 \citep[][]{Nayyeri2017}  have been used to confirm the absence 
 of significant biases in our final sample, cut at $[3.6]<25$.  
 
 A comparison with the literature showed the improvements in terms of statistics 
 with respect  to HST surveys at  high $z$ \citep[e.g.][]{Santini2012,Duncan2014}. 
 The large volume of COSMOS has allowed us to collect  rare massive galaxies, 
 although most of those at $z>2$ are   severely reddened by dust, 
 increasing the uncertainties in their $z_\mathrm{phot}$ and $\mathcal{M}$. 
 Comparing to the SMF of LBGs, where such massive and dusty galaxies may be totally missing, we 
 stressed the importance of high-quality MIR data to build mass-selected galaxy samples at $z>4$. 
  Now in its final phase, eight years after the end of the cryogenic mission, the Spitzer program 
 will soon be superseded by the James Webb Space Telescope (JWST),  
 to shed light on this peculiar galaxy population.

 Besides the emphasis on the early universe,  
 we remark that our SMF covers in a coherent way a time interval larger than previous work, 
 providing an overview of the last $\sim\!13$\,Gyr of galaxy evolution.  
 In addition we robustly selected (via the $NUVrJ$ diagram) active and passive galaxies, 
 deriving their SMF  up to $z=4$. At higher redshifts the contribution of the passive sample to 
 the SMF becomes negligible, and however the few passive galaxies we found need  
 follow-up observations to be confirmed. 
 Concerning the whole evolutionary  path from $z\sim6$ to $\sim\!0$, 
 our results are summarised in the following.  
 
 \begin{enumerate}
 
\item  Considering the growth of the SMF with cosmic time, 
we marked $z\simeq3$ as a key moment of galaxy evolution. At lower redshifts the best fit to 
the SMF is a double Schechter function. At $z>3$ the SMF shows a smoother profile  (especially at the knee of the function) 
and at $z>5$ it can be fit  also by a power-law function with a cut-off at 
$\sim\!3\times10^{11}\,\mathcal{M}_\odot$.  
To a first approximation the emergence of an additional component in the low-$z$ SMF is 
related to the assembly of the passive galaxy sample. However also the active SMF 
is fit by a double Schechter (a feature often ignored e.g.~in some phenomenological models). 
The fact that star formation starts fading in the core of galaxies 
already at $z\simeq2\upto2.5$ \citep{Tacchella2015,Tacchella2016} 
may be a hint that the change of SMF shape at $z\simeq3$ is related to such  
inside-out quenching. Further evidence  
is needed to verify this hypothesis.

 \vspace{10pt}
 \item  
 At $z \gtrsim3$  we also found a change in the relation between stellar mass and halo mass functions.  
 While at $z\lesssim2$  the SMF shape diverges from the HMF in both mass regimes (above and below $\mathcal{M}_\star$), 
 at higher redshifts the massive end of the SMF  has the same slope of the HMF. 
 This implies that the $\mathcal{M}/\mathcal{M}_\mathrm{h}$ ratio is roughly constant at $\mathcal{M}\gtrsim\mathcal{M}_\star$. 
 We interpret this trend as evidence of a reduced  quenching for massive galaxies at $z\geqslant3$, 
 such that the star formation process becomes dominated by simple baryonic cooling. 
 Thus, according to our observations, AGN do not trigger significant feedback  
 (either in radio or quasar mode) during the first $\sim2$\,Gyr after the Big Bang. 
 This is consistent to hydrodynamical simulations in which  
 AGN ejecta at high-$z$ can hardly stop (or prevent) star formation.

\vspace{10pt} 
 \item   There is a progressive flattening of the SMF low-mass end since $z\sim6$. 
 For massive galaxies ($\gtrsim5\times10^{10}\,\mathcal{M}_\odot$),  
  number density  
 increases  about one order of magnitude more than for $\sim\!10^{9}\,\mathcal{M}_\odot$ galaxies. 
 Fitting our data points with a Schechter function, we found $\alpha$ ranging  
 from $-2.11_{-0.13}^{+0.30}$ to $-1.47_{-0.02}^{+0.02}$, from $z\simeq5$ to $0.1$.  
 A similar slope at $z>4$ has been found by other authors \citep[e.g.][]{Song2016} and in 
 hydrodynamical simulations  \citep[]{Garel2016}. Other SMFs are shallower, with $\alpha\simeq-1.6$ 
  up to  $z\sim6\upto7$ \citep[see e.g.][]{Grazian2015}. 
  Such a disagreement is mainly due to the Eddington bias correction:  
  depending on  the characterisation of stellar mass errors made by the authors, 
  the effect of the bias correction on the Schechter function can vary significantly.

\end{enumerate}
 Our work shall gain additional momentum from the next wave of  
 UltraVISTA  (as the survey is still ongoing) and Spitzer images, 
 and also from the development of new tools for data analysis 
 \citep[e.g.][]{Masters2015,Speagle2016,Morrison2017}. 
 Spectro-photometric data from JWST will certainly be beneficial to 
 answer some of the questions raised here, 
 but to probe the massive end of high-$z$ SMFs the role of 
 large-area surveys like COSMOS2015 will remain fundamental.

 \begin{acknowledgements}
 The authors warmly thank the anonymous referee for her/his 
 constructive comments. 
The authors thank Shoubaneh Hemmati and Hooshang Nayyeri for providing us 
with the CANDELS Multiwavelength Catalog in the COSMOS field, and 
Andrea Grazian and Thibault Garel for sending their results in a convenient 
digitalised format. 
ID thanks Marta Volonteri, Jeremy Blaizot, Yohan Dubois, Andrea Grazian, Roberto Maiolino 
for very useful discussions. 
ID and OI acknowledge funding of the French Agence Nationale de la Recherche for the SAGACE project. 
CL acknowledges support from a Beecroft fellowship. 
ID acknowledges the European Union's Seventh Framework programme under
grant agreement 337595 (ERC Starting Grant, ``CoSMass'')
AF acknowledges support from the Swiss National Science Foundation. 
The COSMOS team in France acknowledges support from the Centre National d'\'{E}tudes Spatiales. 
\end{acknowledgements}

\bibliography{library}

\begin{appendix}
\section{Photometric redshifts in COSMOS2015}
\label{APP1}

Despite the $z_\mathrm{phot}$ accuracy 
we reached in L16, in the present work we  refined our SED fitting 
procedure with specific improvement for the high-$z$ analysis (see Sect.~\ref{SED fitting}).  
It is worth reminding that 
the photometry used here is exactly the same published in L16. 
   
\begin{figure}
 \includegraphics[width=0.9\columnwidth]{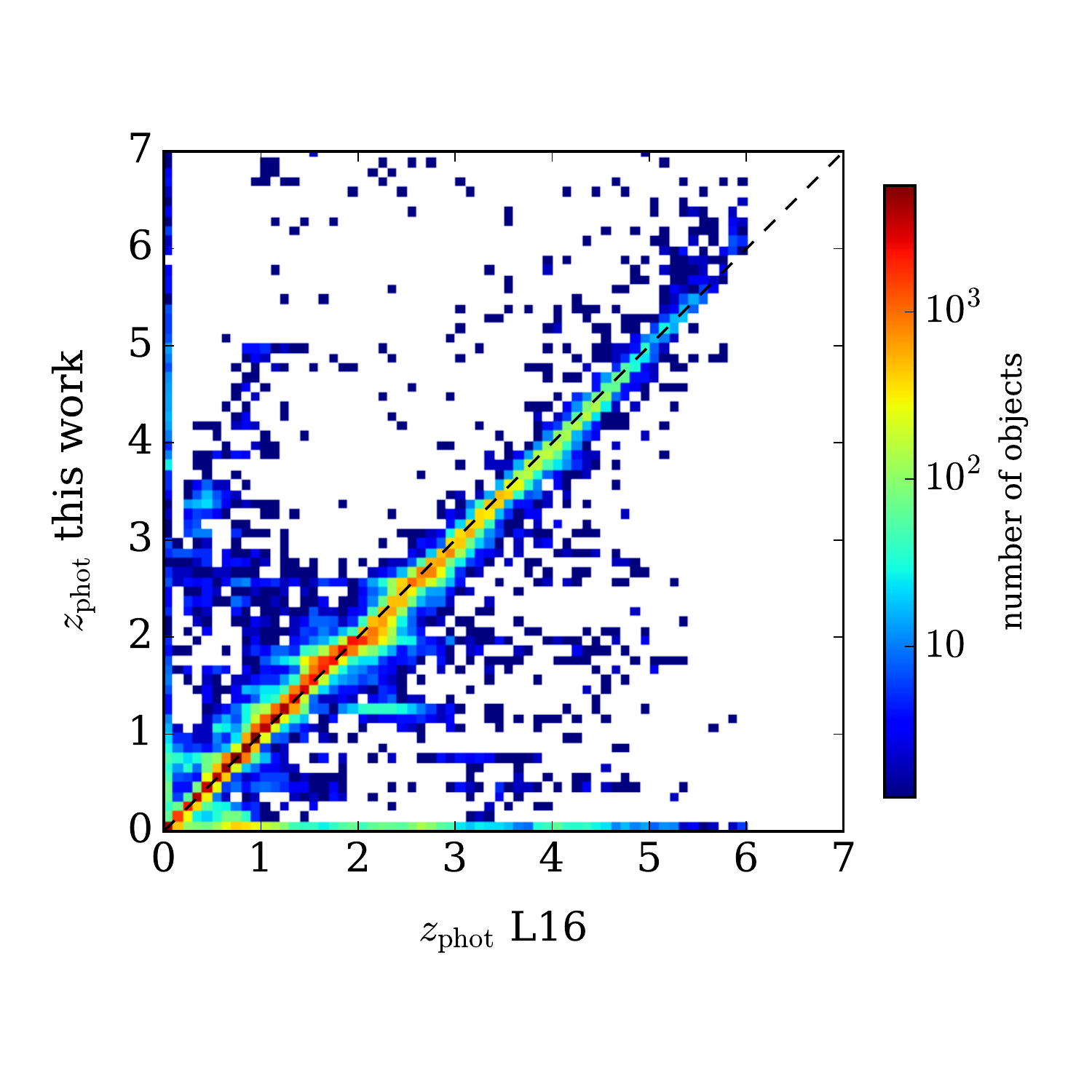}
 \caption{Official photometric redshifts  \citep[$z_\mathrm{phot,L16}$,][]{Laigle2016} of  COSMOS2015 
 compared to the new estimates from the present work. The sources have $[3.6]<25$ and are 
 selected in the UltraDeep  area. 
 Red-orange pixels include 90\% of the objects  (whose total number is 125\,578). A dashed line 
 is a reference for the 1:1 correspondence. 
  }
 \label{fig_L16-D16}
 \end{figure}

Figure~\ref{fig_L16-D16} shows the comparison between photometric 
redshifts computed in L16 ($z_\mathrm{phot,L16}$) 
and the new ones  (which 
we will name in the following simply $z_\mathrm{phot}$). 
The  comparison includes 125\,578 objects brighter than 25 in the IRAC $[3.6]$ channel, from the UltraDeep region of the COSMOS2015 catalogue. 
Excluding stars, we find that for 68\% of them the difference $\Delta z\equiv\vert z_\mathrm{phot,L16} - z_\mathrm{phot} \vert$ is smaller than 0.05 (and $\Delta z<1$ for 99\% of the galaxies). 
In addition to  10\,013 photometric objects already classified as stars in L16,  
we  identified further 11\,231 stars in the new SED fitting run 
(there are also 2\,116 galaxies that were previously labelled as stars). 

Despite the overall agreement, one can note in Fig.~\ref{fig_L16-D16} a subsample of objects that moved from $z_\mathrm{phot,L16}<1$ to $z_\mathrm{phot}\simeq3$. 
These sources, with the larger $E(B-V)$ range and the new SFHs we adopted, are now classified  as dusty galaxies at high redshift. 
The $B$ drop-out in their photometry can be interpreted either 
as a Balmer break (according to L16) or a strongly attenuated UV slope (in the new computation). 
Other groups of galaxies that changed significantly their redshift (e.g.~from $z\sim 4$ to $\sim\!1$) 
have no statistical impact on our analysis. 

The most interesting galaxies, which also represent a challenge for SED fitting techniques, 
are  those at $z_\mathrm{phot}>2.5$, more massive than $\log(\mathcal{M/M}_\odot)>10.5$, 
with a strong FIR re-emission ($>\!90\,\mu$Jy in MIPS $24\,\mu$m, $S/N>3$). 
They belong to the subsample of dusty galaxies:  more than 80\% of them have $E(B-V)>0.3$,
  and an uncertain photometric redshift as mentioned above 
\citep[see][for an insight on massive passive galaxies at high $z$]{Spitler2014,Casey2014a,Martis2016}. 
Although there are only 124 of them  in the UltraDeep region, lying in the exponential tail of the SMF they can 
produce a non-negligible offset if their $z_\mathrm{phot}$ is wrong. 
We emphasise that their $z_\mathrm{phot}$ distribution agrees with that of the $24\,\mu$m 
 emitters studied by \citet{Wang2016}  in CANDELS (i.e.~in both cases thare is 
 a sharp drop of  detections at $z\sim3$). 
  For twelve of these  ``MIPS-bright'' sources, for which we have also a 
  spectroscopic measurement, 
   $z_\mathrm{phot}$ is within  $\pm0.1(1+z_\mathrm{spec})$.  
 Most of the MIPS-bright galaxies are fit  by  templates with $\mathrm{SFR}>100\,\mathcal{M}_\odot\,\mathrm{yr}^{-1}$, 
  although in reality their  IR emission may be caused not only by heated dust, but also (at least partly) by an active galactic nucleus \citep[e.g.][]{Casey2014b,Marsan2015,Marsan2016}. 

\begin{figure}
 \includegraphics[width=0.9\columnwidth]{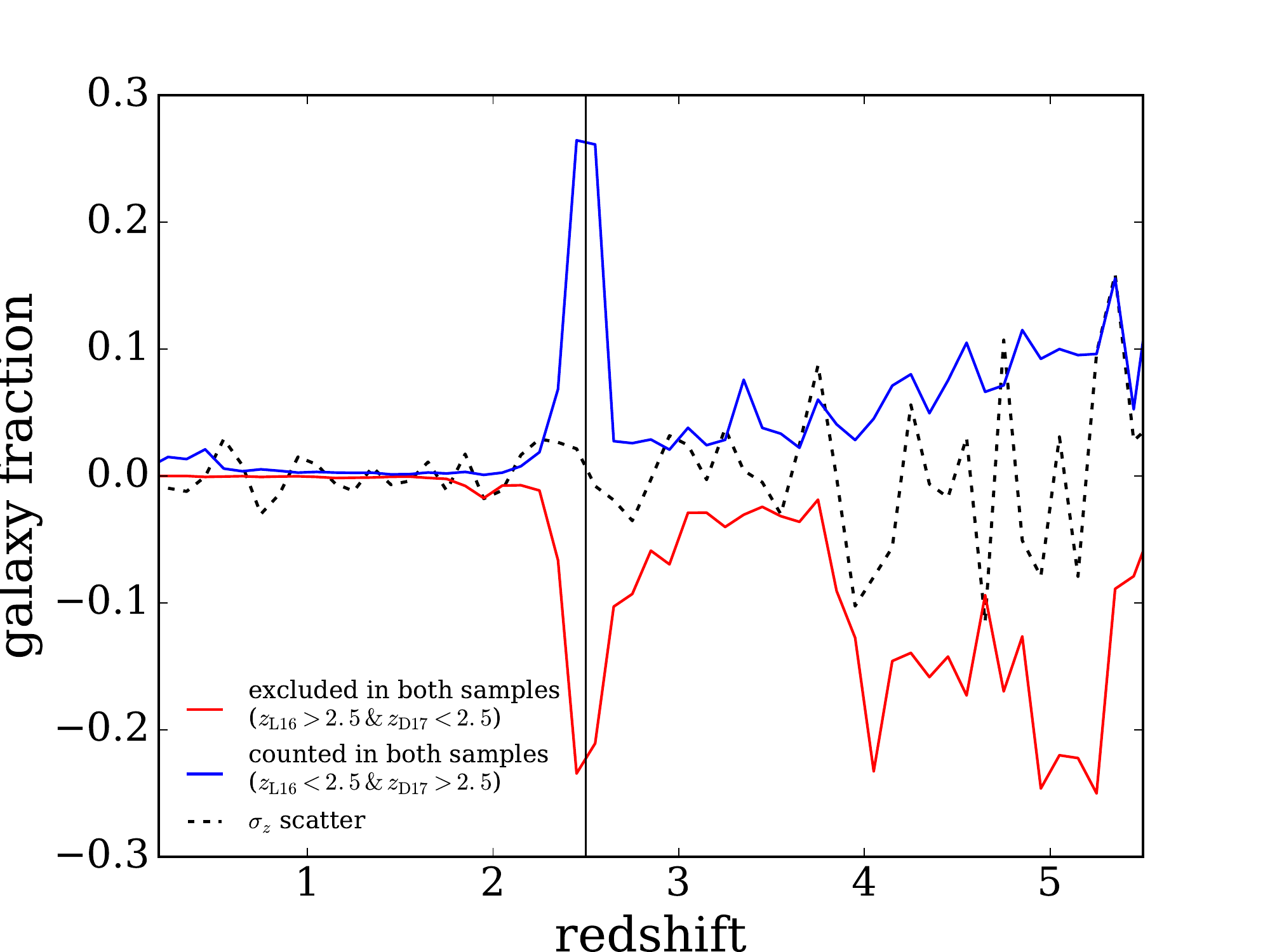}
 \caption{Fraction of galaxies neglected in the SMF computation (red line, negative values)  
 and those included both in the L16 sample at low $z$ and in the 
 revised computation at $z_\mathrm{phot}>2.5$ (blue line, positive values). 
 A vertical solid line separates the two samples. Variations due to $\sigma_z$ are shown for comparison (dotted line).  
 }
 \label{fig_zphot_inout}
 \end{figure}

  Figure~\ref{fig_zphot_inout} summarises the changes described above, showing as a function of $z$ the fraction of galaxies either excluded from the SMF computation (i.e., 
 $z_\mathrm{phot,L16}>2.5$ and  $z_\mathrm{phot}<2.5$) or counted twice 
 ($z_\mathrm{phot,L16}<2.5$ and  $z_\mathrm{phot}>2.5$). These galaxies  have a negligible impact at $z<2$. At $z\simeq2.5$, where we join the two samples, we observe the scatter due to random errors
  (i.e.~$z_\mathrm{phot,L16}$ and $z_\mathrm{phot}$ are compatible within  $1\sigma_z$); the number of galaxies  totally neglected and the number of ``duplicated'' galaxies balance out in this $z$-bin. 
At $z\gtrsim3$ the fraction of low-$z$ galaxies (according to L16) that have been relocated at $z>2.5$ in our analysis is $5\!-\!10\%$. There is a larger number of objects with  $z_\mathrm{phot,L16}>2.5$  that we ruled out as interlopers from our high-$z$ sample, as discussed above. 
   In Fig.~\ref{fig_zphot_inout} we also plot an estimate of the  scatter  due to $\sigma_z$, obtained recomputing $N(z)$ from our Monte Carlo simulation (Sect.~\ref{SED fitting 2-stellar mass}). 
 This comparison shows that changes related to the replacement of $z_\mathrm{phot,L16}$ with the new estimates are of the same order of magnitude of typical fluctuations due to $z_\mathrm{phot}$ statistical errors.

\section{Implications of a multi-band detection technique}
\label{APP3}

  To compute the SMF, first we have  to asses the minimum stellar mass ($\mathcal{M}_\mathrm{lim}$) 
  below which 
  the  sample incompleteness can impair our measurement. 
  We mentioned in Sect.~\ref{SED fitting 2-stellar mass limit}  two possibilities
   that are feasible in COSMOS2015: 
  a mass complete sample derived from a $K_\mathrm{s}$-band selection, 
  or one derived after a cut in $[3.6]$. 
  A selection in IRAC, rather than in a single NIR band, 
  is  motivated by the fact that our detections 
   are made by  combining images from multiple bands, some of them deeper than $K_\mathrm{s}$.

    We can directly compare  
    to \citet{Stefanon2015}, who work  with the UltraVISTA dataset of \citet[][]{Muzzin2013a}, 
    with the same limiting magnitude of COSMOS2015, but 
    in the shallower $\mathcal{A}_\mathrm{D}$ region.   Their source extraction is 
     similar to ours although they used  $K_\mathrm{s}$ only as a prior image. 
   By the inspection of  the IRAC residual maps, they find that 
   their $z_\mathrm{phot}>4$ sample increases by up to 38\% (depending 
   on the method used to estimate photometric redshifts). 
   Among the 408 sources they recover from the residual maps (in the whole UltraVISTA area) 
   48\% are naturally included in COSMOS2015 (M.~Stefanon, private communication). 
   \citet{Caputi2015} make another test by    
   comparing the IRAC sources detected in UltraVISTA DR1  to those from 
   DR2 in the UltraDeep 
   stripes, whose depth increased by 0.7\,mag.  Their work shows 
   the difference between  $\mathcal{A}_\mathrm{D}$ and $\mathcal{A}_\mathrm{UD}$ regions.  
   \citeauthor{Caputi2015} find  574 IR-bright 
   ($[4.5]<23$) sources detected in DR2 but not in DR1 (i.e., with $K_\mathrm{s}>24$).  
   We note that about 75\% of them are detected in $z^{++}$, confirming the convenience  of adding this band 
   in the $\chi^2$-stacked image not to suffer for such an incompleteness (see L16).

  \begin{figure}
  \includegraphics[width=0.9\columnwidth]{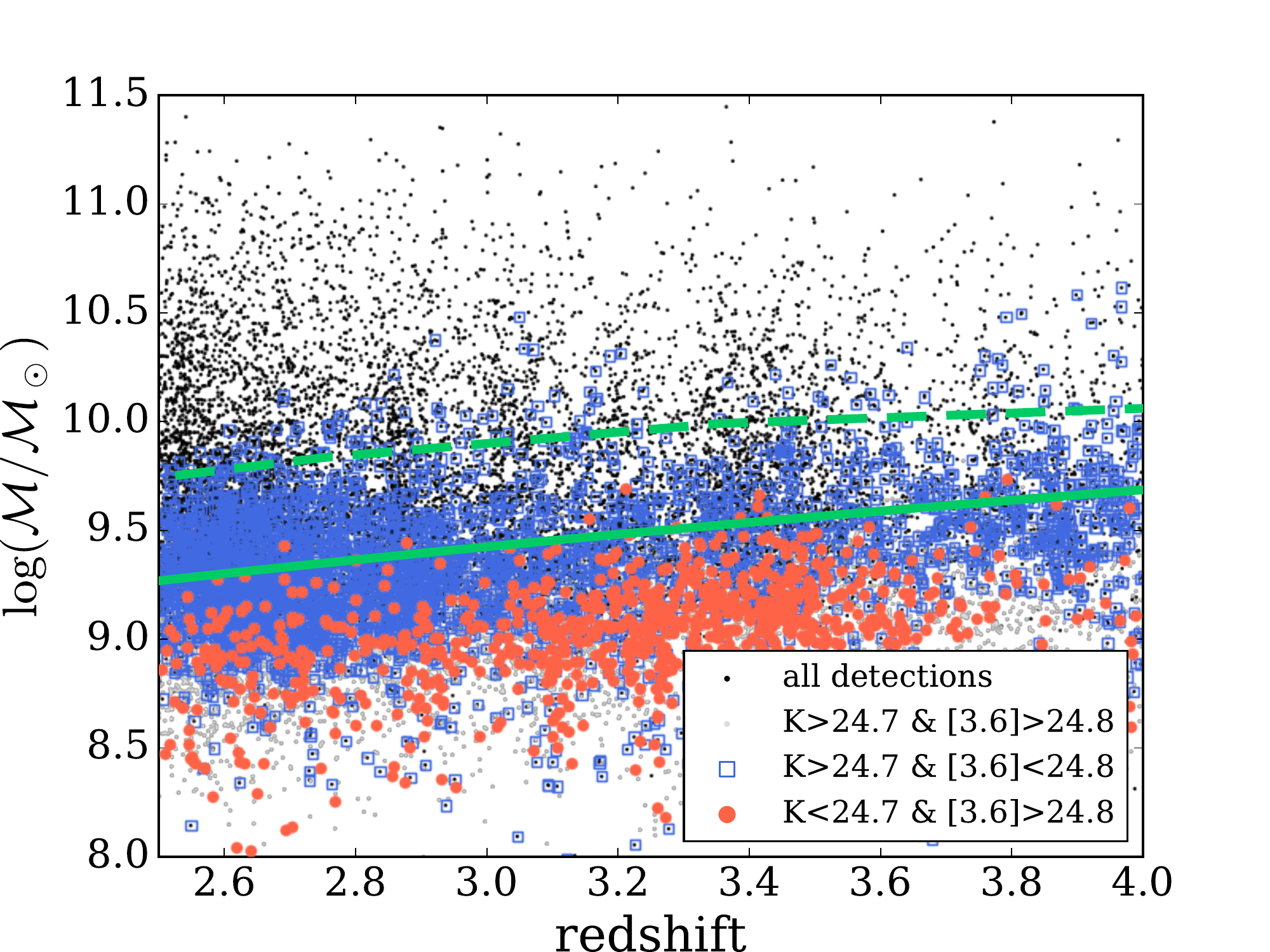}
  \caption{Galaxy stellar mass  as a function of redshift, in a range where a cut in either  $K_\mathrm{s}$ or  
  $[3.6]$ are applicable, in order to build a flux limited sample. 
  In the plot, small circles are all the COSMOS2015 galaxies 
   in $\mathcal{A}_\mathrm{UD}$, those detected in both $K_\mathrm{s}$ and $[3.6]$ (black dots) and the ones 
   that are not (grey dots).  
  Red filled circle (blue empty squares) show galaxies fainter than the $[3.6]$ ($K_\mathrm{s}$) magnitude limit, 
   but detected in $K_\mathrm{s}$ ($[3.6]$).  
   Green lines represent the stellar mass 
   completeness limits $\mathcal{M}_\mathrm{lim}(z)$ (see Sect.~\ref{SED fitting 2-stellar mass limit})
   resulting   from either the  $[3.6]$-selected sample (solid lines) or the $K_\mathrm{s}$-selected (dashed line). 
}
  \label{fig_miss_mass}
  \end{figure}

  As already emphasised, our  flux-limited sample with a cut at $[3.6]<25$ 
  results in a lower $\mathcal{M}_\mathrm{lim}(z)$ 
  with respect to $K_\mathrm{s}<24.7$. 
  In the $zYJHK_\mathrm{s}$ image we detect 17\,319 galaxies 
  between $z=2.5$ and 4  within $\mathcal{A}_\mathrm{UD}$. Among them, 954 (9\%) have $K_\mathrm{s}<24.7$ but 
   are not in the IRAC selected sample, being 
  fainter than $[3.6]_\mathrm{lim,UD}$. On the contrary, more than double (2\,182 objects) would be excluded 
  by the cut in $K_\mathrm{s}$, despite  their $[3.6]<[3.6]_\mathrm{lim,UD}$. 
  Besides the percentage of missing objects, we stress that galaxies 
  faint in IRAC  are on average 
  less massive than $10^{9.4}\,\mathcal{M}_\odot$,  while 
  a selection in $K_\mathrm{s}$ would remove more massive objects (see Fig.~\ref{fig_miss_mass}). 
  
  Why the two selections are so different in terms of stellar mass completeness? 
  First, the $(K_\mathrm{s}-[3.6])$ colour of low- and intermediate-mass galaxies at $2.5<z<4$ 
  is on average  red  (median $K_\mathrm{s}-[3.6]=0.2$) and tends to be  redder moving to higher masses. 
  For several galaxies the difference 
   is sufficient to include them among the $[3.6]$ detections but not in $K_\mathrm{s}$. 
   Another reason is that in such a faint regime the average 
   $S/N$  of $K_\mathrm{s}$ is lower than $[3.6]$, also because flux extraction in the latter is 
   improved by the $\chi^2$-stacking strategy. Therefore the $K_\mathrm{s}$ measurement for  
   low-mass galaxies  is more scattered, with higher chances to be far below the threshold we imposed.

  \section{COSMOS2015 stellar mass function at $0<z<4$}
    \label{APP2} 
   
 \begin{figure}
 \includegraphics[width=0.9\columnwidth]{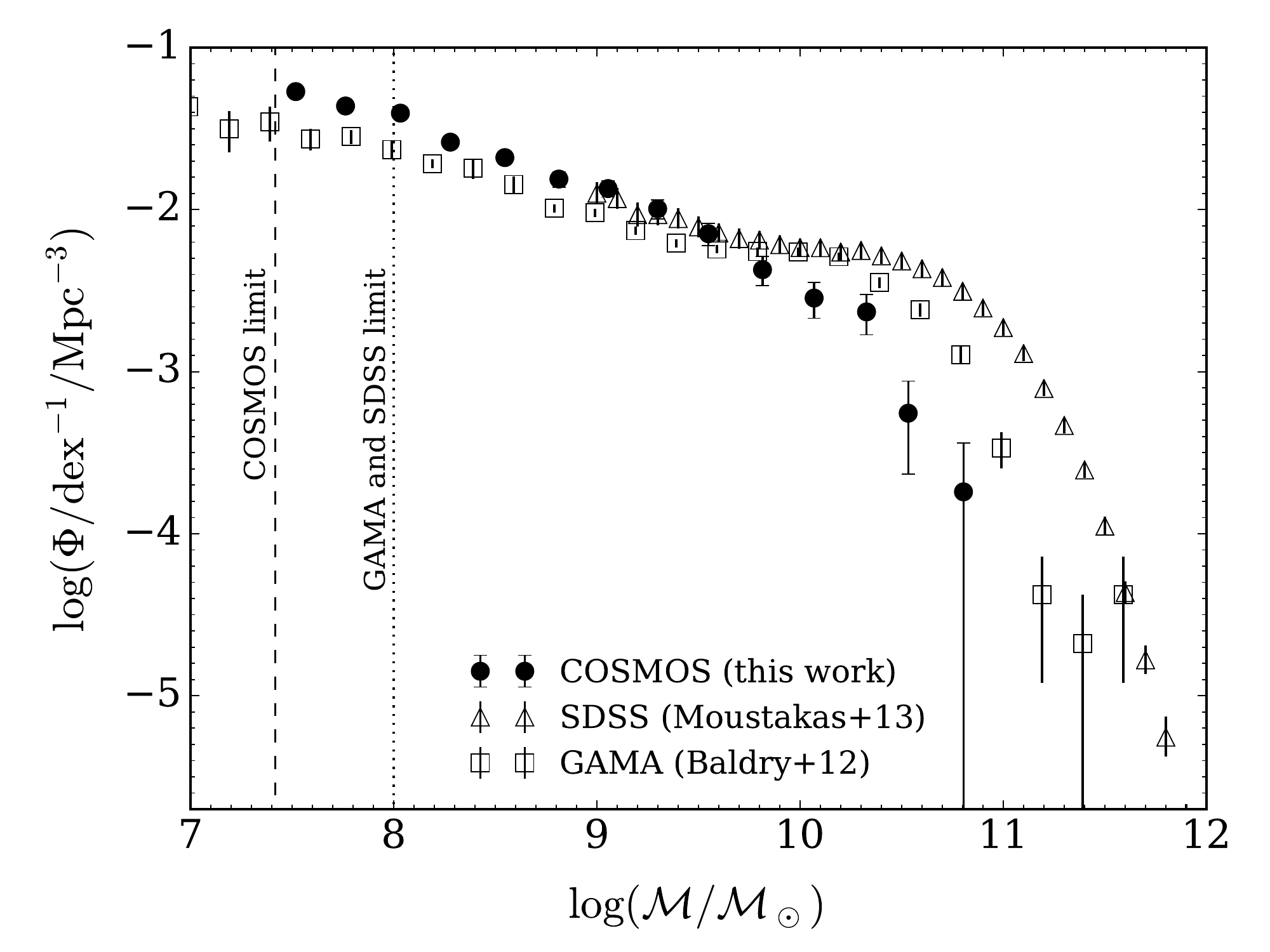}
 \caption{The galaxy SMF of COSMOS2015  (filled circles), SDSS \citep[][$0<z<0.2$, empty triangles]{Moustakas2013}, 
 GAMA \citep[][$0<z<0.06$, empty squares]{Baldry2012}. Vertical lines show the stellar mass limit of each survey.  
 A  $10^8\,\mathcal{M}_\odot$ limit is shown also for the SDSS survey,  considering the lowest value \citep{Li&White2009} 
 among the  ones used in  various studies. Other authors 
 (like \citeauthor{Moustakas2013})  adopt a more conservative threshold for their estimates ($\sim\!10^{9}\,\mathcal{M}_\odot$). }
 \label{fig_smf_z0}
 \end{figure}

  We show in this Appendix the SMF of galaxies at $z<3$ derived from the L16 
  SED fitting estimates, along to the 
  new results at higher redshifts (see Sect.~\ref{Results}). 
  The evolution of these SMFs has been already shown in Fig.~\ref{fig_smfevo} and \ref{fig_smfevo_typ}. 
  Here we plot the SMF in each $z$-bin separately. 
   
   The SMF at $0<z<0.15$ (median redshift $\langle z \rangle = 0.12$)  
   is presented in Fig.~\ref{fig_smf_z0}, together with 
   the SMF of the Sloan Digital Sky Survey \citep[SDSS, ][$\langle z \rangle=0.1$]{Moustakas2013} 
   and the Galaxy And Mass Assembly survey \citep[GAMA,][$z<0.06$]{Baldry2012}. 
   Our survey probes a small volume in the local universe, so the   high-mass end   
   is highly incomplete  (also considering the bias due to saturated sources). 
   On the other hand the deeper exposure of  COSMOS allows us to probe the SMF  at lower masses  than 
   SDSS and GAMA: while the sample of \citet{Baldry2012} is complete above $10^8\,\mathcal{M}_\odot$ 
   \citep[the limit is the same for SDSS, see][]{Li&White2009} 
   our SMF extends down to $\sim\!2.5\times10^{7}\,\mathcal{M}_\odot$.  
   Fitting a Schechter function to the SMF we find that the  low-mass end has a slope $\alpha=-1.47\pm0.02$ 
   (note that \citealp{Baldry2012} find $-1.47\pm0.05$).

\begin{figure*}
 \includegraphics[width=0.96\textwidth]{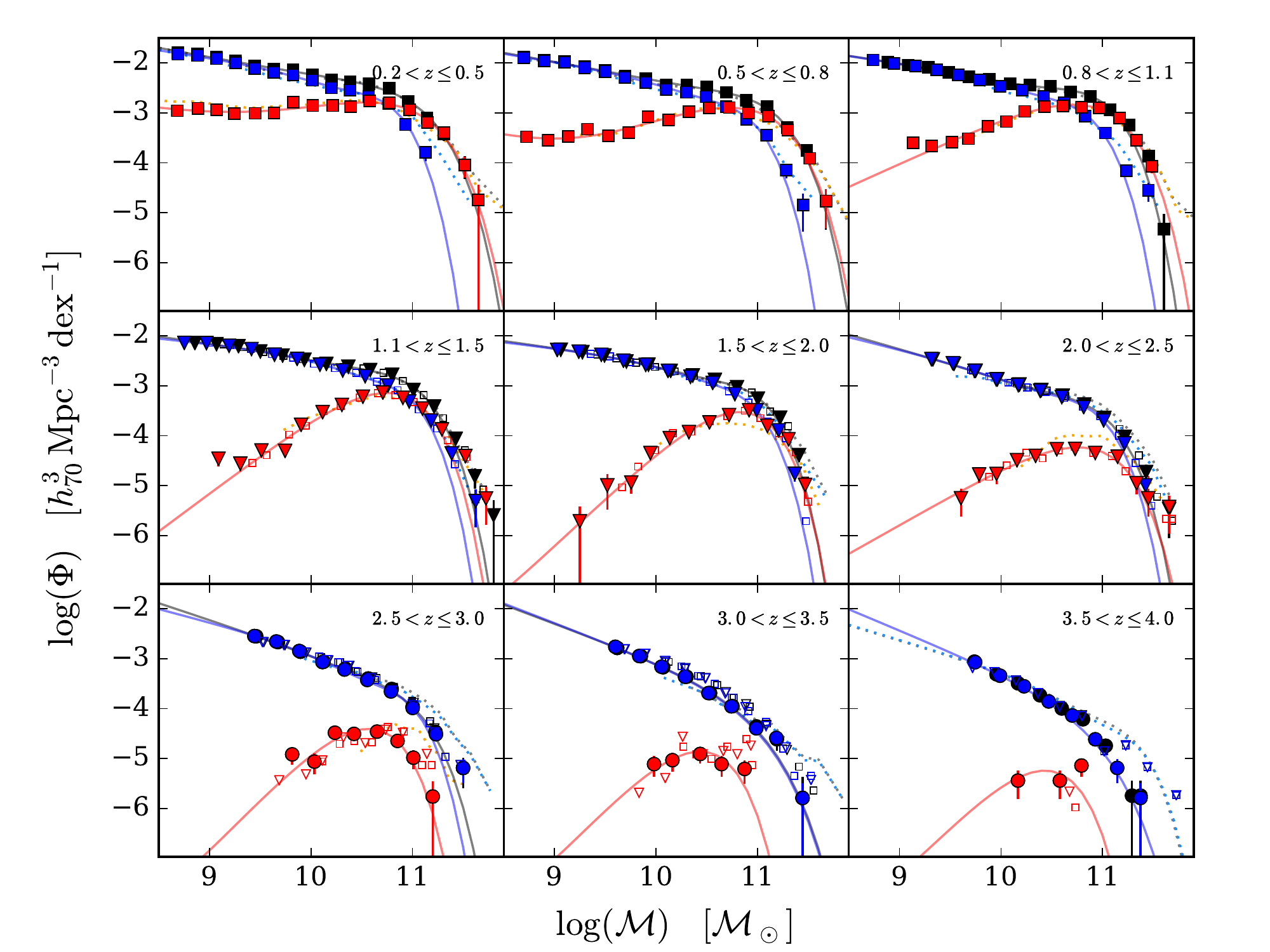}
 \caption{Three different $1/V_\mathrm{max}$ estimates of the  COSMOS2015 galaxy SMF, from $z=0.2$ to 4.
 The estimates are derived from: a $K_\mathrm{s}$-selected sample 
 using the official SED fitting from L16 (squares); a $[3.6]$-selected sample 
 that also relies on $z_\mathrm{phot,L16}$ and $\mathcal{M}$ of L16 (triangles); the  high-$z$ $[3.6]$-selected sample  
 presented in this paper (circles). Active and passive galaxies, classified  in Sect.~\ref{SED fitting 2-Galaxy type} 
 by means of $(NUV-r)$ and $(r-J)$ colours, are shown with blue and red symbols respectively. 
 The SMFs used in Sect.~\ref{Discussion} to discuss $\sim\!10$ billion years of galaxy evolution are shown 
 with filled symbols, while smaller empty symbols are used in the $z$-bins where the reference SMF changed. 
 In each bin, Schechter functions (solid lines, same colours for total, active, and passive galaxies) fit the 
 data points of the reference sample (i.e., the filled symbols). 
 We also plot  the Schechter functions fitting 
 the SMF of UltraVISTA DR1 galaxies \citep{Ilbert2013} with dotted lines. All the fits are corrected for the Eddington bias.  }
 \label{fig_mf_3x3}
 \end{figure*}

   In Fig.~\ref{fig_mf_3x3} we show   the SMF of both 
   $K_\mathrm{s}$- and $[3.6]$-selected galaxies from $z=0.2$ to 4. 
  The various estimates are overall in good agreement. In particular  at $2.5<z\leqslant3$ the L16 estimate 
    joins well with the  SMF derived from the new SED fitting run.    
   Moving to higher redshifts there are small discrepancies (always $\leqslant\!0.2$\,dex)
    in the comoving number density. The overestimate  
    (observed mainly in the high-mass end) of the L16 galaxy SMF is likely due 
    to the systematics (interlopers, different set of template) discussed in Appendix \ref{APP3}. 
    However, it is difficult  to pinpoint a single cause for such a difference as 
    several effects may act in combination. For example  age-metallicity degeneracy, 
   after removing subsolar templates, forces  to choose   younger galaxy ages, 
   underestimating the  $\mathcal{M}/L$ ratio  \citep[see][]{Bell&deJong2001}; 
   however the offset goes in the opposite direction when dust attenuation adds further degeneracy 
   \citep[see][]{Davidzon2013}. 
 
   In Fig.~\ref{fig_mf_3x3} we also show the SMF of the UltraVISTA DR1 galaxies \citep{Ilbert2013}. 
   Their estimates are in good agreement with COSMOS2015 
   up to $z=2$. 
   At higher redshifts we observe the same systematic effect pointed out in \citet{Faisst2016b}: 
   without the SPLASH coverage, many high-$z$ objects in UltraVISTA DR1 were not provided with 
   an accurate MIR photometry (if not at all), and their mass was underestimated \citep[see Fig.~4 of][]{Faisst2016b}. 

\end{appendix}

\end{document}